\documentclass[10pt]{iopart}

\usepackage{iopams}
\expandafter\let\csname equation*\endcsname\relax
\expandafter\let\csname endequation*\endcsname\relax
\usepackage{amsmath} 
\usepackage{amssymb,bbold,color}  
\usepackage{graphicx}
\usepackage{multirow}
\usepackage{pifont}

\let\originalleft\left
\let\originalright\right
\renewcommand{\left}{\mathopen{}\mathclose\bgroup\originalleft}
\renewcommand{\right}{\aftergroup\egroup\originalright}

\newcommand{\bra}[1]{\ensuremath{\left\langle #1\right|}}
\newcommand{\ket}[1]{\ensuremath{\left|#1\right\rangle}}

\providecommand{\abs}[1]{\left\lvert#1\right\rvert}
\providecommand{\norm}[1]{\lVert#1\rVert}

\providecommand{\tr}{{\rm Tr}}
\renewcommand{\phi}{\varphi}

\begin{document}

\title[Signatures of Many-Particle Interference]{Signatures of Many-Particle Interference}

\author{Mattia Walschaers}

\address{Laboratoire Kastler Brossel, Sorbonne Universit\'e, CNRS, ENS-PSL Research University, Coll\`ege de France}

\ead{mattia.walschaers@lkb.upmc.fr}
\vspace{10pt}
\begin{indented}
\item[]Septembre 2019
\end{indented}

\begin{abstract}
Quantum systems with many constituents give rise to a range of conceptual, analytical and computational challenges, hence, the label ``complex systems''. In the first place, one can think of interactions, described by a many-body Hamiltonian, as the source of such complexity. However, it has gradually become clear that, even in absence of interactions, many-body systems are more than just the sum of their parts. This feature is due to many-body interference.

One of the most well-known interference phenomena is the Hong-Ou-Mandel effect, where total destructive interference is observed for a pair of (non-interacting) identical photons. 
This two-photon interference effect can be generalised to systems of many particles which can be either fermionic or bosonic. The resulting many-particle interference goes beyond quantum statistical effects that are contained in the Bose-Einstein or Fermi-Dirac distributions, and is dynamical in nature.

This Tutorial will introduce the mathematical framework for describing systems of identical particles, and explain the notion of indistinguishability. We will then focus our attention on dynamical systems of free particles and formally introduce the concept of many-particle interference. Its impact on many-particle transition probabilities is computationally challenging to evaluate, and it becomes rapidly intractable for systems with large numbers of identical particles. Hence, this Tutorial will build up towards alternative, more efficient methods for observing signatures of many-particle interference. A first type of signatures relies on the detection of a highly sensitive -but also highly fragile- processes of total destructive interference that occurs in interferometers with a high degree of symmetry. A second class of signatures is based on the statistical features that arise when we study the typical behaviour of correlations between a small number of the interferometer's output ports. We will ultimately show how these statistical signatures of many-particle interference lead us to a statistical version of the Hong-Ou-Mandel effect.

The work presented in this Tutorial was one of the four shortlisted finalists of the 2018 DPG SAMOP dissertation prize.
\end{abstract}

\submitto{\JPB}
%\maketitle

% Hello, @QuantPhComments!

\newpage
\section{From Interference to Boson Sampling}\label{sec:intro}

Interference has been a corner stone of quantum physics since its very beginning, as can be read in Dirac's {\em The Principles of Quantum Mechanics} \cite{Dirac-1930} from 1930, where an entire section is devoted to interference of photons. In this book, Dirac writes that {\em ``Interference between two different photons never occurs.''}, a claim which would prove to be controversial. Several decades later, we find that Dirac's claim is invoked as a motivation for a series of works \cite{Mandel-1963,mandel-1964,Pfleegor-1967,mandel-1983} to show the interference between independent laser beams. 

The advent of parametric downconversion \cite{Klyshko-1970, Burnham:1970aa} enabled the first interference experiments with actual photon pairs \cite{Grangier:1986aa,hong_measurement_1987,Shih:1988aa}. In the light of this Tutorial, the most notable of these experiments is the one by Hong, Ou, and Mandel, where a pair of photons, entering a balanced beamsplitter from different input ports, are shown to bunch together in one of the two output ports. In the ideal case, this interference effect occurs with an interferometric visibility of 100$\%$, which greatly surpasses the 50$\%$ limit \cite{ou-1988} that is attainable with classical light (i.e.~coherent states). However, it should be noted that this $50\%$ limit has recently been disputed \cite{Sadana-2018}.

The observation of the Hong-Ou-Mandel effect sparked an interest in multi-photon interferometry \cite{belinskii_interference_1992,Banaszek:1997aa,tichy_four-photon_2011,Pan:2012aa,tillmann_generalized_2015}, which ultimately led to interference experiments with multiple photons \cite{ra_nonmonotonic_2013,Spagnolo:2013aa,PhysRevLett.118.153603,Giordani-2018,Agresti:2019aa,Zhong:2018aa}. The destructive Hong-Ou-Mandel interference has been generalised to arbitrarily many particles in multiport interferometers by the derivation of suppression laws \cite{crespi-suppression-2016,PhysRevA.91.013811,2058-9565-2-1-015003,tichy_many-particle_2012,tichy_zero-transmission_2010,10.1088/1367-2630/aaad92,dittel-2017,dittel2-2017}, and by the identification of more general principles of bunching \cite{ou_photon_1999,carolan_experimental_2014,shchesnovich_universality_2015}. As such, many-particle interference has become a prominent branch of research in quantum optics.

From the point of view of quantum computation, many-particle interference has attracted growing attention due to two milestone protocols: First, the Knill-Laflamme-Milburn scheme \cite{knill_scheme_2001} was developed in 2001, which uses a controllable number of photons, passive linear optics, and post-selection to construct a universal quantum computer. Due to the limited number of photons that can be generated with modern-day technology, the experimental realisation has remained limited to modestly sized quantum computation with only a few qubits \cite{Qiang:2018aa}. A second spike of interest came with the proposal of Boson Sampling by Aaronson and Arkhipov in 2010 \cite{aaronson_computational_2013}, and the associated line of research towards quantum supremacy. In this case, the goal is not to perform an actual computation, but rather to implement a task that cannot be simulated by a classical computer (in polynomial time). It turns out that gathering detector clicks from the output ports of a randomly chosen multiport interferometer is such a task, once we inject sufficiently many photons. This work has then sprouted a wide range of proof-of-principle experiments \cite{broome_photonic_2013,crespi_integrated_2013,spring_boson_2013,tillmann_experimental_2013,Wang:2017aa}.\\

Developments in photonics, as described above, are highly promising in the sense that photons are easy to manipulate while keeping decoherence effects at bay. However, this platform currently suffers from a significant limitation: it is extremely difficult to generate quantum states with a high number of photons. This issue has sparked interest in other directions, such as using Gaussian quantum states of light instead of single photons at the input of the interferometer \cite{lund_boson_2014,bentivegna_experimental_2015, PhysRevLett.119.170501}. Other approaches to achieve many-particle quantum interference have diverted away from light all together. As will be discussed throughout this Tutorial, the true nature of many-boson interference lies within the bosonic commutation relations, which can just as well be achieved with bosonic atoms \cite{urbina_multiparticle_2014,Lopes:2015aa,Islam:2015aa,Roos:2017aa}.

Since many-particle interference is tied to the inherent indistinguishable of photons and atoms, it should come as no surprise that many-particle interference effects can also be unveiled for fermionic particles \cite{tichy_many-particle_2012,tichy_zero-transmission_2010,10.1088/1367-2630/aaad92,dittel-2017,dittel2-2017}. The most natural candidates for implementing such interference effects are fermionic atoms \cite{Rom:2006aa,Henny:1999aa,Jeltes:2007aa,Kiesel:2002aa,Preiss:2019aa}, but recent developments in electron quantum optics also bring electrons into the picture \cite{Bocquillon:2012aa,Dubois:2013aa}. Finally, one may also manipulate the internal degrees of freedom (DOF) in photons to mimic fermionic interference effects \cite{sansoni_two-particle_2012}. We will see that many-boson and many-fermion interference are actually similar in many ways, even though the latter is in fact much more tractable to simulate on a classical computer \cite{aaronson_computational_2013}.

Using matter rather than light is often unpractical due to the unavoidable decoherence effects, but there is also a more fundamental difference: matter interacts. Much as interactions between cold atoms can be tuned, or electrons can be manipulated, one cannot get around the fact that these particles can interact with each other. This possibility opens many new questions about the interplay between interactions and many-particle interference, which are currently just beginning to be explored in a systematic way \cite{1367-2630-19-12-125015,Brunner:2018aa}.\\

In this Tutorial, we focus on a specific question related to all the above many-particle interference effects: how can they be observed? Given that bosonic many-particle interference is hard to simulate, it is also hard to know what exactly it should look like in a laboratory. How can we, for example, distinguish genuine 50-particle quantum interference from partial interference between 25 pairs of particles upon injection in an arbitrary interferometer? These and related questions have formalised as the so-called validation problem for Boson Sampling \cite{aolita_reliable_2015,gogolin_boson-sampling_2013,tichy_stringent_2014,spagnolo_experimental_2014,aaronson_bosonsampling_2014,Flamini:2019aa,Agresti:2019aa}. It was recently proven that an absolute certification of a sampling problem such as Boson Sampling is generally impossible \cite{Hangleiter:2019aa}, which implies that reasonable experimental assumptions are important. In particular, it is feasible to identify certain signatures of many-particle interference and to compare them to other sampling models, such as for example fully distinguishable particles, which serve to model possible errors in experiments. Such a comparison provides a way to falsify potential Boson Samplers. This Tutorial will build towards explaining two specific types of signatures of many-particle interference: general suppression laws \cite{dittel-2017,dittel2-2017}, and statistical signatures \cite{walschaers_statistical_2014,Giordani-2018}. These signatures provide accessible benchmarks that can be used to validate Boson Sampling, in the sense that they can exclude a range of possible alternative models (such models can either be experimental errors or known ways of fabricating sampling data that mimic certain features of bosonic interference). The crucial difference with certification is that validation does not guarantee that the data were generated by a genuine Boson Sampler, but it attests that the data are more likely to be generated by a boson sampler than by a certain set of alternatives.\\

The Tutorial aims to provide the broader theoretical framework of many-particle systems, and many-particle interference in particular. The goal is to provide a detailed look under the hood of this field of research, and elucidate the mathematical framework that underpins these developments. In Section \ref{sec:formalism}, we start out by introducing the mathematical frameworks of first and second quantisation with the ultimate goal of introducing multimode Fock spaces. Near the end of this mathematical journey, we arrive at the concept of {\em distinguishability} in Section \ref{sec:Distinguish}, one of the most crucial notions in this Tutorial. The appearance of distinguishability in a universe built up out of identical particles is often considered confusing. In this section, we aim to clarify it by introducing the notions of internal and external DOF, and by emphasising the importance of the measurements that are performed on particles. In Section \ref{eq:ManyPartInt}, we use the developed mathematical framework to introduce the concept of many-particle interference. We show how this phenomenon arises from the indistinguishability of particles, and how it degrades when particles gradually become more distinguishable. Internal DOF can provide and implicit type of ``which-particle information'', the effect of which is similar to that of ``which-way information'' in single-particle interference experiments. We then devote Section \ref{sec:Signature} to exploring observable signatures of many-particle interference, which ultimately lead to a statistical analog for the Hong-Ou-Mandel distinguishability transition for many particles in multimode interferometers.
 
\section{Many-particle systems}\label{sec:formalism}

Any quantum particle --be it a boson or a fermion-- can be described by its {\em wave function} $\ket{\psi}$, which is an element of the Hilbert space ${\cal H}$. This Hilbert space encompasses all the DOF of the particle, which can range from the position in space to its internal DOF such as spin or polarisation. ${\cal H}$ allows to define the algebra of observables ${\cal B}({\cal H}),$ i.e.~the set of bounded operators on ${\cal H}$. For any {\em physical observable} $O \in {\cal B}({\cal H})$, we demand that $O^{\dag} = O$ to ensure that the measurement statistics is real. With these definition, we can connect the theory to measurements of the observable $O$, because the $k$th statistical moment $m_k$ of these measurements is given by $m_k = \bra{\psi}O^k\ket{\psi}$.

The Hilbert space ${\cal H}$ provides us with a sufficient mathematical framework to describe a single particle. When a second particle is added to the game, one could naively be tempted to combine the DOF of both particles by forming the tensor product ${\cal H} \otimes {\cal H}$. This approach is valid when the particles are distinct (e.g.~when we are considering a photon and an electron), but it fails to take into account a particularity of quantum physics: quantum particles can be identical. Indeed, nature has made sure that two electrons with the same internal DOF are really the same in every possible sense. We will now describe the mathematical framework that was developed to deal with systems that comprise many of such identical particles.

This section combines elements from various textbooks in mathematical physics \cite{alicki_field-theoretical_2010,alicki_quantum_2001,bratteli_operator_1987,bratteli_operator_1997,petz_invitation_1990}.
 
\subsection{Towards Fock space}\label{sec:ToFock}

In this Tutorial, we will mainly restrict ourselves to systems which can effectively be described by a finite number of DOF. This avoids the need to dwell in the more complicated mathematical frameworks of C$^*$-algebras and their representations, which is commonly used in quantum statistical mechanics \cite{bratteli_operator_1987,bratteli_operator_1997,petz_invitation_1990}. The mathematical framework is based on the structure of {\em Fock space}, which we gradually introduce here. In Section \ref{sec:twoPart}, we review the well-known two-particle case, which serves as a basis for the generalisation to and arbitrary number of particles in Section \ref{sec:NPart}. These results can then, in turn, be generalised in Section \ref{sec:Fock} to spaces where the number of particles is not fixed.

\subsubsection{Two identical particles}\label{sec:twoPart}

Let us start by considering two identical particles with wave functions $\ket{\psi_1}, \ket{\psi_2} \in {\cal H}$. Note that ${\cal H}$ describes the DOF of each individual particle, and, thus, we will refer to is as the {\em single-particle Hilbert space}. Because both particles are identical, we cannot simply combine both wave functions in a normal tensor product (see \ref{sec:appSumProd} for more on tensor products). Physically, any measurement of the system should remain completely unchanged when all the DOF of each particle are swapped. In mathematical terms, this implies that the system is invariant under the permutation of particles. Such a permutation can be implemented by the unitary permutation operator $P$, which acts as 
\begin{equation}\label{eq:twoParticleSwap}
P\, \ket{\psi_1} \otimes \ket{\psi_2} =  \ket{\psi_2} \otimes \ket{\psi_1}.
\end{equation}
The required invariance of experimental measurements under such transformations must be encrypted in the two-particle quantum state. Hence, the two-particle wave function $\ket{\Psi} \in {\cal H}\otimes {\cal H}$ must fulfil the following property:
\begin{equation}\label{eq:twoParticle}
P \ket{\Psi} = e^{i \theta} \ket{\Psi},
\end{equation}
where $\theta \in \mathbb{R}$ is a global phase (note that global phases of the quantum state have no impact on quantum measurements). What we get from (\ref{eq:twoParticle}) is actually an eigenvalue equation. Indeed, the two-particle wave functions are actually the eigenvectors of the permutation operator $P$.

The eigenvalue (\ref{eq:twoParticle}) may remind some physicists of reflection symmetries, where $P$ is a type of parity operator. Indeed, we can simply observe that $P^2 = \mathbb{1}$, which directly leads to the condition that $e^{2i\theta} = 1$. In other words, we find that, either $\theta = 0$, or $\theta = \pi$. This leads us to two big classes of two-particle states:
\begin{align}\label{eq:PsiBPsiF}
P \ket{\Psi_B} = \ket{\Psi_B}, \quad \text{ and } \quad P \ket{\Psi_F} = - \ket{\Psi_F},
\end{align}
where $\ket{\Psi_B}$ are the {\em bosons}, and $\ket{\Psi_F}$ are the {\em fermions}. One can then use (\ref{eq:twoParticleSwap}) to obtain state vectors that satisfy (\ref{eq:PsiBPsiF}) to describe a state of two identical particles with wave functions $\ket{\psi_1}, \ket{\psi_2} \in {\cal H}$
\begin{align}
&\ket{\Psi_B} = \ket{\psi_1} \vee \ket{\psi_2} \equiv \frac{1}{\sqrt{2}} \big( \ket{\psi_1} \otimes \ket{\psi_2} +  \ket{\psi_2} \otimes \ket{\psi_1} \big),\label{eq:twoBoson}\\
&\ket{\Psi_F} =\ket{\psi_1} \wedge \ket{\psi_2} \equiv \frac{1}{\sqrt{2}} \big( \ket{\psi_1} \otimes \ket{\psi_2} -  \ket{\psi_2} \otimes \ket{\psi_1}, \big).\label{eq:twoFermions}
\end{align}
 where (\ref{eq:twoBoson}) holds when the particles are identical bosons, and (\ref{eq:twoFermions}) when they are identical fermions. A direct consequence of the anti-symmertrisation for fermions is {\em Pauli's exclusion principle}: two fermions cannot occupy the same single-particle wave function, since $\ket{\psi} \wedge \ket{\psi} = 0$.

This also implies that the actual Hilbert space of a two-boson or a two-fermion system is smaller than ${\cal H}\otimes {\cal H}$. All the relevant physics for such systems can actually be described with the Hilbert spaces
\begin{align}
&{\cal H}^{(2)}_{B} \equiv {\rm span}\{\ket{\psi_1}\vee \ket{\psi_2} \mid \psi_1, \psi_2 \in {\cal H}\},\\
&{\cal H}^{(2)}_{F} \equiv {\rm span}\{\ket{\psi_1}\wedge \ket{\psi_2} \mid \psi_1, \psi_2 \in {\cal H}\},
\end{align}
for bosons and fermions, respectively. The ``{\rm span}'' refers to the space generated by all linear combinations of elements within the set. Let us remark that 
\begin{equation}{\cal H}^{(2)}_{B} \oplus {\cal H}^{(2)}_{F} \cong {\cal H}\otimes {\cal H},\label{eq:HilbIdentityTwo},\end{equation}
where ``$\cong$'' indicates an isomorphism, meaning that both spaces have essentially the same structure. This particular isomorphism implies that we can obtain the full space ${\cal H}\otimes {\cal H}$ by combining the fermionic and bosonic subspaces.

Note, finally, that when we consider an arbitrary observable $O$ on such a two-particle space, we can implement the permutation of the particles in the Heisenberg picture: $O \mapsto P^{\dag}OP$. The demand that the physics be independent under such permutations is verified by evaluating the $k$th moment
\begin{align}
&\bra{\Psi_B}(P^{\dag}OP)^k\ket{\Psi_B} =  \bra{\Psi_B}P^{\dag}O^kP\ket{\Psi_B} =  \bra{\Psi_B}O^k\ket{\Psi_B},\label{eq:measureTwoB}\\
&\bra{\Psi_F}(P^{\dag}OP)^k\ket{\Psi_F} =  \bra{\Psi_F}P^{\dag}O^kP\ket{\Psi_F} =  (-1)^2 \bra{\Psi_F}O^k\ket{\Psi_F} = \bra{\Psi_F}O^k\ket{\Psi_F}.\label{eq:measureTwoF}
\end{align}
Because this holds for every $k \in \mathbb{N},$ we can, indeed, conclude that the measurement statistics is independent under the permutation of paricles.

\subsubsection{$n$-particle space}\label{sec:NPart}

When we consider $n$ identical particles with wave functions $\psi_1, \dots, \psi_n \in {\cal H}$, the situation becomes considerably more complicated. At the root of this complication lies the fact that there are many different ways to permute $n > 2$ particles, whereas for the case $n=2$ there is only one non-trivial permutation -- given by (\ref{eq:twoParticleSwap}).

In the $n$-particle scenario, we have to consider all possible permutations, and, hence, cover all possible $\sigma \in S_n$ (where $S_n$ represents the symmetry group). We define the permutation operator $P_{\sigma}$, which implements the permutation $\sigma$ as
\begin{equation}\label{eq:nPartPerm}
P_{\sigma} \ket{\psi_1}\otimes \dots \otimes \ket{\psi_n} = \ket{\psi_{\sigma(1)}}\otimes \dots \otimes \ket{\psi_{\sigma(n)}}. 
\end{equation}
The fundamental demand is the same as in Section \ref{sec:twoPart}: the statistics of measurements must remain the same upon the permutation of particles. In analogy with the two-particle case, this  suggests that the $n$-particle state $\Psi \in {\cal H}^{\otimes n}$ must fulfil the criterion
\begin{equation}\label{eq:permute}
P_{\sigma} \ket{\Psi} = e^{i \theta_{\sigma}}\ket{\Psi},
\end{equation}
where the phase $\theta_{\sigma}$ can vary with $\sigma$. However, equation (\ref{eq:permute}) must hold for all possible choices of $\sigma \in S_n$. Therefore, the $n$-particle wave function $\ket{\Psi}$ is now an eigenvector of all possible permutation operators $P_{\sigma}$.

To proceed, some insight in the eigenvectors of the permutation operators of the type (\ref{eq:nPartPerm}) is required. The study of such operators is narrowly related to the representation theory of groups. The interested reader is invited to delve into the literature \cite{hamermesh_group_1989,fulton_young_1997} on the Schur-Weyl duality and Young tableaux to explore the rich features of these mathematical objects. In the present Tutorial, we simply distill the important result: $\ket{\Psi}$ must either be fully symmetrised or fully anti-symmetrised, which is deduced from the framework of Young stabilisers. 

Hence, the state of $n$ identical particles with wave functions $\psi_1, \dots, \psi_n \in {\cal H}$ can be described by
\begin{align}\label{eq:PsiB}
&\ket{\Psi_B} = \ket{\psi_1} \vee \dots \vee \ket{\psi_n} \equiv \frac{1}{\sqrt{n!}} \sum_{\sigma \in S_n} \ket{\psi_{\sigma(1)}} \otimes \dots \otimes \ket{\psi_{\sigma(n)}},\\
&\ket{\Psi_F} = \ket{\psi_1} \wedge \dots \wedge \ket{\psi_n} \equiv \frac{1}{\sqrt{n!}} \sum_{\sigma \in S_n} {\rm sign}(\sigma)  \ket{\psi_{\sigma(1)}} \otimes \dots \otimes \ket{\psi_{\sigma(n)}},\label{eq:PsiF}
\end{align}
for bosons and fermions, respectively. In literature, one often refers to many-fermion wave functions of this type as {\em Slater determinants}. These vectors can be used to construct the Hilbert spaces 
\begin{align}
&{\cal H}^{(n)}_{B} \equiv {\rm span}\{\ket{\psi_1}\vee\dots \vee \ket{\psi_n} \mid \psi_1, \psi_2 \in {\cal H}\},\label{eq:hilB}\\
&{\cal H}^{(n)}_{F} \equiv {\rm span}\{\ket{\psi_1}\wedge\dots \wedge \ket{\psi_n} \mid \psi_1, \psi_2 \in {\cal H}\},\label{eq:hilF}
\end{align}
which describe systems of $n$ identical bosons (\ref{eq:hilB}) or fermions (\ref{eq:hilF}). These spaces have the special property that any wave function $\Psi \in {\cal H}^{(n)}_{B/F}$ fulfils the identity (\ref{eq:permute}). However, there is no generalisation of (\ref{eq:HilbIdentityTwo}) to systems with more than two particles, as ${\cal H}^{(n)}_{B} \oplus {\cal H}^{(n)}_{F}$ is of much lower dimension than ${\cal H}^{\otimes n}$. 

Finally, we note that the identities (\ref{eq:measureTwoB}, \ref{eq:measureTwoF}) still hold in the $n$-particle scenario, which implies that the measurements of physical observables are, indeed, independent under permutations.

\subsubsection{Fock space}\label{sec:Fock}

As a final step in the description of many-particle systems, we equip the mathematical framework with the possibility to describe fluctuating particle numbers. In this section, we will focus on the mathematical space, the Fock space, that is required to deal with different particle numbers. The operators that describe changes in particle numbers will be discussed in Section \ref{sec:SecQuant}.

The bosonic and fermionic Fock spaces, built on the single-particle subspace ${\cal H}$, are defined as
\begin{align}
&{\cal F}_{B}({\cal H})\equiv {\cal H}^{(0)}_{B} \oplus  {\cal H}^{(1)}_{B} \oplus {\cal H}^{(2)}_{B} \oplus \dots,\label{eq:FockB}\\
&{\cal F}_{F}({\cal H})\equiv {\cal H}^{(0)}_{F} \oplus  {\cal H}^{(1)}_{F} \oplus {\cal H}^{(2)}_{F} \oplus \dots,\label{eq:FockF}
\end{align}
where we consider an infinite number of terms. We have introduced the direct sum ``$\oplus$'', which is introduced in more detail in \ref{sec:appSumProd}. First, note that ${\cal H}^{(1)}_{B/F}$ is simply the single-particle Hilbert space ${\cal H}$. Furthermore, we encounter the space ${\cal H}^{(0)}_{B/F}$, which represents the vacuum state, i.e.~the state with out any particles. Because there is only one possible state, denoted $\ket{0}$, that describes the system without any particles in it, the Hilbert space ${\cal H}^{(0)}_{B/F}$ is one-dimensional. A one-dimensional complex Hilbert space is just the set of complex numbers, and, thus, ${\cal H}^{(0)}_{B/F} \cong \mathbb{C}$. To understand this structure, it is instructive to consider a general many-particle wave function $\ket{\Psi}$, which takes the form 
\begin{equation}\label{eq:PsiInFock}
\ket{\Psi} = \Psi^{(0)}\oplus \lvert{\Psi^{(1)}}\rangle \oplus \lvert{\Psi^{(2)}}\rangle \oplus \dots, \quad \text{with }\lvert{\Psi^{(n)}}\rangle \in  {\cal H}^{(n)}_{B/F},
\end{equation}
which can be interpreted as a superposition of states with different particle numbers. Note that $\Psi^{(0)} \in \mathbb{C}$ is a complex number that describes the vacuum contribution in this superposition; when we measure the system, there is a probability $\abs{\Psi^{(0)}}^2$ observing that the system contains no particles. Due to the probabilistic interpretation of the wave function in quantum theory, it is crucial that $\Psi$ is normalised, i.e.
\begin{equation}
\norm{\Psi}^2 = \sum_{n=0}^{\infty} \norm{\Psi^{(n)}}^2 = 1.
\end{equation}
An important object for Section \ref{sec:SecQuant} is the {\em vacuum state}. This pure state is described by a wave function that contains no particles at all:
\begin{equation}\label{eq:vacuum}
\ket{0} = 1 \oplus 0 \oplus 0 \oplus \dots, 
\end{equation}
where the $0$'s represent the zero vectors in each of the $n$-particle spaces. 

As a less trivial {\em example}, let us consider the coherent state $\ket{\alpha}$ for a single bosonic mode. Because we consider a single mode system there is only one mode that can be populated by particles, which means that the single-particle Hilbert space ${\cal H}$ is one-dimensional, i.e.~${\cal H} \cong \mathbb{C}$. As a consequence, the single-mode coherent state $\ket{\alpha} \in {\cal F}_{B}(\mathbb{C})$. The single-mode bosonic Fock space has the special property of only having a single state $\ket{n}$ with $n$ particles, for each value of $n \in \mathbb{N}$. In a quantum optics textbook, one will find that for any $\alpha \in \mathbb{C}$ a coherent state can be constructed as
\begin{equation}
\ket{\alpha} = e^{-\frac{\abs{\alpha}^2}{2}}\sum_{n=0}^{\infty}\frac{\alpha^n}{\sqrt{n!}}\ket{n},
\end{equation}
In the the Fock space formalism as introduced in (\ref{eq:FockB}), we can express $\ket{\alpha}$ in the form (\ref{eq:PsiInFock}) as
\begin{equation}\label{eq:coh}
\ket{\alpha} = e^{-\frac{\abs{\alpha}^2}{2}} \oplus \alpha e^{-\frac{\abs{\alpha}^2}{2}} \oplus \frac{\alpha^2}{2} e^{-\frac{\abs{\alpha}^2}{2}} \oplus \dots \oplus \frac{\alpha^n}{\sqrt{n!}}e^{-\frac{\abs{\alpha}^2}{2}}\oplus \dots
\end{equation}
Due to our choice of a single-mode system, we find that every term in the direct sum is just a complex number, which would not have been the case for a multimode systems. Our choice of a coherent state as an example highlights the narrow connection between a single-mode bosonic Fock space and such a harmonic oscillator, which we will now explore more formally.
\\ 

Now that we have defined the bosonic and fermionic Fock spaces, we can focus our attention on their structures. In particular, one could wonder how structures from the single-particle Hilbert space ${\cal H}$ translate to the Fock space ${\cal F}({\cal H})$. It turns out that Hilbert spaces with a direct sum structure, i.e., ${\cal H} = {\cal G} \oplus {\cal K}$ (where ${\cal G}$ and ${\cal K}$ are Hilbert spaces), leads to interesting features of the Fock space ${\cal F} ({\cal G} \oplus {\cal K})$. The direct sum, here, is completely unrelated to the direct sum that appears in (\ref{eq:FockB}, \ref{eq:FockF}) to separate layers of the Fock space with different particles. 

Physically, the direct sum structure $({\cal G} \oplus {\cal K})$ can be used to break up a system in different parts. For example, for ultra-cold atoms, trapped in an optical lattice, the single-particle Hilbert space can represent the lattice. The direct sum structure may be used to break up the lattice in sub-lattices, or we may even go down to a direct sum of all individual lattice sites \cite{bloch_many-body_2008,Mosonyi:2008aa}. As a second example, one can consider photons in quantum optics. In this case, the single-particle Hilbert space is equivalent to the set optical mode space. The direct sum can then be used to break up the system in a specific modes basis \cite{Treps:2005aa}. It should be emphasised that, here, we refer to a direct sum structure in the single-particle Hilbert space (or mode space in an optics jargon). 

To structure Fock space, we use the core idea that a direct sum structure in the single-particle Hilbert space induces a tensor product structure in the Fock space. This fact is mathematically formalised in the following isomorphism
\begin{equation}\label{eq:isoHilbert}
{\cal F}_{B/F}({\cal G} \oplus {\cal K}) \cong {\cal F}_{B/F}({\cal G})\otimes {\cal F}_{B/F}({\cal K}).
\end{equation}
What (\ref{eq:isoHilbert}) tells us, is that we can take the whole single-particle Hilbert space (the whole lattice or the whole mode space) and build a Fock space on it to accommodate the particles. Equivalently, we can break the system down in substructures (e.g.~individual lattice sites or individual modes) and construct a Fock space for each of them, which are then combined by means of a normal tensor product. The result (\ref{eq:isoHilbert}) is rather tedious to prove with our present toolbox, hence, we will come back to this point in Section \ref{sec:SecQuant} where we explicitly construct the isomorphism in the language of second quantisation.\\

In the light of (\ref{eq:isoHilbert}), it should be pointed out that any discrete Hilbert space ${\cal H}$ is isomorphic to a direct sum structure (see \ref{sec:appSumProd} for further details)
\begin{equation}
{\cal H} \cong \mathbb{C}\oplus \mathbb{C} \oplus \mathbb{C}\oplus \mathbb{C} \oplus \dots. 
\end{equation}
Because of (\ref{eq:isoHilbert}), this implies that their Fock spaces can be written as
\begin{equation}\label{eq:isoHilbert2}
{\cal F}_{B/F}({\cal H}) \cong {\cal F}_{B/F}(\mathbb{C})\otimes {\cal F}_{B/F}(\mathbb{C})\otimes {\cal F}_{B/F}(\mathbb{C}) \otimes \dots .
\end{equation}
It becomes immediately apparent that we can learn a lot about many particle systems by analysing the properties of the smaller (and relatively simple) Fock spaces ${\cal F}_{B/F}(\mathbb{C})$, which we will refer to as the {\em single-mode Fock spaces}. At this point, we see a first major difference between bosons and fermions.

As mentioned in our discussion of (\ref{eq:coh}), a single-mode bosonic system only has one single $n$-particle state $\ket{n}$ for every possible number of particles. In other words, for bosons, the $n$-particle space constructed on the set complex numbers is simply the set of complex numbers itself (since a Hilbert space generate by one single vector is equivalent to the complex numbers). More formally, in the light of (\ref{eq:hilB}), we find that $\mathbb{C}^{(n)}_B = \mathbb{C}$. Hence, by inserting this identity in (\ref{eq:FockB}) we obtain
\begin{equation}
{\cal F}_{B}(\mathbb{C}) = \mathbb{C} \oplus \mathbb{C} \oplus \mathbb{C} \oplus \dots  \cong {\cal L}^2(\mathbb{R}),
\end{equation}
where the direct sum is of infinite length. In physical terms, this means that the single-mode Fock space is equivalent to a {\em quantum harmonic oscillator}. This consideration supports the fact that optical modes are treated as harmonic oscillators in quantum optics. Equation (\ref{eq:isoHilbert2}) shows that the space that describes photons in an $m$-mode optical system is equivalent to a system of $m$ quantum harmonic oscillators.

The single-mode Fock space for fermions has a very different structure, because $\mathbb{C}^{(n)}_F = 0$ for all $n > 1$. This is a direct consequence of Pauli's exclusion principle, since there is only one wave function in the single-particle Hilbert space. This single-particle wave function can only be occupied by a single fermion. Equation  (\ref{eq:FockB}) then reduces to
\begin{equation}
{\cal F}_{F}(\mathbb{C}) = \mathbb{C} \oplus \mathbb{C}  \cong \mathbb{C}^2,
\end{equation}
 which is the Hilbert space that describes a two-level system. This highlights a fundamental connection between fermions and {\em spin systems}. The identity (\ref{eq:isoHilbert2}) then implies that we can map a fermionic systems to a spin chain (and vice versa), which is formalised by the Jordan-Wigner transformation \cite{alicki_quantum_2001,jordan_uber_1928}.

\subsection{Second quantisation}\label{sec:SecQuant}
The formalism in Section \ref{sec:ToFock}, which is usually referred to as first quantisation, has the inconvenience that it requires the use of symmetrisation or anti-symmetrisation of tensor product structures, which tend to mask the fundamental structures of the many-particle system. {\em Second quantisation} provides a more insightful framework that focuses more on how states are populated with particles. In quantum field theory, this framework represents the particles' nature as fundamental excitation of a physical field. In quantum optics, for example, photons represent the excitations of the electromagnetic field.

\subsubsection{Creation and annihilation operators}

The basis of the second quantisation formalism, is the {\em creation operator}, typically denoted with $a^{\dag}(\phi)$, which describes the act of adding a single particle with wave function $\phi \in {\cal H}$ to a quantum state in Fock space. We can define the creation operator by its action on the many-particle wave function $\ket{\Psi} \in {\cal F}_{B/F}({\cal H}),$ given by (\ref{eq:PsiInFock}),
\begin{align}\label{eq:adagB}
&a^{\dag} (\phi) \ket {\Psi} = 0 \oplus \Psi^{(0)} \ket{\phi} \oplus \ket{\phi} \vee \lvert{\Psi^{(1)}}\rangle \oplus \ket{\phi} \vee \lvert{\Psi^{(2)}}\rangle \oplus \dots, \qquad \text{for bosons,}\\
&a^{\dag} (\phi) \ket {\Psi} = 0 \oplus \Psi^{(0)} \ket{\phi} \oplus \ket{\phi} \wedge \lvert{\Psi^{(1)}}\rangle \oplus \ket{\phi} \wedge \lvert{\Psi^{(2)}}\rangle \oplus \dots. \qquad \text{for fermions.} \label{eq:adagF}
\end{align}
We see that the creation operator depletes the vacuum and maps every $n$-particle contribution to the $(n+1)$-particle sector. Hence, we can construct the wave functions (\ref{eq:PsiB}, \ref{eq:PsiF}) using this formalism:
\begin{align}\label{eq:fundTensorB}
&\ket{\psi_1} \vee \dots \vee \ket{\psi_n} = a^{\dag}(\psi_1)\dots a^{\dag}(\psi_n)\ket{0}, \qquad \text{for bosons}\\
&\ket{\psi_1} \wedge \dots \wedge \ket{\psi_n} = a^{\dag}(\psi_1)\dots a^{\dag}(\psi_n)\ket{0}, \qquad \text{for fermions}\label{eq:fundTensorF}
\end{align}
where one must not forget that the bosonic and fermionic creation operators are different objects which act on different spaces. It must be pointed out that any wave function $\ket{\Psi}$ in the Fock space ${\cal F}({\cal H})$ can be represented by a linear combination of vectors of the type $a^{\dag}(\psi_1)\dots a^{\dag}(\psi_n)\ket{0}$, where the different vectors in this linear combination may contain different numbers of creation operators. Furthermore, one can directly deduce the linear property of creation operators:
\begin{equation}\label{eq:linear}
a^{\dag}(x\psi + y\phi) = x\,a^{\dag}(\psi) + y\, a^{\dag}(\phi), 
\end{equation}
which holds for all $\ket{\phi},\ket{\psi} \in {\cal H}$ and all $x,y \in \mathbb{C}$.

A narrowly related operator, that is of importance in many-particle physics, is the {\em number operator} $\hat N$. This operator can also be defined in terms of its action on an arbitrary $\ket{\Psi} \in {\cal F}_{B/F}({\cal H}):$
\begin{equation}\label{eq:N}
\hat{N} \ket{\Psi} = 0 \oplus  \lvert{\Psi^{(1)}}\rangle \oplus 2 \lvert{\Psi^{(2)}}\rangle \oplus 3 \lvert{\Psi^{(3)}}\rangle \oplus \dots,
\end{equation}
It is not difficult to see that wave functions that are fully defined within the $n$-particle space, i.e., those of the form $0\oplus \dots \oplus 0 \oplus \lvert{\Psi^{(n)}}\rangle 0 \oplus \dots \oplus 0$, are eigenvectors of the number operator $\hat N$, with associated eigenvalue $n$. They are commonly referred to as {\em number states} or {\em Fock states}.

The adjoint operation of the creation operator is known as the annihilation operator $a(\phi)$. This operator is easiest to understand in terms of its action on wave functions of the form (\ref{eq:PsiB}, \ref{eq:PsiF}). We find that {\em for bosons}
\begin{equation}\label{eq:aB}
a(\phi) \big[ \ket{\psi_1} \vee \dots \vee \ket{\psi_n} \big] = \sum_{j=1}^n \langle \phi \mid \psi_j \rangle\, \ket{\psi_1} \vee\dots \vee \ket{\psi_{j-1}}\vee \ket{\psi_{j+1}} \vee \dots \vee \ket{\psi_n} ,
\end{equation}
while {\em for fermions} we obtain
\begin{equation}\label{eq:aF}
a(\phi) \big[ \ket{\psi_1} \wedge \dots \wedge \ket{\psi_n} \big] = \sum_{j=1}^n \langle \phi \mid \psi_j \rangle\, \ket{\psi_1} \wedge \dots \wedge \ket{\psi_{j-1}}\wedge \ket{\psi_{j+1}} \wedge  \dots \wedge \ket{\psi_n}.
\end{equation}
Observe that wave functions in the $n$-particle are now mapped to the $(n-1)$-particle sector. Also note that when $\ket{\Psi}$ is normalised, this is typically not the case for $a^{\dag}(\phi)\ket{\Psi}$, neither for $a(\phi)\ket{\Psi}$. This aspect highlights the non-unitary nature of these operators, implying that one must generally renormalise the state after applying the creation/annihilation operator. For fermions, this procedure never poses any problems. For bosons, however, one may encounter difficulties. Indeed, one must guarantee that 
\begin{equation}\label{eq:condExistence}
\bra{\Psi} a(\phi) a^{\dag}(\phi) \ket{\Psi} < \infty,
\end{equation}
which imposes additional constraints on the wave functions $\ket{\Psi} \in {\cal F}_{B}({\cal H})$ that have a physical meaning. Because (\ref{eq:condExistence}) must be fulfilled for all $\phi \in {\cal H}$, it can be shown that the important condition for the many-particle wave function (\ref{eq:PsiInFock}) to fulfil is 
\begin{equation}
\bra{\Psi} \hat N \ket{\Psi} = \sum_{n=0}^{\infty} n \norm{\Psi^{(n)}}^2 < \infty. 
\end{equation}
The left-hand side represents the expected result for a measurement of the total particle number in the state $\ket{\Psi}$. In other words, the {\em bosonic Fock space can only accommodate states with a finite amount of particles}.\\

The virtue of the second quantisation formalism lies in the calculus of creation and annihilation operators. The final element necessary to understand this calculus are the {\em canonical (anti-)commutation relations}.\footnote{Remember that the {\em commutator} of two operators $A$ and $B$ is defined as $[A,B] = AB - BA$, whereas their anti-commutator is given by $\{A,B\} = AB + BA$.} With the tools introduced above, it should not be too hard to verify that
\begin{equation}\label{eq:CCR}
[a^{\dag}(\psi_1), a^{\dag}(\psi_2)] = 0\quad \text{and }\quad   [a(\psi_1), a^{\dag}(\psi_2)] = \langle \psi_1 \mid \psi_2 \rangle\mathbb{1} \quad \text{for bosons,}
\end{equation}
and 
\begin{equation}\label{eq:CAR}
\{a^{\dag}(\psi_1), a^{\dag}(\psi_2)\} = 0\quad \text{and }\quad   \{a(\psi_1), a^{\dag}(\psi_2)\} = \langle \psi_1 \mid \psi_2 \rangle\mathbb{1} \quad \text{for fermions.}
\end{equation}
In particular, the fact that fermionic creation operators fulfil $\{a^{\dag}(\psi), a^{\dag}(\psi\} = 0$ means that we can never create two fermions with the same single-particle wave function $\psi \in {\cal H}.$\\

We can now use the creation and annihilation operators to revisit the identity ${\cal F}({\cal G}\oplus{\cal K}) \cong {\cal F}({\cal G})\otimes{\cal F}({\cal K})$ in (\ref{eq:isoHilbert}). At the basis of this important identity lies the isomorphism $U$, which was rather intricate to define in Section \ref{sec:Fock}. However, in second quantisation, we can define the action of the isomorphism on the creation operators:
\begin{align}
&U a^{\dag}(\psi_1\oplus \psi_2) U^{\dag} =  a^{\dag}(\psi_1) \otimes \mathbb{1} + \mathbb{1} \otimes a^{\dag}(\psi_2) \quad \text{for bosons,}\\
&U a^{\dag}(\psi_1\oplus \psi_2) U^{\dag} =  a^{\dag}(\psi_1) \otimes \mathbb{1} + (-\mathbb{1})^{\hat N} \otimes a^{\dag}(\psi_2) \quad \text{for fermions,}\label{eq:ImpIdFerm}
\end{align}
where $(-\mathbb{1})^{\hat N}$ is known as the parity operator (it returns $1$ for wave functions with an even number of particles and $-1$ for states with an odd number of particles). To get the isomorphism (\ref{eq:isoHilbert}), all that remains to be done, is to define the action of $U$ on the vacuum:
\begin{equation}
U \ket{0}_{{\cal G} \oplus {\cal K}} = \ket{0}_{{\cal G}} \otimes \ket{0}_{{\cal K}}.
\end{equation}
With these definitions, and with (\ref{eq:fundTensorB}, \ref{eq:fundTensorF}) we can now understand the isomorphism (\ref{eq:isoHilbert}) in a much more elegant way.

It is also insightful to revisit the single-mode spaces ${\cal F}_{B}(\mathbb{C})$ and ${\cal F}_{F}(\mathbb{C})$ in the light of second quantisation. First of all, it should be emphasised that the single-mode space only has a single creation (and annihilation) operator $a^{\dag}$. As we stressed before, the mathematical framework is essentially defined by the calculus of creation and annihilation operators. For the bosonic single-mode Fock space, we find that (\ref{eq:CCR}) reduces to $[a,a^{\dag}] = \mathbb{1},$ which is exactly the commutation relation that describes the {\em ladder operators of a harmonic oscillator}. For the fermionic case, we find that (\ref{eq:CCR}) describes an operator with properties $\{a,a^{\dag}\} = \mathbb{1}$ and $(a^{\dag})^2 = 0$. This is exactly the recipe for the Pauli operator $\sigma^+$, given by a matrix
\begin{equation}
\sigma^+ = \begin{pmatrix} 0 & 1 \\ 0 & 0\end{pmatrix},
\end{equation}
which solidifies the connection between fermionic systems and spin chains.\\

This concludes our description of how second quantisation is used to describe states. However, the full potential of the formalism stems from its possibility to also describe observables, as we will see in the next section.

\subsubsection{Single-particle observables}

Second quantisation can describe not only many-particle wave functions, but also many-particle observables.
To this end, we start by introducing the important framework of single-particle observables. To lower notational overhead, we will omit the ``$B/F$'' subscripts when the results are valid for both fermionic and bosonic systems.\\

A single-particle observable $A$ is an operator that acts on the single-particle Hilbert space, i.e.~$A \in {\cal B}({\cal H}$. In a sense, it represents an attribute of each individual particle particle. Such an observable can be embedded in the space of observables acting on the $n$-particle space, by constructing the operator\footnote{Technically, since we are only considering spaces of either symmetrised or anti-symmetrised vectors, we must formally define $A^{(n)}$ as a restriction of the expression in (\ref{eq:npartsinglepartobs}) to the space ${\cal H}^{(n)}_{B/F}$. For example, one may encounter expressions as $(A\otimes \mathbb{1} + \mathbb{1}\otimes A)\mid_{{\cal H}^{(n)}_{B/F}}$. For the sake of limiting notational overhead, these restrictions are not explicitly mentioned.}
\begin{equation}\label{eq:npartsinglepartobs}
A^{(n)} = A \otimes \mathbb{1}\otimes\dots \otimes \mathbb{1} + \mathbb{1}\otimes A \otimes \mathbb{1}\otimes\dots \otimes\mathbb{1} +\dots + \mathbb{1}\otimes \dots \otimes \mathbb{1}\otimes A.
\end{equation}
We can then lift the single-particle observable to the level of the Fock space by defining
\begin{equation}
{\cal F}(A) \equiv 0 \oplus A \oplus A^{(2)} \oplus \dots.
\end{equation}
Note that (\ref{eq:npartsinglepartobs}) implies that single-particle observables are additive in many-particle systems. An important example of such an observable is the Hamiltonian of a system of {\em non-interacting particles}. We can describe such as system by a single-particle Hamiltonian $H \in {\cal B}({\cal H})$, which acts on the many-particle Fock space as ${\cal F}(H)$. The additivity now makes sense, as we can simply add the energy of each individual particle to obtain the total energy of the system.
Another important example is the number operator (\ref{eq:N}) is also a single-particle observable. Indeed, it can be verified that $N= {\cal F}(\mathbb{1})$.

Single-particle observables have useful features, such as their behaviour with respect to commutators. When we consider two observables $A, B \in {\cal B}({\cal H})$, we find that \begin{equation}
[{\cal F}(A), {\cal F}(B)] = {\cal F}\big([A,B]\big). 
\end{equation}
This result is particularly important when dealing with the dynamics of non-interacting (or quasi-free) particles. In the Heisenberg picture, we find that the dynamics of an observable $\hat X \in {\cal B} \big({\cal F} ({\cal H})\big)$ is generated by a Hamiltonian ${\cal F}(H)$, as prescribed by
\begin{equation}\label{eq:dynamics1particle}
\frac{\rm d}{{\rm d}t} \hat X = i [{\cal F}(H), \hat X], \quad \text{with } \hat X(0) = \hat X_0 
\end{equation}
When $\hat X = {\cal F}(X)$ is a single particle observable, we find that
\begin{equation}
{\cal F} \bigg( \frac{\rm d}{{\rm d}t}  X - i [H, X] \bigg) = 0,
\end{equation}
which has a solution ${\cal F}[X(t)]$, where
\begin{equation}
X(t) = e^{itH} X_0 \,e^{-itH}.
\end{equation}
In other words, when one deals with many-particle systems, but only single-particle observables are considered and there is no interaction between the particles, everything can be solved on the level of the single-particle Hilbert space ${\cal H}$.\\

It will probably not come as a surprise that here, too, creation and annihilation operators can play an important role to simplify calculations, as well as to gain a deeper insight on the structure of these single-particle observables. By evaluating the action of observables of the form ${\cal F}(A)$ on wave functions of the type (\ref{eq:fundTensorB}), one can eventually derive the identity
\begin{equation}\label{eq:singlePartOpa}
{\cal F}(A) = \sum_{i,j} \bra{e_i}A\ket{e_j} a^{\dag}(e_i)a(e_j),
\end{equation}
where the vectors $\ket{e_i} \in {\cal H}$ form a basis of the Hilbert space. Again, let us highlight the simple example of the number operator
\begin{equation}
\hat N = {\cal F}(\mathbb{1}) = \sum_i a^{\dag}(e_i)a(e_i),
\end{equation}
which is independent of the chosen basis $\{\ket{e_i}\}$. We can understand this result a little better by introducing the {\em single-mode number operator} 
\begin{equation}\label{eq:local Number}
\hat n(\psi) = {\cal F}(\ket{\psi}\bra{\psi}) = a^{\dag}(\psi)a(\psi),
\end{equation}
which counts the number of particles that occupy the single-particle wave function $\ket{\psi} \in {\cal H}$ (or in terms of quantum optics, the number of particles in the mode associated with $a^{\dag}(\psi)$).\\

%This framework can straightforwardly be generalised to observables that act on $n$- particle spaces. Say that we consider $B \in {\cal H}^{(n)}_{B/F}$, we can lift this observable to the Fock space by defining {\em for bosons}
%\begin{equation}\label{eq:nPartObsB}
%{\cal F}_B(B) = \sum_{\substack{i_1, \dots, i_n,\\ j_1, \dots, j_n}} \bra{e_{i_1}}\vee \dots \vee \bra{e_{i_n}} B \ket{e_{j_1}} \vee \dots \vee \ket{e_{j_n}} a^{\dag}(e_{i_1})\dots a^{\dag}(e_{i_n})a(e_{j_1})\dots a(e_{j_n}),
%\end{equation}
%and {\em for fermions}
%\begin{equation}\label{eq:nPartObsF}
%{\cal F}_F(B) = \sum_{\substack{i_1, \dots, i_n,\\ j_1, \dots, j_n}} \bra{e_{i_1}}\wedge \dots \wedge \bra{e_{i_n}} B \ket{e_{j_1}} \wedge \dots \wedge \ket{e_{j_n}} a^{\dag}(e_{i_1})\dots a^{\dag}(e_{i_n})a(e_{j_1})\dots a(e_{j_n}).
%\end{equation}
%It is a useful exercise to verify that for $B \in {\cal H}^{(n)}_{B/F}$, we find that
%\begin{equation}
%\bra{0}a(\psi_1) \dots a(\psi_{n'}) {\cal F}(B) a^{\dag}(\psi_{n'})\dots a^{\dag}(\psi_{1}) \ket{0} = 0 \quad \text{ for $n'<n$}.
%\end{equation}
%In other words, $n$-particle observables just return for wave functions that contain fewer than $n$ particles.

To introduce a final important class of operators that is narrowly related to the single-particle observables, let us return to (\ref{eq:dynamics1particle}). In general, the dynamics is described by
\begin{equation}
\hat X(t) = e^{it {\cal F}(H)} \hat X_0 e^{-it {\cal F}(H)},
\end{equation}
where we are confronted with the propagator $\exp [-it {\cal F}(H)]$. It turns out that this propagator has an appealing form:\footnote{Again, the $n$-fold tensor products must formally be restricted to the spaces of symmetrised or anti-symmetrised ${\cal H}^{(n)}_{B/F}$ wave functions.}
\begin{equation}
e^{-it {\cal F}(H)} = 1 \oplus e^{-itH} \oplus (e^{-itH} \otimes e^{-itH}) \oplus \dots.
\end{equation}
This is an example of a so-called {\em exponential element} in the set of operators on the Fock space. In general, for $A \in {\cal B}({\cal H})$, we define these objects as
\begin{equation}\label{eq:EA}
E(A) = 1 \oplus A \oplus (A \otimes A) \oplus \dots.
\end{equation}
These objects have a list of interesting properties which will be used (though sometimes implicitly) throughout the remainder of the text:
\begin{align}
&E(A^{\dag}) = E(A)^{\dag},\\
&E(A)E(B) = E(AB),\label{eq:EProd}\\
&E(A)a^{\dag}(\psi) = a^{\dag}(A\psi) E(A),\label{eq:Ea}\\
&a(\psi)E(A) = E(A)a(A \psi),\label{eq:EAnnil}\\
&E(e^A) = e^{{\cal F}(A)},\\
&E(A \oplus B) \cong E(A)\otimes E(B).
\end{align}
For the reader who wants to get acquainted with the many-particle formalism, proving these identities may be a fruitful exercise.\\

The construction of single-particle observables can in principle be generated to describe more general classes of $n$-particle observables. For such observables, we find a generalisation of (\ref{eq:singlePartOpa}) with $n$ creation operators and $n$ annihilation operator. The most general observables on the Fock space can contain many different terms with varying numbers of creation and annihilation operators. Take, for example, the Hamiltonian that describes a Hubbard model \cite{Hub, Hub2, Hub3}. This Hamiltonian contains a single-particle term which describes the tunnelling between different sites, and a two-particle term that accounts for interaction between particles. A general observable $\hat X$ on the Fock space is actually polynomials in creation and annihilation operators. From a mathematical perspective, this is where the real importance of creation and annihilation operator lies: they are the {\em generators of the algebra of many-particle observables}. As such, one could argue that they are the most fundamental objects this mathematical framework.

\subsubsection{Gaussian states and quasi-free states}\label{sec:GaussianStates}

One last piece of technical machinery that is useful to introduce, is the notion of Gaussian -- and quasi-free -- states. These states are, in a way, the most controllable quantum states in many-particle systems. They are commonly associated with ground- or thermal states of systems of non-interacting particles. In quantum optics, the states that describe the coherent light emitted by a laser and squeezed light also belong to the class of Gaussian states.\\

A general quantum state need not be pure, a consideration that commonly leads us to the framework of density matrices. A density matrix $\rho$ is a trace-class operator on the Fock space ${\cal F}({\cal H})$, which has to be positive semi-definite (i.e., have positive eigenvalues) and normalised (ie., $\tr \rho = 1$). The class of {\em Gaussian states} can be entirely described in terms of how their density matrix behaves with respect to products of creation- and annihilation operators.

We start by considering the monomial $a^{\#}(\psi_1) \dots a^{\#}(\psi_{n})$, where $a^{\#}$ is either a creation or annihilation operator (there is no reason to specify). A Gaussian state is completely defined by the expectation values of these monomials. To simplify matters considerably, we will restrict ourselves to the case where $\tr [\rho a^{\#}(\psi)] = 0$,\footnote{It turns out to be completely straightforward to include this case for bosons, but nearly impossible to include it for fermions.} where we find that {\em for bosons} a Gaussian state is any state that behaves as follows
\begin{align}
&\tr [a^{\#}(\psi_1) \dots a^{\#}(\psi_{2n + 1}) \rho] = 0,\\
&\tr [a^{\#}(\psi_1) \dots a^{\#}(\psi_{2n}) \rho] = \sum_{{\cal P}} \tr[a^{\#}(\psi_{i_1})a^{\#}(\psi_{j_1})]\dots \tr[a^{\#}(\psi_{i_n})a^{\#}(\psi_{j_n})],\label{eq:pefectMatch}
\end{align}
where ${\cal P}$ represents the so-called perfect matchings (or pair-partitions). To get such a perfect matching, we break up the set of indices $\{1, \dots, 2n\}$ up in pairs. An example of such a perfect matching is $\{\{1,2\},\{3,4\}, \dots, \{2n-1, 2n\}\}$, but, obviously, there are many other possible ways to divide the set of indices in pairs (the number of possible ways is given by the product of all the odd numbers up to $2n-1$). In (\ref{eq:pefectMatch}), we denote a generic perfect matching as $\{\{i_1,j_1\}, \dots, \{i_n,j_n\}\}$, with $i_1 \leq i_2 \leq \dots \leq i_n$ and $j_1 \leq j_2 \leq \dots \leq j_n$.

Fermionic Gaussian states, which are more commonly known as quasi-free states, have an additional complication due to the anti-commutation relations. Not only must we do bookkeeping of perfect matchings, we must also keep track of signs. This forces us to include the sign $\epsilon$ of the perfect matching. For a given perfect matching, $\epsilon = {\rm sign}(\sigma),$ where $\sigma$ is the permutation that maps the set $\{1,\dots, 2n \}$ to the set $\{i_1,j_1,i_2,j_2, \dots,i_n,j_n\}$. With this notation, we find {\em for fermions}
\begin{align}
&\tr [a^{\#}(\psi_1) \dots a^{\#}(\psi_{2n + 1}) \rho] = 0,\\
&\tr [a^{\#}(\psi_1) \dots a^{\#}(\psi_{2n}) \rho] = \sum_{{\cal P}} \epsilon \tr[a^{\#}(\psi_{i_1})a^{\#}(\psi_{j_1})\rho]\dots \tr[a^{\#}(\psi_{i_n})a^{\#}(\psi_{j_n})\rho].\label{eq:pefectMatchF}
\end{align}
where ${\cal P}$ again represents all the possible perfect matchings.

The above definitions of Gaussian (or quasi-free) states may look a little tedious, and in practice they often, indeed, turn out to be quite hard to evaluate for large $n$. Nevertheless, these states have a profound advantage as compared to vast majority of other states: they are easy to understand and interpret. In particular, note that the states are completely determined by expectation values of single-particle operators $ \tr[a^{\#}(\psi_{1})a^{\#}(\psi_{2})\rho].$ When these expectation values are known, the can be used to evaluate the expectation value of an arbitrary observable via (\ref{eq:pefectMatch}, \ref{eq:pefectMatchF}).

It is useful to describe two matrices $Q, S \in {\cal B}({\cal H}),$ that act on the single particle Hilbert space to characterise a Gaussian state. It is most convenient to define these matrices component-wise
\begin{equation}\label{eq:QS}
\bra{\psi}Q\ket{\phi} = \tr[a^{\dag}(\psi) a(\phi) \rho], \quad \text{and } \quad \bra{\psi}S\ket{\phi} = \tr[a(\psi) a(\phi) \rho].
\end{equation}
The matrix $Q$ is positive semidefinite (i.e.~$Q \geq 0$), and is often called the {\em coherence matrix}. In many ways, this object behaves as a non-normalised density matrix that describes the single-particle behaviour of the system. The matrix $S$, on the other hand, has a special feature: it is {\em conjugate-linear}, which means that
\begin{equation}
S(x \ket{\phi} + y \ket{\psi}) = x^* S\ket{\phi} + y^* S\ket{\psi},
\end{equation} 
where $x^*$ is the complex conjugate of $x$. It is not hard to see from (\ref{eq:pefectMatch}, \ref{eq:pefectMatchF}) that these matrices suffice to characterise the Gaussian state entirely.

Of course, $Q$ and $S$ cannot simply be chosen freely, they have to fulfil some conditions to make sure that the Gaussian state is normalised, positive, and reflects the correct bosonic (\ref{eq:CCR}) and fermionic (\ref{eq:CAR}) features.  These conditions tend to be quite different for bosons and fermions. {\em For fermions,} we find first of all that $S^{\dag} = - S,$
due to the anti-commutation relation. Furthermore, we must guarantee the positivity of the state, which is most generally achieved by the condition
\begin{equation}\label{eq:positivityGaussianStates}
\tr[\rho (a^{\dag}(\psi) + a(\phi))(a^{\dag}(\phi) + a(\psi)) ] \geq 0.
\end{equation}
Using the Schur complement, (\ref{eq:positivityGaussianStates}) can be shown to lead to the condition
\begin{equation}\label{eq:condQ}
Q \geq 0 \quad \text{and} \quad S^{\dag}Q^{-1}S + Q \leq \mathbb{1}.
\end{equation}
The matrix $Q$ can be understood as a single-particle density matrix, which is not normalised to one, but rather does $\tr Q$ give the number of particles. The latter condition in (\ref{eq:condQ}) also implies that $Q \leq \mathbb{1}$, which is a manifestation of Pauli's exclusion principle. Indeed, the matrix elements $\bra{\psi}Q\ket{\psi}$ denoted the number of particles that occupy single-particle wave function (i.e.~mode) $\ket{\psi} \in {\cal H}$. Therefore, the condition that $Q \leqslant \mathbb{1}$ implies that $\bra{\psi}Q\ket{\psi} \leqslant 1$ for any single-particle wave function. Hence, the condition directly implies that there is never more than one particle that occupies the same single-particle wave function.

{\em For bosons,} a different condition must to be imposed for $S$, since the commutation relation implies that
$
S^{\dag} = S.
$
When the positivity condition (\ref{eq:positivityGaussianStates}) is enforced for fermionic Gaussian states, we find that
\begin{equation}
Q \geq 0 \quad \text{and} \quad S^{\dag}Q^{-1}S - Q \leq \mathbb{1}.
\end{equation}
Note that the differences with the fermionic case appear small, but the physical implications are huge. Most notably, bosons can bunch together and occupy the same state, which can ultimately lead to exotic phenomena such as Bose-Einstein condensation.\\

Much of the remainder of this work will deal with number states, and in particular with those of the form (\ref{eq:fundTensorB}) and (\ref{eq:fundTensorF}). However, it should be stressed that for fermions these Slater determinants are, in fact, Gaussian states. It is an excellent exercise to verify this. To do so, start by taking a set of vectors $\psi_1, \dots, \psi_n$ such that $\langle \psi_i \mid \psi_j \rangle =\delta_{ij}$, and defining
\begin{equation}\label{eq:SlaterDet}
\ket{\Psi} = \frac{1}{\cal N}a^{\dag}(\psi_1)\dots a^{\dag}(\psi_n)\ket{0},
\end{equation}
one must now check that (\ref{eq:pefectMatchF}) holds, and more specifically that
\begin{equation}\label{eq:FermGaussNumber}
\bra{\Psi}a^{\#}(\phi_1) \dots a^{\#}(\phi_{2m}) \ket{\Psi} = \sum_{{\cal P}} \epsilon  \bra{\Psi }a^{\#}(\phi_{i_1})a^{\#}(\phi_{j_1})\ket{\Psi}\dots \bra{\Psi}a^{\#}(\phi_{i_n})a^{\#}(\phi_{j_m})\ket{\Psi},
\end{equation}
where ${\cal P}$ again represents all the possible perfect matchings.
Life is considerably simplified since $\bra{\Psi}a(\psi)a(\phi)\ket{\Psi} = 0,$ so that using (\ref{eq:QS}) we find that $S=0$. Hence, the state is completely described by $Q$. It turns out that the fermionic coherence matrix for these states is given by
\begin{equation}
Q = \sum_{i=1}^n \ket{\phi_i}\bra{\phi_i}.
\end{equation}
To simplify the expression (\ref{eq:FermGaussNumber}), let us define the submatrix $Q^{\{\phi\}}$ of $Q$ with matrix elements $Q^{\{\phi\}}_{ij} = [\bra{\phi_i}Q\ket{\phi_j}]_{ij}$. Using this matrix, we then obtain
\begin{equation}
\bra{\Psi}a^{\#}(\phi_1) \dots a^{\#}(\phi_{2m}) \ket{\Psi}  = \det Q^{\{\phi\}}.
\end{equation}
In a slightly more general sense, one can prove that fermionic Gaussian states with $S=0$ are pure if and only if $Q$ is a projection operator, i.e.~$Q^2 = Q$.

Gaussian states are typically considered to be the many-particle states that are closest to classical physics, even though they can already describe some non-classical features such as squeezing. Hence, their Gaussianity means that Slater determinants (\ref{eq:SlaterDet}) are not expected to induce exotic quantum phenomena. Much to the contrary, bosonic states of the form (\ref{eq:fundTensorB}), sometimes known as Fock states, are {\em non-Gaussian states}, which seems to imply that they are more non-classical than their fermionic counterparts. For a framework where fermions and bosons seems to be so much alike, this may appear odd. However, this profound difference is believed to lie at the basis of the computational hardness of bosonic many-particle interference, which is a topic of Section \ref{eq:ManyPartInt}.
 
\subsection{Distinguishability}  \label{sec:Distinguish}
Before we move on to discussing the topic of many-particle interference, we take a moment to address an important issue in many-particle systems. This apparent paradox is related to the concept of indistinguishability: {\em how can there be distinguishable particles?} Distinguishable particles are typically associated with tensor product structures, i.e.~the two-particle state of two distinguishable particles with wave functions $\ket{\phi}, \ket{\psi} \in {\cal H},$ is simply given by $\ket{\phi} \otimes \ket{\psi}$. However, if these particles are both electrons with the same spin, this wave function should be $\ket{\phi} \wedge \ket{\psi}$---according to the framework of the previous sections, as seen in (\ref{eq:twoBoson}, \ref{eq:twoFermions}). Nevertheless, when these electrons are very far away from eachother, shouldn't they actually be distinguishable? This type of questions often causes confusion to those who just start working with many-particle systems, and here we will clarify the issue. The solution to this apparent paradox is deeply ingrained in the identity (\ref{eq:isoHilbert}).

To address this issue, let us consider a box $\Lambda \subset \mathbb{R}^3$, such that the Hilbert space for a quantum particle in such a box is given by ${\cal L}^2(\Lambda)$. We can now populate the box with two particles, with wave functions $\ket{\psi},\ket{\phi} \in {\cal L}^2(\Lambda).$ When we {\em assume that these particles are fermions}, we find that their two-particle state is given by
\begin{equation}
\ket{\Psi} = a^{\dag}(\psi)a^{\dag}(\phi)\ket{0}.
\end{equation}
Obvious, these creation and annihilation operators fulfil the anti-commutation relation (\ref{eq:CAR}), implying that the particles are indistinguishable. Now, let us split the box in two parts, $\Lambda_1$ and $\Lambda_2$, such that $\Lambda_1 \cup \Lambda_2 = \Lambda$. The Hilbert space can now be written as ${\cal L}^2(\Lambda) \cong {\cal L}^2(\Lambda_1) \oplus {\cal L}^2(\Lambda_2)$. The wave functions inherit this structure, such that we may write $\psi \mapsto \psi_1 \oplus \psi_2$ and $\phi \mapsto \phi_1 \oplus \phi_2$. Note, that this is still the same system, we just gave each half of the box a different name. We can now use the identity  (\ref{eq:ImpIdFerm}), that lies at the basis of (\ref{eq:isoHilbert}), to rewrite the state $\ket{\Psi}$ of the two particles in the box:
\begin{align}
\ket{\Psi} &= \big[a^{\dag}(\psi_1) \otimes \mathbb{1} + (-\mathbb{1})^{\hat N} \otimes a^{\dag}(\psi_2)\big]\big[a^{\dag}(\phi_1) \otimes \mathbb{1} + (-\mathbb{1})^{\hat N} \otimes a^{\dag}(\phi_2)\big] \ket{0}\otimes \ket{0}\\
&= a^{\dag}(\psi_1)a^{\dag}(\phi_1)\ket{0}\otimes \ket{0} - a^{\dag}(\phi_1)\ket{0}\otimes a^{\dag}(\psi_2)\ket{0}\label{eq:blabla} \\
&\quad+ a^{\dag}(\psi_1)\ket{0}\otimes a^{\dag}(\phi_2)\ket{0} + \ket{0}\otimes a^{\dag}(\psi_2) a^{\dag}(\phi_2)\ket{0}.\nonumber
\end{align}
Thus far, this is merely a rewriting of that state $\ket{\Psi} \in {\cal F}[{\cal L}^2(\Lambda)]$ in terms of the equivalent space $ {\cal F}[{\cal L}^2(\Lambda_1)] \otimes  {\cal F}[{\cal L}^2(\Lambda_2)],$ and the particles still seem to have their indistinguishable character. Let us now assume that one particle is fully localised in one part of the box ${\Lambda_1},$ whereas the other particle lingers in the other side of the box ${\Lambda_2}$. This implies that $\psi \mapsto \psi_1 \oplus 0,$ and $\phi \mapsto 0 \oplus \phi_2$ (i.e., $\psi_2 = \phi_1 = 0$). When this is inserted in (\ref{eq:blabla}), we find
\begin{equation}\label{eq:distStructure}
\ket{\Psi} \cong a^{\dag}(\psi_1)\ket{0}\otimes a^{\dag}(\phi_2)\ket{0},
\end{equation}
and, thus, we find the tensor product structure that is associated with distinguishable particles. In particular, the particles can be distinguished from one another by virtue of the different spatial structure of their wave functions. Note that a completely equivalent argument can be given for bosonic particles.

In the reasoning that led towards (\ref{eq:distStructure}), the capability of identifying $\Lambda_1$ and $\Lambda_2$ is crucial. This highlight that {\em distinguishability cannot be understood independent of the measurement.} If the partition of the box were chosen differently, the structure (\ref{eq:distStructure}) would not be found. When we assume that the two-particle state is initially given by (\ref{eq:distStructure}), the particles' wave functions may evolve over time to give rise to a many-particle wave function of the form (\ref{eq:blabla}). Hence, dynamics may influence the capability to distinguish particles.\\

Up to this point, we used a particular spatial structure of the particles' wave functions, which are  {\em external DOF}, to distinguish them. However, particles can also be rendered distinguishable because of {\em internal DOF}, which will de facto make them non-identical. Common examples of such internal DOF might be the frequency (or time-frequency mode) of a photon, the spin of an electron, et cetera.

Mathematically, the separations of internal and external DOF can be achieved on the level of the single-particle Hilbert space ${\cal H} = {\cal H}_E \otimes {\cal H}_I$\footnote{Be careful, the tensor product in ${\cal H}_E \otimes {\cal H}_I$ is a tensor product between different single-particle DOF, and it is completely unrelated to the tensor product between Fock spaces in (\ref{eq:isoHilbert}) the we have come to associate with disntinguishability.}, where ${\cal H}_E$ denotes the Hilbert space of external DOF, and ${\cal H}_I$ describes the internal DOF. To show how these internal DOF can make particles distinguishable, we will again put two particles in a box $\Lambda \subset \mathbb{R}^3$. To vary a little bit compared to the previous example, let us assume that this time the particles are photons---and, thus, bosons---with their own polarisations. Hence, we must set the Hilbert spaces ${\cal H}_E = {\cal L}^2(\Lambda)$ and ${\cal H}_I = \mathbb{C}^2$. This implies that
\begin{equation}\label{eq:isomIntExt}
{\cal H} = {\cal L}^2(\Lambda) \otimes \mathbb{C}^2 \cong {\cal L}^2(\Lambda) \oplus {\cal L}^2(\Lambda),
\end{equation}
where the latter isomorphism should be straightforward to check mathematically. Physically, we must remember that the direct sum in $ {\cal L}^2(\Lambda) \oplus {\cal L}^2(\Lambda)$ implies a chosen basis for the polarisation modes, e.g., horizontal and vertical polarisation. We then populate the box with two particles, with wave functions $\ket{\psi} \otimes \ket{p}$ and $\ket{\phi} \otimes \ket{s} \in {\cal H}$. We can now write the polarisation states in terms of the chosen mode basis, i.e., $\ket{p} = p_1 \ket{H} + p_2 \ket{V}$ and $\ket{s} = s_1 \ket{H} + s_2 \ket{V}$. In terms of the isomorphism in (\ref{eq:isomIntExt}), we then find that $\ket{\psi} \otimes \ket{p} \mapsto p_1\ket{\psi} \oplus p_2\ket{\psi}$ and $\ket{\phi} \otimes \ket{s} \mapsto s_1\ket{\phi} \oplus s_2\ket{\phi}$. By virtue of (\ref{eq:ImpIdFerm}), we can than write our two-particle state in the Fock space as
\begin{align}
\ket{\Psi}& = a^{\dag}(\psi \otimes p)a^{\dag}(\phi \otimes s)\ket{0} \\
&=a^{\dag}( p_1 \psi \oplus p_2 \psi)a^{\dag}( s_1 \phi \oplus s_2\phi)\ket{0}\\
&=p_1s_1a^{\dag}(\psi)a^{\dag}(\phi)\ket{0}_H\otimes \ket{0}_V + p_2s_1a^{\dag}(\phi)\ket{0}_H\otimes a^{\dag}(\psi)\ket{0}_V\label{eq:blabla2} \\
&\quad+ p_1s_2a^{\dag}(\psi)\ket{0}_H\otimes a^{\dag}(\phi)\ket{0}_V + p_2s_2\ket{0}_H\otimes a^{\dag}(\psi) a^{\dag}(\phi)\ket{0}_V.\nonumber
\end{align}
Notice that we essentially describe the system as a Fock space for particles with polarisation state $\ket{H}$ ``tensored'' to a Fock space for particles with polarisation state $\ket{V}$. Again, we up to this point, this is just a matter of rewriting. However, when we set $\ket{p} = \ket{H}$ and $\ket{s} = \ket{V}$, we observe that the state reduces to
\begin{align}
\ket{\Psi}& = a^{\dag}(\psi)\ket{0}_H\otimes a^{\dag}(\phi)\ket{0}_V,
\end{align}
and we again uncover the tensor project structure associated with distinguishable particles. Again, notice that our capability of observing this distinguishability hinges from the chosen polarisation basis to measure. A crucial difference to the previously discussed distinguishability based on spatial DOF, is that the internal DOF are not expected to change via a simple free evolution.\\

The acute reader may have realised that identity (\ref{eq:isoHilbert2}) implies that we can distinguish particles in orthogonal single-particle wave functions (or ``modes'' in the optics jargon). This is a correct observation, but it relies strongly on the capacity to measure exactly the right set of single-particle wave functions. Since fermions cannot occupy the same single-mode wave function, we can in principle always find a way to distinguish them. The electrons in an atom can, for example, be distinguished by the energy levels, and orbitals they occupy in combination with their spins. In this sense, bosons can be more ``truly indistinguishable'', because bosons can occupy the same single-particle wave function. In optics, for example, one could argue that the only truly indistinguishable photons are the ones that occupy exactly the same mode.

However, as was stressed several times, what really determines whether or not particles behave in a distinguishable or indistinguishable manned, is the measurement. This concept lies at the basis of the phenomenon of many-particle interference, which will be extensively discussed in the remainder of this Tutorial.

\section{Many-particle interference}\label{eq:ManyPartInt}

We concluded Section \ref{sec:Distinguish} with a discussion on the subtle subject of distinguishability, where is was emphasised that distinguishability of particles does not only depend on the state of the particles, but also on the measurement setup. This idea is reminiscent of the wave-particle duality, where the experimental setup determines whether we will observe wave-like or particle-like features. There is a parallel to our framework, where the experimental setup will determine whether particles show their indistinguishable nature, or rather behave as distinguishable particles.

The wave-particle duality can famously be tested in interferometers, like Young's double slit setup. In absence of decoherence effects, we will a priori observe interference fringes in the output of the experiment, associated with wave-like behaviour. However, which-way information can completely destroy these interference fringes and lead to particle-like measurement statistics. In this section, we will explore a remarkably similar feature of many-particle systems: when many identical particles are jointly injected into an interferometer, we will observe many-particle interference effects that are associated with their mutual indistinguishability. However, when we posses some form of ``which-particle'' information, we can destroy these interference fringes and recover the statistics associated with many particles. 

We start exploring the interference phenomenon in the two-particle scenario, known as the Hong-Ou-Mandel effect. Subsequently, we explore the many-particle extension, and explain what it means for particles to be partially distinguishable.

\subsection{The Hong-Ou-Mandel effect}\label{sec:HOM}
It is instructive to start by exploring many-particle interference for the two-particle case. Let us start by introducing the interferometric setup, involving two separate beams of particles. Each one of these beams is simply represented by one single mode, such that the single-particle Hilbert space ${\cal H}$ associated with the two beams is of dimension two, i.e. ${\cal H} \cong \mathbb{C}^2$. We can describe this space in terms of a basis $\{\ket{e_1}, \ket{e_2}\}$, where $\ket{e_j}$ is associate with the $j$th beam. 

Recalling (\ref{eq:isoHilbert2}), the Fock space of the system is given by ${\cal F}({\cal H}) \cong {\cal F}(\mathbb{C})\otimes{\cal F}(\mathbb{C})$, where each ${\cal F}(\mathbb{C})$ represents the Fock space of particles in one of the beams. This implies that a particle in the first beam can effectively be distinguished from a particle in the second beam, as described in Section \ref{sec:Distinguish}. Indeed, the beam gives as a form of ``which-particle'' information. Like in any interferometer, we will now scramble this information by mixing the two beams (or, more generally, the modes).

The easiest way to mix two beams is using a passive linear element called {\em beamsplitter} (following the jargon in optics) \cite{Zeilinger:1981aa}. On the level of the single-particle Hilbert space ${\cal H}$, the beamsplitter is a unitary operator, given by 
\begin{equation}\label{eq:50-50-BS}
U = \frac{1}{\sqrt{2}}\begin{pmatrix} 1& 1 \\ -1& 1
\end{pmatrix}
\end{equation}
As such, we find that the each input beam is mixed between the two-output beams, i.e. 
\begin{equation}\label{eq:actionBeamsplitterH}
\ket{e_1}\mapsto U\ket{e_1} = \frac{1}{\sqrt{2}}(\ket{e_1} + \ket{e_2}),\quad \text{and } \quad
\ket{e_2}\mapsto U\ket{e_2} = \frac{1}{\sqrt{2}}(\ket{e_1} - \ket{e_2}),
\end{equation}
such that the beams are clearly mixed, and we can subsequently no longer obtain ``which-particle'' information by measuring the individual output modes. This unitary operator provides us with a description of the beamplitter on the level of the single-particle Hilbert space. To lift it to the many-particle Fock space, we will employ the exponential element $E(U)$, as introduced in (\ref{eq:EA}).

Now that we described the beamsplitter, we can select the initial state
\begin{equation}
\ket{\Psi} = a^{\dag}(e_1)a^{\dag}(e_2)\ket{0},
\end{equation}
such that we have exactly one particle in each of the two separate beams. Note that this mode occupation is the only possible case for which we can compare fermions and bosons, since for fermions $a^{\dag}(e_1)a^{\dag}(e_1) = 0$. When we study the evolution of the two particles in the Schr\"odinger picture, we can apply the unitary transformation of the beamsplitter to the state. By applying (\ref{eq:Ea}), we find that
\begin{equation}
\ket{\Psi} \mapsto E(U)\ket{\Psi} = a^{\dag}(Ue_1)a^{\dag}(Ue_2)\ket{0}.
\end{equation}
By virtue of (\ref{eq:linear}) and (\ref{eq:actionBeamsplitterH}), we can rewrite this as
\begin{equation}
E(U)\ket{\Psi} = \frac{1}{2}\big[a^{\dag}(e_1)a^{\dag}(e_1) - a^{\dag}(e_2)a^{\dag}(e_2) + a^{\dag}(e_1)a^{\dag}(e_2) - a^{\dag}(e_2)a^{\dag}(e_1)\big]\ket{0}.
\end{equation}
Through the commutation relations (\ref{eq:CCR}) for bosons and the anti-commutation relations (\ref{eq:CAR}) for fermions, this expression simplifies considerably:
\begin{align}
&E(U) \ket{\Psi} = \frac{1}{2} \big[a^{\dag}(e_1)a^{\dag}(e_1)\ket{0} - a^{\dag}(e_2)a^{\dag}(e_2)\ket{0}\big] \quad \text{for bosons,}\label{eq:HOMB}\\
&E(U) \ket{\Psi} = a^{\dag}(e_1)a^{\dag}(e_2)\ket{0} \quad \text{for fermions.}\label{eq:HOMF}
\end{align}
Observe that the seemingly simple difference between commutation and anti-commutation completely changes the quantum state that exits the beamsplitter. For bosons, we find that the two particles are either both in the first or both in the second output beam, which is referred to as {\em bosonic bunching}. The fermions, instead, are found each in a different beam, and, thus, they are said to {\em anti-bunch}. Both effects can be understood as an interference phenomenon at the level of probability amplitudes, due to the particles' indistinguishability. Nevertheless, one can alternatively describe the fermionic anti-bunching as a manifestation of Pauli's exclusion principle. After all, there is only one fermionic two-particle state in a two-mode setup, and thus there is no other state that fermions could populate. Note that in a dynamical sense, these particles (both bosons and fermions) are non-interacting and, as such, one could describe the beamsplitter in terms of a single-particle Hamiltonian. Given that the particles are not physically interacting with each other, the observed phenomenon can only be explained through their indistinguishability. The observed effect was already hinted at in Section \ref{sec:Distinguish}: the beamsplitter causes a mismatch between the modes that are measured and the modes that are occupied by the particles. Because the particles are otherwise identical, this mismatch induces a behaviour of indistinguishable particle, which causes interference effects.\\

To really interpret (\ref{eq:HOMB}, \ref{eq:HOMF}) as an interference phenomenon, we must include the measurement stage of the output beams in our description. The standard measurement setup in this experiment consists of a particle detector on each of the two output beams. For bosons, these experiments were first carried out with photons, since they are readily available, non-interacting particles and the optical elements necessary for the manipulation are easily accessible. Photo-detection usually cannot resolve the number of detected photons (conventional detectors click if at least one photon hits the sensor). However, since one typically has good control over the initial number of photons, when both detectors click in our two-particle setup, we be confident that there was exactly one photon in each output beam. This is why it is common to perform a coincidence measurement, where one evaluates the probability of both detectors clicking simultaneously. Mathematically, this probability is obtained by projecting onto a measurement state $\ket{M} \in {\cal F}({\cal H})$ that is associated with the positive-operator valued measure (POVM) of the pair of detectors. For a coincidence count of both detectors, we must choose
\begin{equation}
\ket{M} = a^{\dag}(e_1)a^{\dag}(e_2)\ket{0},
\end{equation}
the wave function with one particle in each beam. The probability $p_{\Psi \rightarrow M}$ to detect the output state $\ket{M}$, given that we injected the wave function $\ket{\Psi}$ into the beamsplitter, is then given by
\begin{align}\label{eq:probcoince}
p_{\Psi \rightarrow M} &= \abs{\bra{ M } E(U) \ket{\Psi}}^2\\
& = 0 \qquad \text{for bosons,}\label{eq:probcoinceB}\\
& = 1 \qquad \text{for fermions,}\label{eq:probcoinceF}
\end{align}
where we used (\ref{eq:HOMB},\ref{eq:HOMF}), and applied the commutation relations (\ref{eq:CCR}) for bosons and the anti-commutation relations (\ref{eq:CAR}) for fermions. For bosons, the experimental outcome associated with $\ket{M}$ never occurs, thus, we say that this measurement outcome is {\em suppressed}.

To further the interpretation of this suppression phenomenon as a quantum interference effect, we must understand what happens to the system when we have some ``which-particle'' information. The simplest way of approaching this scenario is to assume that both particles are completely different, and following them through the interferometer. Because the particles are distinguishable, they evolve independently of each other, and we can use the standard rules for composition of probabilities. This allows us to briefly forget about the whole quantum mechanical treatment. In this case, we can simply look at the probability that the particle in the $j$th input beam is detected in the $k$th output beam, $p_{j \rightarrow k}$, which is given by
\begin{equation}\label{eq:singlePartProb}
p_{j \rightarrow k} = \abs{\bra{e_k}U \ket{e_j}}^2.
\end{equation}
Because fully distinguishable particles in different beams are completely uncorrelated, we can simply evaluate the two-particle probability with combinatorics:
\begin{equation}\label{eq:prob_dist}
p_{\{1,2\} \rightarrow \{1,2\}} =  p_{1 \rightarrow 1}p_{2 \rightarrow 2} + p_{1 \rightarrow 2}p_{2 \rightarrow 1} = \frac{1}{2},
\end{equation} 
where we see that the probability of a coincidence count for distinguishable particles is different from both to bosonic and the fermionic case. In the context of quantum interference, we see destructive interference for this output event in the bosonic case, whereas the interference is constructive for fermions.\\

To finalise our understanding of this interference effect, we develop a quantum mechanical derivation of the probability $p_{\{1,2\} \rightarrow \{1,2\}}$ for distinguishable particles. For this derivation, we go back to Section \ref{sec:Distinguish}, where it was argued that internal DOF are sufficient to render particles distinguishable. As such, we extend the single-particle Hilbert space to ${\cal H} = \mathbb{C}^2 \otimes {\cal H}_I$, where $\mathbb{C}^2$ still represents the two beams that are mixed on the beamsplitter, and ${\cal H}_I$ describes the internal DOF (which we will leave unspecified for the time being). When we assume that there is no entanglement between the beam and the internal DOF, we define the single-particle wave functions for the particles' internal DOF $\phi, \psi \in {\cal H}_I$, such that the full many-particle wave function $\ket{\Psi} \in {\cal F}({\cal H})$ is now given by
\begin{equation}
\ket{\Psi} = a^{\dag}(e_1 \otimes \phi) a^{\dag}(e_2 \otimes \psi) \ket {0}.
\end{equation}
We then assume that the beamplitter only mixes the beams, leaving the internal degrees untouched, such that its operation on the many-particle wave function can be described by $E(U\otimes \mathbb{1})$, and
\begin{equation}
E(U\otimes \mathbb{1}) = a^{\dag}(U e_1 \otimes \phi) a^{\dag}(U e_2 \otimes \psi) \ket {0}.
\end{equation}
To add generality to this treatment, we leave $U$ unspecified for the moment. The measurement stage is the most subtle one, since we want the detectors to resolve the number of particles in the output beams, and make a detection regardless of the internal DOF. However, we still need to define the POVM on the full Hilbert space ${\cal F}({\cal H})$. To this end, we define a basis ${\cal I} = \{f_1, f_2, \dots\}$ of ${\cal H}_I$, measurement states of the form\footnote{For simplicity, we choose the basis for the internal DOF to be countable, but the presented argumentation also carries through for continuous bases.}
\begin{equation}
\ket{M; q,r} = a^{\dag}(e_1 \otimes f_q) a^{\dag}(e_2 \otimes f_r) \ket {0},
\end{equation}
and the POVM element, associated with a detection of one particle in each beam, as
\begin{equation}
P_M = \sum_{q,r} \ket{M; q,r}\bra{M; q,r}.
\end{equation}
In other words, we sum (or integrate, in the case of a continuous basis) over all internal DOF of each particle. The probability (\ref{eq:probcoince}) is now generalised to
\begin{align}
p_{\Psi \rightarrow M} &= \bra{ \Psi } E(U^{\dag} \otimes \mathbb{1}) P_M E(U\otimes \mathbb {1}) \ket{\Psi}\\
&= \sum_{q,r} \abs{\bra{0} a(e_2 \otimes f_r)a(e_1 \otimes f_q) a^{\dag}(U e_1 \otimes \phi) a^{\dag}(U e_2 \otimes \psi) \ket {0}}^2.
\end{align}
We can again use (\ref{eq:CCR}, \ref{eq:CAR}) to simplify this expression, and we obtain
\begin{align}\label{eq:HOM1}
p_{\Psi \rightarrow M} &= \sum_{q,r} \abs{U_{11}U_{22} \langle f_q \mid \phi  \rangle \langle f_r \mid \psi  \rangle + U_{12}U_{21} \langle f_r \mid \phi  \rangle \langle f_q \mid \psi  \rangle}^2 \quad \text{for bosons},\\
&= \sum_{q,r} \abs{U_{11}U_{22} \langle f_q \mid \phi  \rangle \langle f_r \mid \psi  \rangle - U_{12}U_{21} \langle f_r \mid \phi  \rangle \langle f_q \mid \psi  \rangle}^2 \quad \text{for fermions},\label{eq:HOM1F}
\end{align}
where we introduce the shorthand notation $U_{ij} = \bra{e_i}U\ket{e_j}$. Because it is useful to understand the more complicated results that will follow, we will now go through the full evaluation of (\ref{eq:HOM1}) and (\ref{eq:HOM1F}):
\begin{align}\label{eq:HOM2}
&\abs{U_{11}U_{22} \langle f_q \mid \phi  \rangle \langle f_r \mid \psi  \rangle \pm U_{12}U_{21} \langle f_r \mid \phi  \rangle \langle f_q \mid \psi  \rangle}^2\\
&\qquad= U_{11}U_{22}U^*_{11}U^*_{22}  \langle \phi \mid f_q \rangle \langle f_q \mid \phi  \rangle  \langle \psi \mid f_r  \rangle \langle f_r \mid \psi  \rangle  \nonumber\\
&\qquad\quad+ U_{12}U_{21}U^*_{12}U^*_{21}  \langle \phi \mid f_r \rangle \langle f_r \mid \phi  \rangle  \langle \psi \mid f_q  \rangle \langle f_q \mid \psi  \rangle \nonumber\\
&\qquad\quad \pm U_{11}U_{22}U^*_{12}U^*_{21}  \langle \phi \mid f_r \rangle \langle f_r \mid \psi  \rangle  \langle \psi \mid f_q  \rangle \langle f_q \mid \phi  \rangle \nonumber\\
&\qquad\quad \pm U_{12}U_{21}U^*_{11}U^*_{22}  \langle \psi \mid f_r \rangle \langle f_r \mid \phi  \rangle  \langle \phi \mid f_q  \rangle \langle f_q \mid \psi  \rangle, \nonumber
\end{align}
where we used the compact notation ``$\pm$'', where ``$+$'' refers to bosons and ``$-$'' fermions.
Furthermore, we use that
\begin{equation}\label{eq:HOM3}
\sum_q \ket{f_q}\bra{f_q} = \sum_r \ket{f_r}\bra{f_r} = \mathbb{1}.
\end{equation}
If we then insert (\ref{eq:HOM2}) and (\ref{eq:HOM3}) in(\ref{eq:HOM1}), we find that {\em for bosons}
\begin{align}\label{eq:HOMB4}
p_{\Psi \rightarrow M} = &\abs{U_{11}}^2 \abs{U_{22}}^2 + \abs{U_{12}}^2 \abs{U_{21}}^2 \\
& + \abs{ \langle \psi \mid \phi \rangle}^2 \big( U_{12}U_{21}U^*_{11}U^*_{22} +  U_{11}U_{22}U^*_{12}U^*_{21} \big),\nonumber
\end{align}
and {\em for fermions}
\begin{align}\label{eq:HOMF4}
p_{\Psi \rightarrow M}  = &\abs{U_{11}}^2 \abs{U_{22}}^2 + \abs{U_{12}}^2 \abs{U_{21}}^2 \\&- \abs{ \langle \psi \mid \phi \rangle}^2 \big( U_{12}U_{21}U^*_{11}U^*_{22} +  U_{11}U_{22}U^*_{12}U^*_{21} \big).\nonumber
\end{align}
Interestingly, the first two terms in both (\ref {eq:HOMB4}) and (\ref{eq:HOMF4}) correspond to the transition probabilities for distinguishable particle, i.e.~$\abs{U_{11}}^2 \abs{U_{22}}^2 + \abs{U_{12}}^2 \abs{U_{21}}^2  = p^{\rm class}_{\{1,2\} \rightarrow \{1,2\}}$ as given by (\ref{eq:prob_dist}). In other words, we can interpret (\ref {eq:HOMB4},\ref{eq:HOMF4}) as
\begin{equation}\label{eq:HOMInterpret}
p_{\Psi \rightarrow M} = p^{\rm class}_{\{1,2\} \rightarrow \{1,2\}} + \abs{ \langle \psi \mid \phi \rangle}^2 \times \text{interference terms}.
\end{equation}
Note in particular the appearance of the factor $\abs{ \langle \psi \mid \phi \rangle}^2$ in front of the interference terms, which represents the overlap of the particles' wave functions for the internal degree of freedom. In Section \ref{sec:Distinguish}, we noted that particles are effectively distinguishable when thee wave functions for their internal degree of freedom are orthogonal, i.e.~when $\abs{ \langle \psi \mid \phi \rangle}^2 = 0$. In this case, (\ref{eq:HOMInterpret}) shows clearly the the interference terms vanish, and that we recover the result (\ref{eq:prob_dist}) for distinguishable particles. On the other hand, when $\abs{ \langle \psi \mid \phi \rangle}^2 = 1$, it is directly verified that we obtain (\ref{eq:HOMB}) for bosons and (\ref{eq:HOMF}) for fermions. In other words, when the wave functions for the internal degree of freedom are the same (up to a phase), we find the result for indistinguishable particles. 

In a more general sense, $\abs{ \langle \psi \mid \phi \rangle}^2$ represents the degree up to which we can distinguish the two particles. As such, it directly captures the amount of ``which-particle'' information that is present in the experiment. Just as for ``which-way'' information in standard interference experiments, we observe that ``which-particle'' information destroys the many-particle interference effects.\\

To conclude this section, we present to Hong-Ou-Mandel interference effect as it is commonly used in experiments, and choose the time-frequency domain as additional internal\footnote{One could of course debate whether the time-frequency domain is really an ``internal'' degree of freedom, but at least it is an additional degree of that often renders identical particles distinguishable.} degree of freedom \cite{TimeFrequency}. 
%Note that this is based on a quantum optics experiment with photons, which means that one must in principle consider relativistic effects. In our discussion, the important impact of relativity is that photons are not necessarily localised in time. In practice this boils down to setting ${\cal H}_{I} = {\cal L}^2(\mathbb{R})$, where we can consider a basis of plane waves $\ket{\omega}$ with well-defined frequency $\omega$, or a basis of localised pulses $\ket{t}$ that arrive at a well-defined time $t$. We then find that $\langle t \mid \omega \rangle = \exp(-it\omega),$ which represents that plane waves are delocalised over time (we cannot be certain at what time a detector will click when it is illuminated by a photon of a well-defined frequency). In general, both of these bases are unphysical, and we will typically find states in superpositions of such time-frequency modes. Furthermore, we note that frequencies $\omega$ can in principle only be positive, which may cause problems in performing Fourier transforms. In practice, this problem can be solves by guaranteeing that the representations of time-frequency modes vanish for negative values of $\omega$.
We can expand a general time-frequency wave function $\ket{\psi} \in {\cal H}_{I} $ as 
\begin{equation}
\ket{\psi} = \int_{\mathbb{R}} {\rm d}t\, F_{\psi}(t) \ket{t} =  \int_{\mathbb{R}} {\rm d}\omega\, \tilde{F}_{\psi}(\omega) \ket{\omega},
\end{equation}
where $F_{\psi}(t)$ is a function that represents the state in the time domain, whereas it Fourier transform $\tilde{F}_{\psi}(\omega)$ describes the wave function $\ket{\psi}$ in the frequency domain. We then find that
\begin{equation}\label{eq:InnerproductTimeFrequency}
\langle \psi \mid \phi \rangle = \int_{\mathbb{R}} {\rm d}t\, F_{\psi}^*(t)F_{\phi}(t) = \int_{\mathbb{R}} {\rm d}\omega\, \tilde{F}^*_{\psi}(\omega)\tilde{F}_{\phi}(\omega),
\end{equation}
where $*$ indicates the complex conjugate. In other words, when we know how to represent the wave functions in either the time or frequency domain, we can use it to calculate $\abs{ \langle \psi \mid \phi \rangle}^2$ in (\ref{eq:HOMB4}) or (\ref{eq:HOMF4}).

\begin{figure}
\centering
\includegraphics[width=0.9 \textwidth]{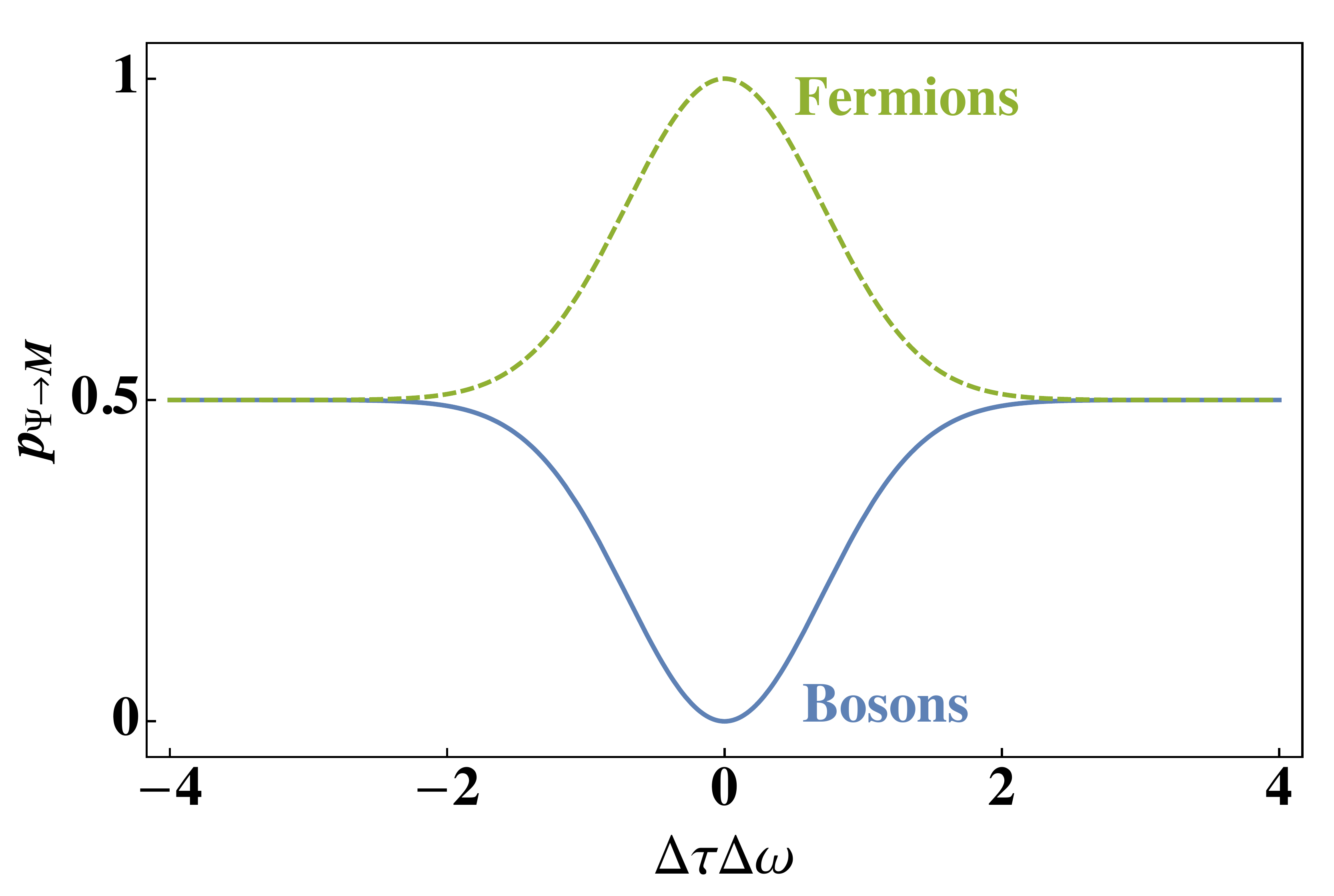}
\caption{Probability of a coincidence measurement at the two different output ports of a balanced beamsplitter, for two particles injected in distinct input ports. When varying the time delay $\Delta \tau$, with $\Delta \tau$ small as compared to the bandwidth $\Delta \omega$, destructive interference is seen for non-interacting bosons (blue solid line), whereas constructive interference, i.e. the Pauli effect, is observed for non-interacting fermions (green dashed line).}
\label{fig:HOM}
\end{figure}

In order to remain close to the the nature of the particles, we treat them as Gaussian wave packets that have a reasonably small uncertainty in time. In optics, one common solution to generate photons is by means of spontaneous parametric down-conversion (SPDC) \cite{Burnham:1970aa,Klyshko-1970}. In general, SPDC is a complicated process that generates (possibly entangled) photon pairs with properties that depend on the details of its implementation (e.g.~pump power, crystal length, the time-frequency mode of the pump, etc.) \cite{Silberhorn}. For simplicity, we will consider a particle with an expected arrival time $\tau$, with central frequency $\omega_0$ and an uncertainty $\Delta \omega$. In the frequency domain, the Gaussian wave packet for this particle is described by
\begin{equation}\label{eq:WavePacketFrequencyDomain}
\tilde{F}_{\psi}(\omega) =\frac{ e^{-\frac{(\omega-\omega_0)^2}{4 \Delta \omega^2}}}{ (2 \pi)^{1/4}  (\Delta \omega)^{1/2}} e^{i \omega \tau}.
\end{equation}
When we assume that both photons in the experiment are generated by the same SPDC process, the values $\omega_0$ and $\Delta \omega$ are the same for both particles. Hence, when we assume that one particle arrives at $\tau_1$ and the other at $\tau_2$, we directly evaluate 
\begin{equation}\label{eq:two-photon-dist}
\abs{\langle \psi \mid \phi \rangle}^2 = e^{-\Delta \omega^2 \Delta \tau^2},
\end{equation}
where $\Delta \tau = \tau_1 - \tau_2$. This result is narrowly related to the so-called Fourier uncertainty relation, which relates the temporal width of a pulse to the spread in the frequency domain (i.e.~the bandwidth). When $\Delta \omega^2 \Delta \tau^2 \rightarrow 0$, and the time-delay between the particle becomes small with respect to the distribution of frequencies, we observe that the particles behave in a distinguishable way. In particular, we can now insert (\ref{eq:50-50-BS}) and (\ref{eq:two-photon-dist}) in (\ref{eq:HOMB4}) to obtain
\begin{equation}
p_{\Psi \rightarrow M} = \frac{1}{2} (1 - e^{-\Delta \omega^2 \Delta \tau^2}) \quad \text{for bosons.}
\end{equation}
If we replace the bosonic photons with a fermionic context, for example by working with matter waves, we find
\begin{equation}
p_{\Psi \rightarrow M} = \frac{1}{2} (1 + e^{-\Delta \omega^2 \Delta \tau^2}) \quad \text{for fermions.}
\end{equation}
As such, we find the well-known Hong-Ou-Mandel dip, and its fermionic equivalent, shown in Fig.~\ref{fig:HOM}. It is clearly seen that we recover also the limiting cases (\ref{eq:probcoinceB}), (\ref{eq:probcoinceF}), and (\ref{eq:prob_dist}). This solidifies the observed phenomenon as a two-particle interference effect, and highlights the importance of distinguishability as a form of ``which-particle'' information that destroys the interference.

\subsection{Determinants and permanents}\label{sec:DetPerm}

The above two-particle interference in the Hong-Ou-Mandel setup can also be observed in many-particle experiments. In this section, we will assume that particles are either fully indistinguishable or fully distinguishable and determine the expressions for the many-particle transition probability $p_{\Psi \rightarrow M}$ in a larger interferometer. In the next section, we present a full quantum treatment of the case where the particle have an internal degree of freedom that allows us to make them (partially) distinguishable.\\

In the Hong-Ou-Mandel scenario, we started by mixing beams in a beamsplitter, which presented us with a simple interferometer. For our many-particle, setup, we generalise this aspect and inject $n$ particles in a large multiport interferometer, with $m$ input ports and $m$ output ports. In analogy to the previous section, we can then treat the single-particle space as being $m$-dimensional, such that ${\cal H} \cong \mathbb{C}^m$. We will assume that the interferometer is simply described by an $m \times m$ unitary matrix $U$, which connects the input ports to the output ports.

We start by considering distinguishable particles, in order to generalise (\ref{eq:prob_dist}). Note that (\ref{eq:singlePartProb}) remains valid in this new scenario: the probability that a particle, which is injected in an input port $j$, is detected in output port $k$ is given by
\begin{equation}\label{eq:singlePartProb2}
p_{j \rightarrow k} = \abs{\bra{e_k}U\ket{e_j}}^2 = \abs{U_{kj}}^2,
\end{equation}
where $e_k$ is the $k$the vector in the standard basis, and thus represents a single-particle wave function that is localised on the $k$ input/output port. We are now interested in a set of $n$ distinguishable particles, that are injected in input ports $i_1, \dots, i_n$ (e.g.~if we inject $n$ particles in the first $n$ input ports, these labels are set to  $i_1 = 1, i_2=2, \dots, i_n=n$), and we evaluate the probability that they are detected in output ports $o_1, \dots, o_n$. This transition probability can be calculated in complete analogy to (\ref{eq:prob_dist}), but we must take all the possible permutations of particles into account. However, when an output port is occupied by several particles, e.g.~$o_1=o_2$, we must make sure to avoid double counting, and therefore we must divide the result by $\sum_{\sigma \in S_n} \delta_{o_1,o_{\sigma(1)}} \dots \delta_{o_n,o_{\sigma(n)}}$. Thus, we obtain
\begin{align}\label{eq:probMultiDist}
p_{\{i_1, \dots, i_n\} \rightarrow \{o_1, \dots, o_n\}} &= \frac{\sum_{\sigma \in S_n} p_{i_1 \rightarrow o_{\sigma(1)}}\dots  p_{i_n \rightarrow o_{\sigma(n)}}}{\sum_{\sigma \in S_n} \delta_{o_1,o_{\sigma(1)}} \dots \delta_{o_n,o_{\sigma(n)}}}\\
& = \frac{\sum_{\sigma \in S_n} \abs{U_{i_1 o_{\sigma(1)}}}^2 \dots  \abs{U_{i_n o_{\sigma(n)}}}^2}{\sum_{\sigma \in S_n} \delta_{o_1,o_{\sigma(1)}} \dots \delta_{o_n,o_{\sigma(n)}}}.\label{eq:prob_dist_n}
\end{align}
The quantity $p_{\{i_1, \dots, i_n\} \rightarrow \{o_1, \dots, o_n\}}$ can be treated as a probability distribution that describes with which probability a certain set of output detectors $o_1, \dots, o_n$ simultaneously click upon the injection of particles in input ports $i_1, \dots, i_n$. Because the number of terms in this sum grows as $n!$ with the number of particles, these probabilities are in generally not easy to calculate. Nevertheless, it is straightforward to sample clicks of output detectors, i.e.~choosing $o_1, \dots, o_n$, which respect the probability distribution (\ref{eq:prob_dist_n}). To do so, one can use (\ref{eq:singlePartProb2}) for each individual particle to select an output port.\\

The derivation becomes considerably more complicated when we consider indistinguishable fermions or bosons. In analogy to the two-particle setting, we define the initial state 
\begin{equation}
\ket{\Psi} = a^{\dag}(e_{i_1})\dots a^{\dag}(e_{i_n})\ket{0},
\end{equation}
and in order to be able to compare fermions and bosons, we assume that all particles are injected in different modes, i.e.~$i_1 \neq i_2 \neq \dots \neq i_n$. The action of the interferometer is then given by
\begin{equation}
\ket{\Psi} \mapsto E(U)\ket{\Psi} =  a^{\dag}(Ue_{i_1})\dots a^{\dag}(Ue_{i_n})\ket{0}.
\end{equation}
We approach the problem via the measurement state associated with the POVM that projects on the output ports $o_1, \dots, o_n$, where the detectors are placed:
\begin{equation}\label{eq:M-manyPart}
\ket{M} = \frac{1}{\cal N}a^{\dag}(e_{o_1})\dots a^{\dag}(e_{o_n})\ket{0},
\end{equation}
where the normalisation constant is different from the case where multiple particles land in the same detector. The quantity of interest is the transition probability 
\begin{equation}\label{eq:transitionProbManyPartNotFinal}
p_{\Psi \rightarrow M} = \abs{\bra{ M } E(U) \ket{\Psi}}^2 = \frac{1}{{\cal N}^2}\abs{\bra{0}a(e_{o_n})\dots a(e_{o_1}) a^{\dag}(Ue_{i_1})\dots a^{\dag}(Ue_{i_n})\ket{0}}^2,
\end{equation}
which can be calculated using Wick's theorem, based on the (anti-)commutation relations for creation and annihilation operators. Alternatively, some reader may find it more convenient to go back to first quantisation and apply (\ref{eq:PsiB}, \ref{eq:PsiF}) to evaluate this quantity. Both methods ultimately lead to the general result that 
\begin{align}
&\label{eq:bosonicInnerProd}\bra{0}a(\phi_n)\dots a(\phi_1) a^{\dag}(\psi_1)\dots a^{\dag}(\psi_n)\ket{0} \\ &\qquad= \sum_{\sigma \in S_n} \langle \phi_1 \mid \psi_{\sigma(1)}\rangle \dots  \langle \phi_n \mid \psi_{\sigma(n)}\rangle \quad \text{for bosons,}\nonumber\\
&\label{eq:fermionicInnerProd}\bra{0}a(\phi_n)\dots a(\phi_1) a^{\dag}(\psi_1)\dots a^{\dag}(\psi_n)\ket{0} \\ &\qquad = \sum_{\sigma \in S_n} {\rm sign}(\sigma) \langle \phi_1 \mid \psi_{\sigma(1)}\rangle \dots  \langle \phi_n \mid \psi_{\sigma(n)}\rangle \quad \text{for fermions.}\nonumber
\end{align}
Note that these results can also be used to determine the normalisation factors for many-particle wave functions such as (\ref{eq:M-manyPart}). In that case, the state is chosen such that $\langle e_{o_k} \mid e_{o_l} \rangle = \delta_{o_k,o_l}$, such that (\ref{eq:bosonicInnerProd}) implies that
\begin{equation}
{\cal N} = \bra{0}a(e_{o_n})\dots a(e_{o_1}) a^{\dag}(e_{o_1})\dots a^{\dag}(e_{o_n})\ket{0} =\sqrt{\sum_{\sigma \in S_n} \delta_{o_1,o_{\sigma(1)}} \dots \delta_{o_n,o_{\sigma(n)}}} \quad \text{for bosons,}
\end{equation}
and, as indicated before, ${\cal N} = 1$ for fermions, since we cannot have more than one fermion in each output port. It should not come as a surprise that the normalisation factor resembles the factor that was introduced in (\ref{eq:prob_dist_n}) to avoid double counting.

The results (\ref{eq:bosonicInnerProd}, \ref{eq:fermionicInnerProd}) can be used to evaluate (\ref{eq:transitionProbManyPartNotFinal}), which leads to
\begin{align}
p_{\Psi \rightarrow M} &= \frac{\abs{\sum_{\sigma \in S_n} U_{ o_{\sigma (1)} i_1} \dots  U_{o_{\sigma (n)} i_n}}^2}{\sum_{\sigma \in S_n} \delta_{o_1,o_{\sigma(1)}} \dots \delta_{o_n,o_{\sigma(n)}}} \quad \text{for bosons,}\label{eq:bosonicExpression}\\
&= \abs{\sum_{\sigma \in S_n} {\rm sign}(\sigma) \, U_{ o_{\sigma (1)}i_1} \dots  U_{o_{\sigma (n)} i_n}}^2 \quad \text{for fermions.}
\end{align}
There are several convenient ways to rewrite these expressions. We start by applying the method that explicitly shows the interference terms:
\begin{align}
\abs{\sum_{\sigma \in S_n} U_{ o_{\sigma (1)} i_1} \dots  U_{o_{\sigma (n)} i_n}}^2 =& \sum_{\sigma, \sigma' \in S_n} U_{o_{\sigma (1)} i_1} \dots  U_{o_{\sigma (n)} i_n}U^*_{o_{\sigma' (1)} i_1} \dots  U^*_{o_{\sigma' (n)} i_n}\\
=& \sum_{\sigma \in S_n} \abs{U_{o_{\sigma (1)} i_1}}^2 \dots  \abs{U_{o_{\sigma (n)} i_n}}^2\\
&+  \sum_{\substack{\sigma, \sigma' \in S_n\\ \sigma \neq \sigma'}} U_{o_{\sigma (1)} i_1 } \dots  U_{ o_{\sigma (n)} i_n}U^*_{o_{\sigma' (1)} i_1} \dots  U^*_{o_{\sigma' (n)} i_n}\nonumber.
\end{align}
Thus, when we focus on output events that are compatible with the Pauli exclusion principle, i.e.~$o_1 \neq o_2 \neq \dots \neq o_n,$ we can rewrite (\ref{eq:bosonicInnerProd}, \ref{eq:fermionicInnerProd}) as
\begin{align}
p_{\Psi \rightarrow M} =& \sum_{\sigma \in S_n} \abs{U_{ o_{\sigma (1)} i_1}}^2 \dots  \abs{U_{ o_{\sigma (n)} i_n}}^2\label{eq:ManyPartIntB}\\
&+  \sum_{\substack{\sigma, \sigma' \in S_n\\ \sigma \neq \sigma'}} U_{ o_{\sigma (1)} i_1} \dots  U_{ o_{\sigma (n)} i_n}U^*_{o_{\sigma' (1)} i_1} \dots  U^*_{o_{\sigma' (n)} i_n} \quad \text{for bosons,}\nonumber\\
p_{\Psi \rightarrow M} =& \sum_{\sigma \in S_n} \abs{U_{o_{\sigma (1)} i_1}}^2 \dots  \abs{U_{o_{\sigma (n)} i_n}}^2\label{eq:ManyPartIntF}\\
&+  \sum_{\substack{\sigma, \sigma' \in S_n\\ \sigma \neq \sigma'}} {\rm sign}(\sigma)\,{\rm sign}(\sigma')\, U_{o_{\sigma (1) i_1 }} \dots  U_{o_{\sigma (n) i_n}}U^*_{ o_{\sigma' (1) i_1}} \dots  U^*_{ o_{\sigma' (n) i_n}} \quad \text{for fermions}\nonumber.
\end{align}
as we can see, we have recovered the transition probability for distinguishable particles (\ref{eq:prob_dist_n}), garnished by additional interference terms. For the two-particle case in (\ref{sec:HOM}), we found that bosonic and fermionic interference behave in a completely opposite way. However, (\ref{eq:ManyPartIntB}) and (\ref{eq:ManyPartIntF}) show that this observation does not generalise the the many-particle case. We find that the interference terms for bosons and fermions are the same up to a sign.%, and the terms for which ${\rm sign}(\sigma){\rm sign}(\sigma') = -1$ are opposite for bosons and fermions. This is contrasted with terms for which ${\rm sign}(\sigma){\rm sign}(\sigma') = 1$, where the bosonic and fermionic interference process described by this particular term is in fact the same.
This means that for many particles, where one must consider a large variety of permutations $\sigma$, we find a behaviour which can be rich and subtle. Furthermore,  (\ref{eq:ManyPartIntB}) and (\ref{eq:ManyPartIntF}) also indicate the the number of interference terms that is to be considered grows roughly as $(n!)^2$ with the number of particles $n$. Hence, it is reasonable to expect that calculating these probabilities is, in general, a challenging task. %A priori, one can therefore expect that calculating these probabilities is a complicated task.\\

To delve deeper into the evaluation of $p_{\Psi \rightarrow M}$, we can recast the probability (\ref{eq:bosonicInnerProd}, \ref{eq:fermionicInnerProd}) in yet another form. First, let us define the matrix $U_{\rm sub}$ -- an $n \times n$ matrix that connects the occupied input ports in $\ket{\Psi}$ to the output ports in $\ket{M}$. We explicitly construct this matrix in terms of its components
\begin{equation}\label{eq:Usub}
(U_{\rm sub})_{jk} = U_{o_j i_k}.
\end{equation}
Now, we use this matrix to rewrite the fermionic result (\ref{eq:fermionicInnerProd}) as follows
\begin{equation}\label{eq:detU}
p_{\Psi \rightarrow M} = \abs{\sum_{\sigma \in S_n} {\rm sign}(\sigma) \, U_{ o_{\sigma (1)}i_1} \dots  U_{o_{\sigma (n)} i_n}}^2 = \abs{\det U_{\rm sub}}^2 \quad \text{for fermions.}
\end{equation}
For bosons, we are confronted with an object that resembles a determinant, but it does not take into account the sign of the permutation. This object is commonly known as the {\em permanent}. For a general $n \times n$ matrix $A$, the permanent is defined as
\begin{equation}
{\rm perm} A = \sum_{\sigma \in S_n} A_{1 \sigma(1)} \dots A_{n \sigma(n)}.
\end{equation}
We can thus rewrite (\ref{eq:bosonicExpression}) as 
\begin{equation}\label{eq:permU}
p_{\Psi \rightarrow M}  = \frac{\abs{{\rm perm}\, U_{\rm sub}}^2}{{\rm perm}\, I},
\end{equation}
where we define $I$ as the matrix with components $I_{jk} = \delta_{o_j, o_k}$. The calculation of ${\rm perm}\, I$ is ultimately a counting exercise where notational overhead is the main difficulty. In the above treatment, we consider the state $\ket{M}$ in terms of its particle nature, and describe it by associating an output port $o_j$ to every particle. However, by virtue of (\ref{eq:isoHilbert2}) we can equivalently describe $\ket{M}$ in terms of the number of particles in each output port. We can define the {\em mode occupation vector} $\vec M$ as an $m$-dimensional vector ($m$ being the number of output ports), and the $k$th component $M_k$ describes the number of particles in the $k$th output port. A good exercise to get insight in permanents is to show that
\begin{equation}
{\rm perm} \, I = \prod_{k = 1}^m M_k!.
\end{equation}
In general, the quantity ${\rm perm}\, U_{\rm sub}$ in (\ref{eq:permU}) is much harder to evaluate, since $U_{\rm sub}$ does not generally have a structure that simplifies the evaluation. As a matter of fact, calculating permanents is a problem that falls in the complexity class $\# P$, making it a notoriously hard computational problem. The determinant in (\ref{eq:detU}) is a much simpler object to evaluate, because of its basis-independence. Hence, for fermionic processes, we can perform a convenient decomposition of $U_{\rm sub}$ and use it to calculate $p_{\Psi \rightarrow M}$ in an efficient way. For bosons, none of these tricks apply. Thus, even though the interference phenomena for bosons (\ref{eq:ManyPartIntB}) and fermions (\ref{eq:ManyPartIntF}) seem highly similar, the bosonic transition probability (\ref{eq:permU}) is much harder to calculate (with present day computers it the calculations become unfeasible around $\approx 60$ bosons).

As we argued before, we can interpret $p_{\Psi \rightarrow M}$ as the probability to measure the state given by mode occupation list $\vec M$, given that we prepared the system in initial state $\Psi$. %In this sense, we can try to pick a set of different detector outputs $\vec M$ in a way that respects the probabilities $p_{\Psi \rightarrow M}$ \cite{Clifford:2018:CCB:3174304.3175276}. 
For bosonic particles, generating detector outputs that respect the probability distribution $p_{\Psi \rightarrow M}$ is known as {\em Boson Sampling}. It turns out that even performing this sampling is hard for a classical computer \cite{aaronson_computational_2013,Clifford:2018:CCB:3174304.3175276}, unless there is a structure in $U_{\rm sub}$ that allows us to calculate permanents in an efficient way (e.g.~when $U_{\rm sub}$ is the identity matrix, everything becomes trivial). This result is strong, in the sense that it does not depend on the type of algorithm that one uses to simulate the sampling. In a sense, this means that -- assuming highly plausible conjectures from computational complexity theory  \cite{aaronson_computational_2013} -- there is no way to circumvent the computational difficulties that are induced by these permanents.

A final interesting remark about the fundamental difference between the computation complexity of fermionic and bosonic interference traces back to Section \ref{sec:GaussianStates}. In the literature of continuous-variable quantum computation, it is quite well-known that Gaussian measurements of Gaussian states can be efficiently simulated with classical computers. Hence, non-Gaussian elements are understood to be a necessary feature to reach a {\em quantum computational advantage}. At present, the exact relation between non-Gaussian features and quantum advantages is not completely understood. Discussions about the importance of non-Gaussian features in either the measurement or the state is typically limited to bosonic systems. In our discussion of many-fermion interference, $\ket{\Psi}$ and $\ket{M}$ were Gaussian states, and the obtained interferences can be simulated efficiently \cite{aaronson_computational_2013}. Hence, one may wonder whether many-fermion sampling can also lead to a quantum advantage when a non-Gaussian element (e.g.~interaction between the particles) is added. At present, this remains an open question.
%Hence, we emphasise here that $\ket{\Psi}$ and $\ket{M}$ are Gaussian states in the fermionic scenario. On the other hand, for bosons $\ket{\Psi}$ and $\ket{M}$ have non-Gaussian features. The importance of non-Gaussian features in either the measurement or the state is typically discussed purely in the context of bosonic systems, but it should be emphasised that non-Gaussian features in quantum computation advantages is also applicable for other particle species.

\subsection{Partial distinguishability} \label{sec:partialdist}
Just like in the two-particle case, we can study the transition from indistinguishable to distinguishable particles. The mechanism by which this process occurs is essentially the same as in Section \ref{sec:HOM}, but the resulting interference phenomena can behave quite differently.\\

As before, we extend the single-particle Hilbert space by adding an internal DOF. We denote this enlarged Hilbert space by ${\cal H}_I$. The full single-particle Hilbert space, upon which we construct the Fock space to describe the many-particle problem, is now described by ${\cal H} = {\cal H}_E \otimes {\cal H}_I$, where ${\cal H}_E$ is an m-dimensional Hilbert space that describes the $m$ input ports. The initial state now becomes
\begin{equation}
\ket{\Psi} = a^{\dag}(e_{i_1}\otimes \psi_1)\dots a^{\dag}(e_{i_n}\otimes \psi_n)\ket{0},
\end{equation}
where the $e_{i_j} \in {\cal H}_E$ are defined as in Section \ref{sec:DetPerm}. Furthermore, we assume that there is no entanglement between internal and external DOF. Moreover, we assume that $i_1 \neq i_2 \neq \dots \neq i_n$. The action of the interferometer is assumed to leave the internal DOF unchanged, and thus it is described by $E(U\otimes \mathbb{1})$. To describe measurements, we fix a basis ${\cal I} = \{f_1, f_2, \dots\}$ of ${\cal H}_I$, and describe the measurement states as
\begin{equation}
\ket{M; r_1,\dots, r_n} =  a^{\dag}(e_{o_1} \otimes f_{r_1})\dots a^{\dag}(e_{o_n} \otimes  f_{r_n})\ket{0}.
\end{equation}
For simplicity, we assume throughout this section that the output ports are different for all particles, i.e.~$o_1 \neq o_2 \neq \dots \neq o_n$. Note that $\ket{M; r_1,\dots, r_n}$ singles out a specific configuration for the internal DOF. However, the actual detectors are assumed to be blind for the internal DOF, which means that the associated POVM element is defined by
\begin{equation}
P_M = \sum_{r_1, \dots, r_n} \ket{M; r_1,\dots, r_n}\bra{M; r_1,\dots, r_n}.
\end{equation}
Here, every $r_k$ is a different index that is summed over, such that we sum (or integrate for continuous bases) over the internal DOF for all particles. As for the two-particle case, we again find that 
\begin{equation}
p_{\Psi \rightarrow M} = \bra{ \Psi } E(U^{\dag} \otimes \mathbb{1}) P_M E(U\otimes \mathbb {1}) \ket{\Psi}.
\end{equation}
To evaluate this probability we must combine the elements that we acquired in Section \ref{sec:DetPerm} via Wick's theorem, and the treatment of internal DOF of Section \ref{sec:HOM}. We first evaluate $\abs{\bra{M; r_1,\dots, r_n}E(U\otimes \mathbb{1})\ket{\Psi}}^2$ by applying (\ref{eq:bosonicInnerProd}) for bosons or (\ref{eq:fermionicInnerProd}) for fermions. Next, we use that $\sum_{r_k} \ket{f_{r_k}}\bra{f_{r_k}} = \mathbb{1}$ to deal with the internal DOF in the detectors, and eventually find that {\em for bosons}
\begin{align}
p_{\Psi \rightarrow M}=&\sum_{\sigma \in S_n} \abs{U_{ o_{\sigma (1)} i_1}}^2 \dots  \abs{U_{ o_{\sigma (n)} i_n}}^2\label{eq:ManyPartIntB-PartDist}\\
&+  \sum_{\substack{\sigma, \sigma' \in S_n\\ \sigma \neq \sigma'}} \langle \psi_{\sigma'(1)} \mid \psi_{\sigma(1)}\rangle \dots  \langle\psi_{\sigma'(n)} \mid \psi_{\sigma(n)} \rangle U_{ o_{\sigma (1)} i_1} \dots  U_{ o_{\sigma (n)} i_n}U^*_{o_{\sigma' (1)} i_1} \dots  U^*_{o_{\sigma' (n)} i_n}, \nonumber
\end{align}
and {\em for fermions}
\begin{align}\label{eq:ManyPartIntF-PartDist}
p_{\Psi \rightarrow M}=&\sum_{\sigma \in S_n} \abs{U_{ o_{\sigma (1)} i_1}}^2 \dots  \abs{U_{ o_{\sigma (n)} i_n}}^2\\
&+  \sum_{\substack{\sigma, \sigma' \in S_n\\ \sigma \neq \sigma'}}\Big( {\rm sign}(\sigma)\,{\rm sign}(\sigma')\, \langle \psi_{\sigma'(1)} \mid \psi_{\sigma(1)}\rangle \dots  \langle\psi_{\sigma'(n)} \mid \psi_{\sigma(n)}\rangle\nonumber\\
&\qquad\qquad\qquad \times U_{ o_{\sigma (1)} i_1} \dots  U_{ o_{\sigma (n)} i_n}U^*_{o_{\sigma' (1)} i_1} \dots  U^*_{o_{\sigma' (n)} i_n}\Big). \nonumber
\end{align}
We observe that, once again, we can interpret the probability the composition of two terms: the probability for distinguishable particles $p_{\{i_1, \dots, i_n\} \rightarrow \{o_1, \dots, o_n\}}$ (\ref{eq:probMultiDist}), and interference terms, like in (\ref{eq:HOMInterpret}). Note that all these interference terms come with a different weight, $ \langle \psi_{\sigma'(1)} \mid \psi_{\sigma(1)}\rangle \dots  \langle\psi_{\sigma'(n)} \mid \psi_{\sigma(n)}\rangle$, which characterises the ``which-particle'' information. In the simple case where $\psi_1 = \psi_2 =\dots =\psi_n$, we recover the expressions (\ref{eq:ManyPartIntB}) and (\ref{eq:ManyPartIntF}) for indistinguishable particles, whereas $\psi_1 \perp \psi_2 \perp \dots \perp\psi_n$ implies that all the factors $ \langle \psi_{\sigma'(1)} \mid \psi_{\sigma(1)}\rangle \dots  \langle\psi_{\sigma'(n)} \mid \psi_{\sigma(n)}\rangle = 0$ for $\sigma \neq \sigma'$, such that we recover the result of distinguishable particles.

In (\ref{eq:HOMB4}) and (\ref{eq:HOMF4}), we saw that two-particle interferences for bosons and fermions manifests itself in completely opposite manners.. The appearance of the factor $\abs{ \langle \psi \mid \phi \rangle}^2$ for the internal DOF moreover gives rise to a {\em monotonous} vanishing of many-particle interference. Much to the contrary, we see that (\ref{eq:ManyPartIntB-PartDist}) and (\ref{eq:ManyPartIntF-PartDist}) differ in a more subtle way, based on the signs of different permutations ${\rm sign}(\sigma){\rm sign}(\sigma')$. Moreover, there is a rich zoo of possible ways of rendering the particles distinguishable, and the factors $ \langle \psi_{\sigma'(1)} \mid \psi_{\sigma(1)}\rangle \dots  \langle\psi_{\sigma'(n)} \mid \psi_{\sigma(n)}\rangle$ certainly do not guarantee a monotonous transition. To some extent, this behaviour has been theoretically and experimentally explored in literature \cite{ra_nonmonotonic_2013,tichy_four-photon_2011}.\\

 %We again emphasise that this approach is more natural for bosons than for fermions. Nevertheless, one may consider fermionic matter waves, or fermionic wave packets traveling at a given speed towards the input ports, such that the dispersion relation translates the uncertainty in arrival to an uncertainty in spatial spread of the wave packet.

\begin{figure}
\centering
\includegraphics[width=0.99 \textwidth]{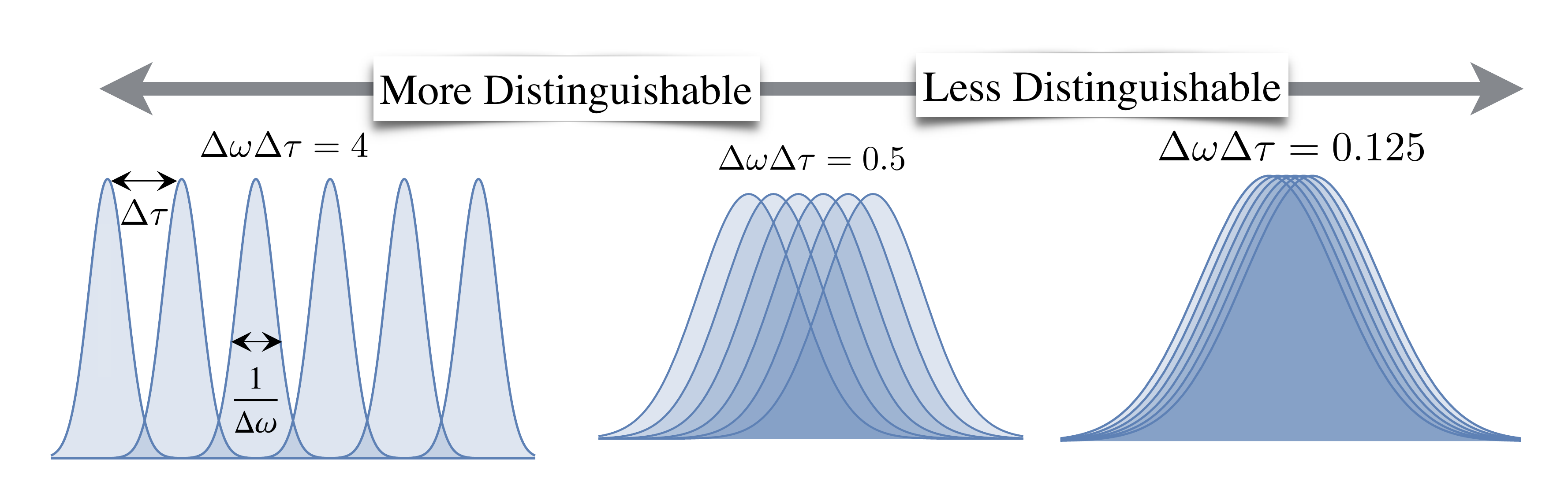}
\caption{Train of wave packets, with fixed time delay $\Delta \tau$ between subsequent wave packets, and fixed temporal width $1/\Delta \omega$ for each wave packets. The degree of distinguishability is shown to be controlled by a single variable: $\Delta \tau \Delta \omega$.}
\label{fig:WavePackets}
\end{figure}

To illustrate the richness of many-particle interference, we consider an example where particles are progressively rendered distinguishable through the time-frequency degree of freedom, as in Section \ref{sec:HOM}. In mathematical terms, we again follow (\ref{eq:WavePacketFrequencyDomain}) and represent the $j$th particle's internal time-frequency DOF by a wave function $\ket{\psi_j}$, with
\begin{equation}\label{eq:WavePacketFrequencyDomain2}
\tilde{F}_{\psi_j}(\omega) =\frac{ e^{-\frac{(\omega-\omega_0)^2}{4 \Delta \omega^2}}}{ \sqrt{(2 \pi)^{1/2}  \Delta \omega}} e^{i \omega \tau_j},
\end{equation}
where $\tau_j$ is the expected arrival time of the particle, $\omega_0$ is the central frequency, and $\Delta \omega$ is the uncertainty in the frequency domain (i.e.~the bandwith). Because we assume that all particles are generated by the same process, but at different times, we assume that $\omega_0$ and $\Delta \omega$ are the same for each particle. Using (\ref{eq:InnerproductTimeFrequency}), we obtain that
\begin{equation}\label{eq:DistParameter}
\langle \psi_j \mid \psi_k \rangle = e^{- \frac{1}{2} \Delta \omega^2 (\tau_j - \tau_k)^2} e^{i \omega_0 (\tau_j - \tau_k)},
\end{equation}
which can be directly inserted in (\ref{eq:ManyPartIntB-PartDist}, \ref{eq:ManyPartIntF-PartDist}) to calculate the transition probability for a specific choice of interferometer $U$ and output ports $\ket{M}$.

\begin{figure}
\centering
\includegraphics[width=0.99 \textwidth]{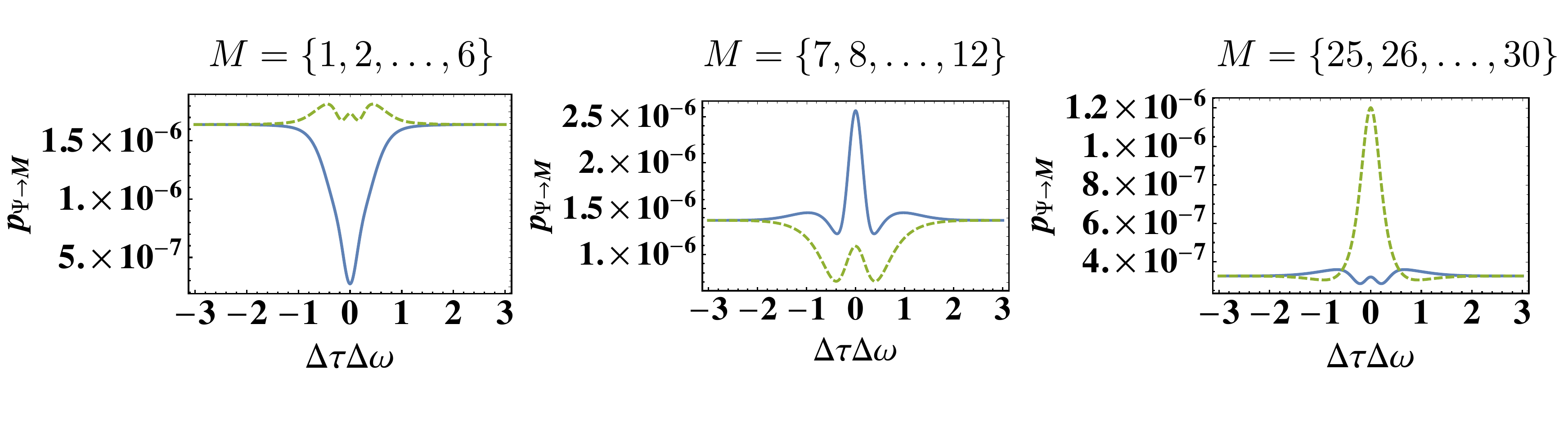}
\caption{Transmission probability $p_{\Psi \rightarrow M}$ (\ref{eq:ManyPartIntB-PartDist}, \ref{eq:ManyPartIntF-PartDist}) for varying distinguishability $\Delta \tau \Delta \omega$. Six bosonic (blue solid line) or fermionic (green dashed line) particles are injected in 30-mode interferometer that implements a randomly chosen (with respect to the Haar measure) unitary transformation $U$. Probabilities for a joint detection event in the detectors $M = \{1, \dots, 6\}$ (left panel), $M = \{7, \dots, 12\}$ (middle panel), and $M = \{25, \dots, 30\}$ (right panel) are shown. }
\label{fig:PartialDist1}
\end{figure}

In Fig.~\ref{fig:PartialDist1} we inject six particles in a 30-mode interferometer that is described by a randomly chosen unitary matrix $U$. The arrival times $\tau_j = j\Delta \tau$ of the particles are chosen such that there is a fixed time-delay $\Delta \tau$ between consecutive particles, as shown in Fig.~\ref{fig:WavePackets}. From (\ref{eq:DistParameter}), we see that the relevant quantity is the time delay in units of the bandwidth, i.e.~ $\Delta \tau \Delta \omega$, which is the parameter that is varied in Fig.~\ref{fig:PartialDist1}. Note that the parameter $\omega_0$ that appears in (\ref{eq:DistParameter}) does not appear in the final expression for $p_{\Psi \rightarrow M}$. From (\ref{eq:ManyPartIntB-PartDist}, \ref{eq:ManyPartIntF-PartDist}) we see that changing the time delays can significantly alter the weight of certain interference terms. Fig.~\ref{fig:PartialDist1} clearly shows that this has a profound and {\em non-monotonous} impact on $p_{\Psi \rightarrow M}$. All three panels are generated with the same input state and unitary interferometer (only the output state is varied) yet we observe a variety of qualitatively different interference phenomena. Note that both bosonic and fermionic many-particle interference can be destructive or constructive. Furthermore, the maximal deviation from the distinguishable-particle limit is not necessarily obtained in $\Delta \tau \Delta \omega = 0,$ which highlights the non-monotonicity. Finally, we also stress that fermionic and bosonic many-particle interference do not necessarily influence the dynamics in opposite directions. %are not necessarily opposites. %as generally both curves behave very differently.

In summary, we have learned that many-particle interference is a rich phenomenon that is hard to predict for growing number of particles. From the example in Fig.~\ref{fig:PartialDist1} we learn that even a single interferometer with a single $n$-particle input state can give rise to a very different behaviour for different output configurations. When we combine this aspect with the large number of possible output configurations and the associated low probabilities for each of these configurations, it becomes clear that many-particle interference effects are hard to unambiguously observe experimentally. This fact lies at the physical basis of the debate on the validation of Boson Sampling. Hence, what is needed to gain deeper insight in this phenomenon is a clear-cut experimental signature of genuine many-particle quantum interference.
 
\section{Signatures of many-particle interference}\label{sec:Signature}
Many-particle interferences are notoriously hard to calculate, especially for bosons, which makes it very hard validate the functionality of devices that use such interferences, e.g.~Boson Sampling. There have been a range of approaches to the problem of verifying Boson Sampling, including techniques from computer science \cite{aaronson_bosonsampling_2014,aolita_reliable_2015} and data science \cite{Agresti:2019aa,Flamini:2019aa}. In this Tutorial, we will focus on a series of rigorous approaches \cite{tichy_stringent_2014,walschaers_statistical_2014} that have a genuinely physical motivation, where the main goal is to identify and observe certain hallmarks of many-particle interference. The different approaches all come with a certain functionality, advantages, and disadvantages. An ideal validation scheme should be {\em robust} (i.e.~it should work for any interferometer regardless of $U$, even when there are errors in its implementation), {\em scalable} (i.e.~it should work for small and large numbers of modes and particles), and {\em versatile} (i.e.~it should be able to distinguish Boson Sampling from several other options). In Table \ref{tab:I}, we provide an overview of validation schemes for Boson Sampling that have been successfully realised in small-scale experiments. We note that the best methods known today are based on finding statistical patterns in sampling data, which can either be done with techniques from data science \cite{Agresti:2019aa,Flamini:2019aa}, or with the help of physical processes \cite{walschaers_statistical_2014}.

\begin{table}
    \begin{tabular}{| c || c | c | c | c | c |}
    \hline
    \multirow{2}{*}{{\bf Method}} &   \multirow{2}{*}{{\bf \em Robust?}} &   \multirow{2}{*}{{\bf \em Scalable?}} & \multicolumn{3}{  c |}{{\bf \em Versatile?}}  \\ \cline{4-6}
     &&& {\em Uniform} \cite{gogolin_boson-sampling_2013} & {\em Distinguishable} & {\em Mean-Field}  \cite{tichy_stringent_2014} \\ \hline
     Row-rank \cite{aaronson_bosonsampling_2014} & \ding{52} & \ding{52} & \ding{52} & \ding{56} & \ding{56}  \\ \hline
     Bunching \cite{carolan_experimental_2014} & \ding{52} & {\bf \em ?} & \ding{52} & \ding{52} & \ding{56} \\ \hline
     Likelihood \cite{spagnolo_experimental_2014} & \ding{56} & \ding{52} & \ding{52} & \ding{52} & {\bf \em ?}  \\ \hline
     Bayesian \cite{Bentivegna:2014aa} & \ding{56} & \ding{52} & \ding{52} & {\bf \em ?} & \ding{52} \\ \hline
     {\bf Suppression laws} \cite{tichy_stringent_2014} & \ding{56} & \ding{52} & \ding{52} & \ding{52} & \ding{52} \\ \hline
     Pattern recognition \cite{Agresti:2019aa} & \ding{52} & \ding{52} & \ding{52} &  \ding{52} & \ding{52} \\ \hline
     {\bf Statistical} \cite{walschaers_statistical_2014} & \ding{52} & \ding{52} & \ding{52} &  \ding{52} & \ding{52} \\ \hline
    \end{tabular}
\caption{List of validation schemes for Boson Sampling (with one important reference) which have been successfully implemented in proof-of-principle experiments. For each method, we indicate whether it is robust (i.e.~it can be used for any interferometer, and can tolerate errors in its implementation), scalable to large-scale implementations, and whether it is versatile. To probe versatility, we indicate whether they can identify the most common alternative sampling models in literature (i.e.~uniform sampling \cite{gogolin_boson-sampling_2013}, sampling of distinguishable particles, and mean-field sampling \cite{tichy_stringent_2014}). The validation methods that are highlighted in bold are discussed in the remainder of this tutorial. See \cite{Agresti:2019aa} for a more detailed table.}\label{tab:I}
\end{table}

%A first approach is based on the observation that bosonic interference generally tends to cause bunching \cite{carolan_experimental_2014}. Thus, one observes that the probability of detecting all particles in distinct output ports of the interferometer is lower for bosons than for distinguishable particles. However, it was shown that such bunching phenomena can be simulated with distinguishable particles and random phase fluctuations, which occurs within a so-called mean-field sampler \cite{tichy_stringent_2014}. Thus, we will explore more sensitive benchmarks for many-particle interference, which are considered to be among the most performant tests of Boson Sampling experiments \cite{Flamini:2018aa}.
%\cite{Wang-2016,Liu:2016aa}
 
\subsection{Suppression laws}\label{sec:SuppressionLaws}

Our first signature is a direct generalisation of the Hong-Ou-Mandel effect of Fig.~\ref{fig:HOM}. %, not in the sense of bunching effects, but rather in the sense of destructive interference. 
When particles in (\ref{eq:HOMB4}) are perfectly indistinguishable, and when the beamsplitter is balanced as in (\ref{eq:50-50-BS}), we observe that a coincidence event is fully suppressed by destructive interference. The concept of suppression laws generalises this destructive interference effect to the case of multiport interferometers in which many particles are injected, and one searches for measurement states $\ket{M}$ that are prohibited by the unitary evolution, i.e.~the probability $p_{\Psi \rightarrow M} = 0$. Note that, since we assume that the particles are fully indistinguishable, we will not consider internal DOF throughout this section.\\

The general bosonic and fermionic suppression laws, as described in \cite{dittel-2017,dittel2-2017}, are strongly related to symmetries that are reflected in, both, the initial state $\ket{\Psi}$ and the unitary transformation $U$. To describe these symmetries, we define the permutation matrix ${\cal P}_{\pi}$ that represents the {\em mode-permutation} $\pi \in S_m$ which acts on the $m$-dimensional single-particle Hilbert space ${\cal H}$ (corresponding to the $m$ input ports of the interferometer). %This permutation should not be confused with the particle-permutations (typically labelled $\sigma \in S_n$) that were introduced in Section \ref{sec:NPart}. 
In our specific context, we consider a permutation of the input ports of the interferometer. Using the notation introduced in Section \ref{sec:DetPerm}, these permutations are given by
%n the notation of Section \ref{sec:DetPerm} he representation ${\cal P}_{\pi}$, in the basis associated with the input ports of the interferometer, can be defined component-wise by
\begin{equation}
{\cal P}_{\pi} e_k = e_{\pi(k)},
\end{equation}
where $e_k$ is the single-particle wave function that is localised on the $k$th input port.
Because ${\cal P}_{\pi}$ is a unitary operator that acts on the single-mode Hilbert space ${\cal H}$, we can make it act on the many-particle state by constructing $E({\cal P}_{\pi})$ as in (\ref{eq:EA}). For simplicity, the input state $\ket{\Psi}$ is chosen such that each input port is populated by at most one particle:
\begin{equation}\label{eq:PsiAgainAndAgain}
\ket{\Psi} = a^{\dag}(e_{i_1})\dots a^{\dag}(e_{i_n})\ket{0}.
\end{equation}
The action of the mode-permutation $\pi \in S_m$ can be evaluated as
\begin{equation}
E({\cal P}_{\pi})\ket{\Psi} =  a^{\dag}(e_{\pi(i_1)})\dots a^{\dag}(e_{\pi(i_n)})\ket{0}.
\end{equation}
To implement the suppression law, we first identify the symmetries of the initial state, i.e.~permutations $\pi \in S_m$ for which
\begin{align}\label{eq:suppressionSymB}
&E({\cal P}_{\pi})\ket{\Psi} = \ket{\Psi} , \quad \text{for bosons,}\\
&E({\cal P}_{\pi})\ket{\Psi} = {\rm sign}(\pi) \ket{\Psi}, \quad \text{for fermions.}\label{eq:suppressionSymF}
\end{align}
As a next step, we must identify the interferometers that will give rise to suppressed output event. To do so, we define the  eigen-decomposition 
\begin{equation}\label{eq:ADA}
{\cal P}_{\pi} = A^{\dag} D A,
\end{equation}
where $A$ is a unitary matrix with the eigenvectors of ${\cal P}_{\pi}$ as columns, and $D$ is a diagonal matrix with the eigenvalues $\{\lambda_1, \dots, \lambda_m\}$ of ${\cal P}_{\pi}$ on the diagonal. To obtain suppressed output events, the interferometer must now be constructed such that 
\begin{equation}\label{eq:UisA}
U = A.
\end{equation}
Note that \cite{dittel-2017,dittel2-2017} provides an extended version where additional phase shifts are added to the input and output ports, but for simplicity we will not restrict to the simplest case (\ref{eq:UisA}). 

We project the output of this interferometer on a particular output state $\ket{M}$, given by 
\begin{equation}\label{eq:M-suppression}
\ket{M} = a^{\dag}(e_{o_1})\dots a^{\dag}(e_{o_n})\ket{0},
\end{equation}
where we assume for simplicity that there is at most one particle per output port, i.e.~$o_1 \neq o_2 \neq \dots \neq o_n$. The transition probability $p_{\Psi \rightarrow M}$ is then given by
\begin{equation}\label{eq:TransitionProbSuppression}
p_{\Psi \rightarrow M} = \abs{\bra{M}E(A)\ket{\Psi}}^2,
\end{equation}
and we can explore the impact of the symmetries (\ref{eq:suppressionSymB}, \ref{eq:suppressionSymF}) on $\bra{M}E(A)\ket{\Psi}$. 

Let us start by considering the case of {\em bosonic particles}. First of all, we use (\ref{eq:suppressionSymB}) to obtain the identity
\begin{equation}\label{eq:thisEqOverHere}
\bra{M}E(A)\ket{\Psi} = \bra{M}E(A)E({\cal P}_{\pi})\ket{\Psi}.
\end{equation}
First we use (\ref{eq:EProd}) to write $E(A)E({\cal P}_{\pi}) = E(A{\cal P}_{\pi})$ and then we insert (\ref{eq:ADA}) to find that $E(A)E({\cal P}_{\pi}) = E(D)E(A)$. Hence, we find that
\begin{equation}
\bra{M}E(A)\ket{\Psi} = \bra{M}E(D)E(A)\ket{\Psi}.
\end{equation}
Now, we can use (\ref{eq:EAnnil}) together with (\ref{eq:M-suppression}) to obtain
\begin{equation}
\bra{M}E(D) = \bra{0} a(De_{o_1})\dots a(De_{o_n}).
\end{equation}
Because $D$ is a diagonal matrix in the basis of localised single-particle wave functions $e_k$, we find that $D e_k = \lambda_k e_k,$ where $\lambda_k$ is the $k$th eigenvalue of ${\cal P}_{\pi},$ associated with the $k$th column of $A$. By virtue of the conjugate-linearity of the annihilation operators, we then find
\begin{equation}
\bra{M}E(D) = \left(\prod_{j = 1}^n \lambda_{o_j}^* \right) \bra{M}.
\end{equation}
By inserting the above identities in (\ref{eq:thisEqOverHere}) we obtain
\begin{equation}
\bra{M}E(A)\ket{\Psi} = \left(\prod_{j = 1}^n \lambda_{o_j}^*\right) \bra{M}E(A)\ket{\Psi},
\end{equation}
which implies
\begin{equation}\label{eq:bosonicSuppression}
\prod_{j = 1}^n \lambda_{o_j} \neq 1 \implies \bra{M}E(A)\ket{\Psi} = 0.
\end{equation}
Returning to (\ref{eq:TransitionProbSuppression}), this conclusion entails the suppression of the detection event associated with $\ket{M}$, i.e.~$p_{\Psi \rightarrow M} = 0$. 

For the {\em fermionic case}, the derivation is analogous. The key difference is that we start from the identity
\begin{equation}\label{eq:thisEqOverThere}
{\rm sign}(\pi)\bra{M}E(A)\ket{\Psi} = \bra{M}E(A)E({\cal P}_{\pi})\ket{\Psi}.
\end{equation}
This identity then leads us to the fermionic equivalent of (\ref{eq:bosonicSuppression}), which reads
\begin{equation}\label{eq:fermionicSuppression}
\prod_{j = 1}^n \lambda_{o_j} \neq {\rm sign}(\pi) \implies \bra{M}E(A)\ket{\Psi} = 0.
\end{equation}
Note that (\ref{eq:fermionicSuppression}) implies that the suppressed output events for bosons and fermions are the same when ${\rm sign}(\pi) = 1$. Nevertheless, it was shown in \cite{dittel-2017,dittel2-2017} that fermions give rise to an {\em extended suppression law} that excludes more output events than (\ref{eq:fermionicSuppression}). The easiest way to prove this extended suppression law is via (\ref{eq:detU}), where we the then use the properties of the determinant, and of $({\cal P}_{\pi})_{\rm sub}$. From (\ref{eq:ADA}), we can derive the identity
\begin{equation}
({\cal P}_{\pi})_{\rm sub} A_{\rm sub} = A_{\rm sub} D_{\rm sub}.
\end{equation}
If $A_{\rm sub}$ is invertible (which is not evident since, we are dealing with a submatrix of a unitary matrix), we can recast this identity into the form 
\begin{equation}
({\cal P}_{\pi})_{\rm sub} = A_{\rm sub} D_{\rm sub} A_{\rm sub}^{-1}.
\end{equation}
This directly implies that $({\cal P}_{\pi})_{\rm sub}$ and $D_{\rm sub}$ must have the same eigenvalues. Because, $D_{\rm sub}$ is still a diagonal matrix, we deduce that these eigenvalues must be $\{\lambda_{o_1}, \dots, \lambda_{o_n}\}$. Notice that the invertibility of $A_{\rm sub}$ is equivalent to demanding that $\det A_{\rm sub} \neq 0$. When we introduce $\Lambda_{\rm sub}$ to indicate the set eigenvalues of of $({\cal P}_{\pi})_{\rm sub}$, we find the {\em extended fermionic suppression law} by contraposition
\begin{equation}\label{eq:extendedFermionicSuppression}
\Lambda_{\rm sub} \neq \{\lambda_{o_1}, \dots, \lambda_{o_n}\} \implies \det A_{\rm sub} = 0,
\end{equation}
and thus, from (\ref{eq:fermionicSuppression}), we find that $p_{\Psi \rightarrow M} = 0$ for the detection event where the particles are detected in output ports $o_1, \dots, o_n$. This proves the extended suppression law for fermions, and it can be shown that the events which are suppressed by condition (\ref{eq:fermionicSuppression}) are also included in the extended suppression law (\ref{eq:extendedFermionicSuppression}).\\

It was shown in \cite{dittel2-2017} that known suppression laws for special interferometers, such as the Fourier matrix \cite{tichy_zero-transmission_2010} and the Sylvester interferometer \cite{10.1088/1367-2630/aaad92,PhysRevA.91.013811,2058-9565-2-1-015003}, fit within the general framework that was presented in this section. Various suppression laws have also been tested experimentally \cite{10.1088/1367-2630/aaad92,crespi-suppression-2016}, and used for the validation of small-scale Boson Sampling experiments. 

The advantage of this benchmark of genuine bosonic many-particle interference (assuming that the particles are fully distinguishable) is that it provides a simple means for falsification. If a state $\ket{M}$ is supposed to be suppressed, but it is nevertheless observed, we know we can reject the claim of genuine many-interference (assuming an ideal detection stage, i.e.~with no dark counts). However, the problem with this method is its sensitivity to imperfections. Suppression laws are based on symmetries, which can be slightly broken, and on full indistinguishability, which can be distorted by the particles' internal DOF. Even though partial distinguishability can be included in the theory of total destructive interference \cite{dittel2-2017}, and experimental imperfections can be mitigated, it still limits the practical use this approach. Hence, alternative methods have been developed that do not exhibit these disadvantages.

\subsection{Statistical signatures and random matrix theory}\label{sec:RMT}
%We will show how two-point correlators between pairs of detectors hold crucial information about the many-particle interferences that take place within the interferometer. In particular, we show that statistical quantifiers (such as the mean and the variance) of the set of all these two-point correlators are robust benchmarks for many-particle interference. Furthermore, these benchmarks can be approximated analytically by using RMT.
As emphasised at the end of the previous section, suppression laws are confronted with several disadvantages. In particular, they only serve as signature of many-particle interference for specific interferometers, given by (\ref{eq:UisA}). These highly symmetric interferometers are generally not of great interest for reaching a quantum advantage through Boson Sampling. In this section we therefore introduce a different method, based on correlations between output detectors. This method has two major advantages. First, the statistical benchmark works regardless of the interferometer that is used to implement the Boson Sampling protocol. Second, very generated sampled output event is used in the validation scheme (in strong contrast with the suppression law, where only suppressed events are capable of falsifying genuine many-particle interference).\\

Up to this point, our analysis aimed at understanding the probability distribution $p_{\Psi \rightarrow M}$ of observing particles in different combinations of output detectors (represented by the measurement-states $\ket{M}$), for a given initial many-particle wave function $\ket{\Psi}$ and a given interferometer $U$. However, here we consider a different type of statistical quantifier: the correlation between output detectors. As we have seen in Section \ref{sec:GaussianStates}, correlations functions of the form $\tr [\rho a^{\dag}(\psi_1) \dots a^{\dag}(\psi_q) a(\phi_1) \dots a(\phi_{q'})]$ are sufficient to fully characterise many-particle state $\rho$, under the condition that these functions are known for all possible monomial lengths $q$ and $q'$, and for all possible choices of single-particle wave functions $\psi_1, \dots, \psi_q,\phi_1 \dots, \phi_{q'} \in {\cal H}$. However, this condition does not imply %is the necessary condition to characterise the full state, which does not necessarily mean 
that we require such a degree of information to identify the presence of many-particle interference. Thus, we can focus on {\em low-order} (in creation and annihilation operators) correlations to extract signatures of many-particle interference. Similar ideas have been implemented for studying many-particle quantum walks \cite{mayer_counting_2011}.

The simplest measurable ``correlation'' is the expectation value of the local number operator for the $o$th output detector, $\hat n_{o} = a^{\dag}(e_o)a(e_o)$. This quantity is obtained by counting the number of photons that are detected by the detector over many runs of the sampling experiment, which allows us to evaluate $\tr [\hat n_{o} \rho]$. When we choose $\rho = E(U) \ket{\Psi}\bra{\Psi} E(U^{\dag})$, with $\ket{\Psi}$ given by (\ref{eq:PsiAgainAndAgain}) and $U$ a unitary that describes an arbitrary interferometer, we find that {\em for fermions and bosons}
\begin{align}
\tr[\hat n_o \rho] &= \bra{0}a(e_{i_n}) \dots a(e_{i_1}) E(U^{\dag}) a^{\dag}(e_o)a(e_o)E(U) a^{\dag}(e_{i_1})\dots a^{\dag}(e_{i_n})\ket{0}\label{eq:and-here}\\
&= \bra{0}a(e_{i_n}) \dots a(e_{i_1}) a^{\dag}( U^{\dag} e_o)a(U^{\dag}e_o) a^{\dag}( e_{i_1})\dots a^{\dag}(e_{i_n})\ket{0}\label{eq:we}\\
&= \sum_{k=1}^n \abs{U_{oi_k}}^2,\label{eq:go}
\end{align}
where we again use the notation $\bra{e_k}U\ket{e_j} =U_{kj}$. The first step follows directly from (\ref{eq:Ea}), whereas the second step requires some combinatorics. The last equality is ultimately an application of the commutation (for bosons) or anti-commutation (for fermions) relations creation and annihilation operators, (\ref{eq:CCR}) and (\ref{eq:CAR}), respectively. The crucial result is that we observe that the average particle number of a single output detector cannot differentiate between fermions and bosons. When the particles are completely distinguishable, we can express
\begin{align}\label{eq:corrN}
\tr[\hat n_o \rho] &= \sum_{k=1}^n p_{i_k \rightarrow o} =  \sum_{k=1}^n \abs{U_{oi_k}}^2,
\end{align}
where we use (\ref{eq:singlePartProb2}). Hence, we also fail to see a difference between distinguishable and indistinguishable particles in the expected number of particles.

The next experimentally feasible possibility is to consider correlations between pairs of detectors: $\tr[\hat n_{o_1}\hat n_{o_2}\rho]$. When we again choose $\rho = E(U) \ket{\Psi}\bra{\Psi} E(U^{\dag})$, with $\ket{\Psi}$ as in  (\ref{eq:PsiAgainAndAgain}), we find that
\begin{align}
&\tr[\hat n_{o_1}\hat n_{o_2}\rho]\label{eq:corGeneral}\\
& =  \bra{0}a(e_{i_n}) \dots a(e_{i_1}) a^{\dag}( U^{\dag} e_{o_1})a^{\dag}(U^{\dag} e_{o_2})a(U^{\dag}e_{o_2})a(U^{\dag}e_{o_1}) a^{\dag}( e_{i_1})\dots a^{\dag}(e_{i_n})\ket{0}.\nonumber
\end{align}
For bosons, the only way to evaluate this quantity is by a direct application of the commutation relations (\ref{eq:CCR}), which results in
\begin{equation}\label{eq:corNNB}
\tr[\hat n_{o_1}\hat n_{o_2}\rho] = \sum_{\substack{k,l = 1\\ k\neq l}}^n \Big( \abs{U_{o_1 i_k}}^2\abs{U_{o_2 i_l}}^2 + U_{o_1 i_k}U_{o_2 i_l}U^*_{o_1 i_l}U^*_{o_2 i_k} \Big), \quad \text{for bosons}.
\end{equation}
For fermions, an analogous calculation is possible, but it is more elegant to use an alternative approach. In Section \ref{sec:GaussianStates}, we stressed that a fermionic state of the form (\ref{eq:PsiAgainAndAgain}), i.e.~a Slater determinant, is a Gaussian state, with correlations given by (\ref{eq:FermGaussNumber}). The matrix $Q$ that characterises the correlations in the initial states $\ket{\Psi}$ is given by 
\begin{equation}\label{eq:Qthermal}
Q = \sum_{k=1}^n \ket{e_{i_k}}\bra{e_{i_k}},
\end{equation}
and with $\rho = E(U) \ket{\Psi}\bra{\Psi} E(U^{\dag})$ and (\ref{eq:FermGaussNumber}) we find
\begin{align}
\tr[\hat n_{o_1}\hat n_{o_2}\rho] &= \bra{\Psi} a^{\dag}( U^{\dag} e_{o_1})a^{\dag}(U^{\dag} e_{o_2})a(U^{\dag}e_{o_2})a(U^{\dag}e_{o_1}) \ket{\Psi}\\
&= \bra{e_{o_1}}UQU^{\dag}\ket{e_{o_1}}\bra{e_{o_2}}UQU^{\dag}\ket{e_{o_2}}\\
&\qquad - \bra{e_{o_1}}UQU^{\dag}\ket{e_{o_2}}\bra{e_{o_2}}UQU^{\dag}\ket{e_{o_1}},\nonumber
\end{align}
where we note that
\begin{align}
&\bra{e_{o_1}}UQU^{\dag}\ket{e_{o_1}}\bra{e_{o_2}}UQU^{\dag}\ket{e_{o_2}} = \sum_{k,l = 1}^n \abs{U_{o_1 i_k}}^2 \abs{U_{o_2 i_l}}^2,\label{eq:blab}\\
& \bra{e_{o_1}}UQU^{\dag}\ket{e_{o_2}}\bra{e_{o_2}}UQU^{\dag}\ket{e_{o_1}} = \sum_{k,l=1}^n U_{o_1 i_k}U_{o_2 i_l} U^{*}_{o_1 i_l} U^*_{o_2 i_k}.\label{eq:blub}
\end{align}
Combining both terms leads to the final result
\begin{equation}\label{eq:corNNF}
\tr[\hat n_{o_1}\hat n_{o_2}\rho] = \sum_{\substack{k,l = 1\\ k\neq l}}^n \Big( \abs{U_{o_1 i_k}}^2\abs{U_{o_2 i_l}}^2 - U_{o_1 i_k}U_{o_2 i_l}U^*_{o_1 i_l}U^*_{o_2 i_k} \Big), \quad \text{for fermions}.
\end{equation}
It should be emphasised that there is a clear difference between the bosonic result (\ref{eq:corNNB}) and the fermionic one (\ref{eq:corNNF}), which is given by the contribution of the {\em interference term} $U_{o_1 i_k}U_{o_2 i_l}U^*_{o_1 i_l}U^*_{o_2 i_k}$. The observant reader has probably noticed the similarity between the obtained correlation functions (\ref{eq:corNNB}, \ref{eq:corNNF}) and the Hong-Ou-Mandel transfer probabilities (\ref{eq:HOMB4}, \ref{eq:HOMF4}), which ultimately lies at the foundation of this statistical signature of many-particle interference: the correlations (\ref{eq:corNNB}) and (\ref{eq:corNNF}) sum over all the possible two-particle processes that connect a pair of input particles to the chosen output detectors. This observation is solidified by considering the correlations for {\em distinguishable particles}, which are given by
\begin{align}\label{eq:corNND}
\tr[\hat n_{o_1}\hat n_{o_2}\rho] &= \sum_{\substack{k,l=1\\ k \neq l}}^n p_{i_k \rightarrow o_1}p_{i_k \rightarrow o_2} =  \sum_{\substack{k,l=1\\ k \neq l}}^n \abs{U_{o_1i_k}}^2\abs{U_{o_2i_l}}^2,
\end{align}
where the absence of the terms $U_{o_1 i_k}U_{o_2 i_l}U^*_{o_1 i_l}U^*_{o_2 i_k}$ indicated the absence of many-particle interference.

The use of correlation functions also provides us with a tool to easily explore the impact of the non-Gaussian statistics when $\ket{\Psi}$ (\ref{eq:PsiAgainAndAgain}) is a bosonic state. To do so, we compare the bosonic number state to a bosonic Gaussian state $\rho$, as introduced in Section \ref{sec:GaussianStates}. For simplicity, we assume that the state is purely thermal, such that $S=0$. In order to have a state which is as close as possible to the number state  (\ref{eq:PsiAgainAndAgain}), we take inspiration from the fermionic Slater determinant and choose $Q$ as in (\ref{eq:Qthermal}). The bosonic case strongly resembles the fermionic Gaussian states:
\begin{align}
\tr[\hat n_{o_1}\hat n_{o_2}\rho] &= \bra{\Psi} a^{\dag}( U^{\dag} e_{o_1})a^{\dag}(U^{\dag} e_{o_2})a(U^{\dag}e_{o_2})a(U^{\dag}e_{o_1}) \ket{\Psi}\\
&= \bra{e_{o_1}}UQU^{\dag}\ket{e_{o_1}}\bra{e_{o_2}}UQU^{\dag}\ket{e_{o_2}}\\
&\qquad + \bra{e_{o_1}}UQU^{\dag}\ket{e_{o_2}}\bra{e_{o_2}}UQU^{\dag}\ket{e_{o_1}}.\nonumber
\end{align}
when we, again, use (\ref{eq:blab}, \ref{eq:blub}), we now find that
\begin{align}\label{eq:corNNG}
\tr[\hat n_{o_1}\hat n_{o_2}\rho] &= \sum_{\substack{k,l = 1\\ k\neq l}}^n \Big( \abs{U_{o_1 i_k}}^2\abs{U_{o_2 i_l}}^2 + U_{o_1 i_k}U_{o_2 i_l}U^*_{o_1 i_l}U^*_{o_2 i_k} \Big)\\
&\qquad +2\sum_{k=1}^n  \abs{U_{o_1 i_k}}^2\abs{U_{o_2 i_k}}^2, \qquad\qquad \text{for thermal bosons},\nonumber
\end{align}
where the final term indicates the difference between a bosonic number state and a bosonic thermal state.\\

\begin{figure}
\centering
\includegraphics[width=0.8 \textwidth]{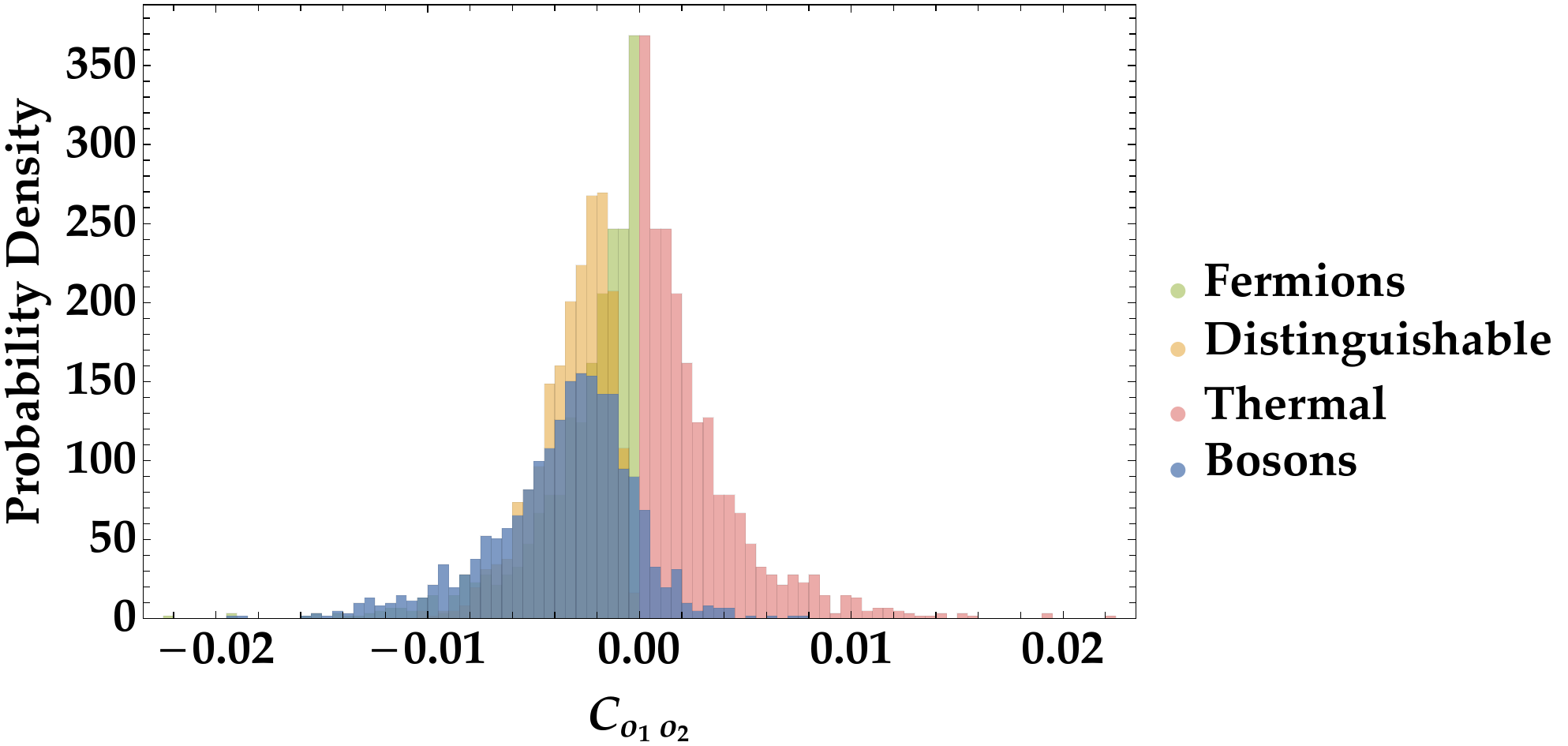}
\caption{Histogram indicating the different values observed for the pair correlations $C_{o_1 o_2}$, obtained for $n=8$ particles in a single, randomly chosen, 50-mode interferometer (with $U$ chosen from the Haar measure). Data are shown for bosonic Fock states (\ref{eq:C2Bosons}), bosonic thermal states (\ref{eq:C2Thermal}), fermionic number states (\ref{eq:C2Fermions}), and distinguishable particles (\ref{eq:distCFinal}).}
\label{fig:HistogramC}
\end{figure}

Note that the terms $\abs{U_{o_1i_k}}^2\abs{U_{o_2i_l}}^2$ appear in the bosonic (\ref{eq:corNNB}), fermionic (\ref{eq:corNNF}), and distinguishable-particle (\ref{eq:corNND}) correlations. On a practical level, this contribution provides no insight on the presence of many-particle interference. To make the signature of many-particle interference more sensitive, we should thus attempt to cancel these terms. Equation (\ref{eq:corrN}) provides a clear inspiration for the statistical quantity that may help achieve this goal: the truncated correlation (also referred as multivariate cumulant), given by
\begin{equation}
C_{o_1 o_2} = \tr[\hat n_{o_1}\hat n_{o_2}\rho] - \tr[\hat n_{o_1}\rho] \tr[\hat n_{o_2}\rho].
\end{equation}
Combining (\ref{eq:corNNB}, \ref{eq:corNNF}, \ref{eq:corNND}, \ref{eq:corNNG}) with (\ref{eq:corrN}) directly leads to the result
\begin{align}
C^B_{o_1 o_2} = -\sum_{k=1}^n  \abs{U_{o_1 i_k}}^2\abs{U_{o_2 i_k}}^2  +   \sum_{\substack{k,l = 1\\ k\neq l}}^n U_{o_1 i_k}U_{o_2 i_l}U^*_{o_1 i_l}U^*_{o_2 i_k} , \quad &\text{(bosons)} \label{eq:C2Bosons}\\
C^T_{o_1 o_2} = \sum_{k=1}^n  \abs{U_{o_1 i_k}}^2\abs{U_{o_2 i_k}}^2  +   \sum_{\substack{k,l = 1\\ k\neq l}}^n U_{o_1 i_k}U_{o_2 i_l}U^*_{o_1 i_l}U^*_{o_2 i_k} , \quad &\text{(thermal)}\label{eq:C2Thermal}\\
C^F_{o_1 o_2} = -\sum_{k=1}^n  \abs{U_{o_1 i_k}}^2\abs{U_{o_2 i_k}}^2  -   \sum_{\substack{k,l = 1\\ k\neq l}}^n U_{o_1 i_k}U_{o_2 i_l}U^*_{o_1 i_l}U^*_{o_2 i_k} , \quad &\text{(fermions)} \label{eq:C2Fermions}\\
C^D_{o_1 o_2} = -\sum_{k=1}^n  \abs{U_{o_1 i_k}}^2\abs{U_{o_2 i_k}}^2, \quad &\text{(distinguishable)}.\label{eq:distCFinal}
\end{align}
It is clear that a given configuration of input ports and interferometer (that implements the unitary $U$) generally lead to different particle number correlations $C_{o_1 o_2}$. However, it is far from clear whether the measurement of a certain value for $C_{o_1 o_2}$ can allow us to identify the particle type. To answer this question, we show a histogram in Fig.~\ref{fig:HistogramC} of the obtained values for a randomly chosen $U$ (from the uniform distribution over all unitary matrices, known as the Haar measure), obtained by considering all possible pairs of output modes $o_1$ and $o_2$. These histograms clearly highlight that $C^D_{o_1 o_2}, C^F_{o_1 o_2} \leqslant 0$ and $C^T_{o_1 o_2}\geqslant 0$. The negativity of $C^D_{o_1 o_2}$ is directly seen from (\ref{eq:distCFinal}), but for $C^F_{o_1 o_2}$ a brief additional analysis is required:
\begin{align}
C^F_{o_1 o_2} &=  -   \sum_{\substack{k,l = 1}}^n U_{o_1 i_k}U_{o_2 i_l}U^*_{o_1 i_l}U^*_{o_2 i_k} \\
&=  -   \sum_{\substack{k = 1}}^n U_{o_1 i_k}U^*_{o_2 i_k}  \left(\sum_{\substack{l = 1}}^nU_{o_1 i_l}U^*_{o_2 i_l}\right)^*\\
&= -\abs{\sum_{k=0}^nU_{o_2 i_k}U^*_{o_1 i_k}}^2 \leqslant 0.
\end{align}
The positivity of $C^T_{o_1 o_2}$ then follows immediately from the observation that $C^T_{o_1 o_2} = - C^F_{o_1 o_2}$. The latter highlights the parallel between fermionic particles and thermal bosons. The most crucial information provide by Fig.~\ref{fig:HistogramC} is that the histogram for a bosonic number state as input in the interferometer overlaps with all other histograms. This implies that we cannot associate a particular range of values to typical bosonic many-particle interference. However, we do observe that the statistical properties of these sets of correlations $C_{o_1 o_2}$ (which we refer to as {\em C-datasets}) differ strongly, depending on the particle type. This observation lies at the basis the statistical signatures for many-particle interference.\\

The C-datasets in Fig.~\ref{fig:HistogramC} are obtained for a single interferometer, and, hence, the statistical features of these distributions can be obtained from a single experimental setup. As a way of quantitatively grasping the properties of these histograms, we evaluate the moments of the distribution, where
\begin{equation}\label{eq:moment}
m_q \equiv \frac{2}{m(m-1)} \sum_{o_1 > o_2} \left(C_{o_1 o_2}\right)^q
\end{equation}
defines the $q$th moment. Essentially, given a linear-optical interferometer, $m_q$ can be estimated by simply looking at all correlations between pairs of output detectors, and averaging over them. Numerically, this is a tractable task, and for some interferometers it is also analytically feasible (see \ref{sec:fourier} for an example). Furthermore, we can acquire some additional understanding of the first moment (see \ref{sec:first_moment}), and we can explicitly derive some relations between the moments obtained for different particle types:
\begin{align}
&\frac{n}{m(m-1)} > m_1^T > m_1^F > m_1^D > m_1^B > -\frac{n}{m(m-1)},\label{eq:ThisEquationHere}\\
&m_1^B = m_1^T + 2 m_1^D.
\end{align}
Furthermore, it is evident from (\ref{eq:ThisEquationHere}) that the size of the interferometer and the number of particles play and important role in determining the order of magnitude of these moments. Therefore, we introduce rescaled quantities that allow us to compare different system sizes more directly. The normalised mean ($NM$) simply rescales the first moment, and the coefficient of variation ($CV$) compares the second to the first moment. These quantities are formally defined as
\begin{align}
&NM = m_1 \frac{m^2}{n},\label{eq:NM}\\
&CV = \frac{\sqrt{m_2 - m_1^2}}{m_1}.\label{eq:CV}
\end{align}
Similar objects can be defined for higher moments (e.g. the Skewness, as in \cite{walschaers_statistical_2014}), but here we will restrict our analysis to $NM$ and $CV$.

\begin{figure}
\centering
\includegraphics[width=0.9 \textwidth]{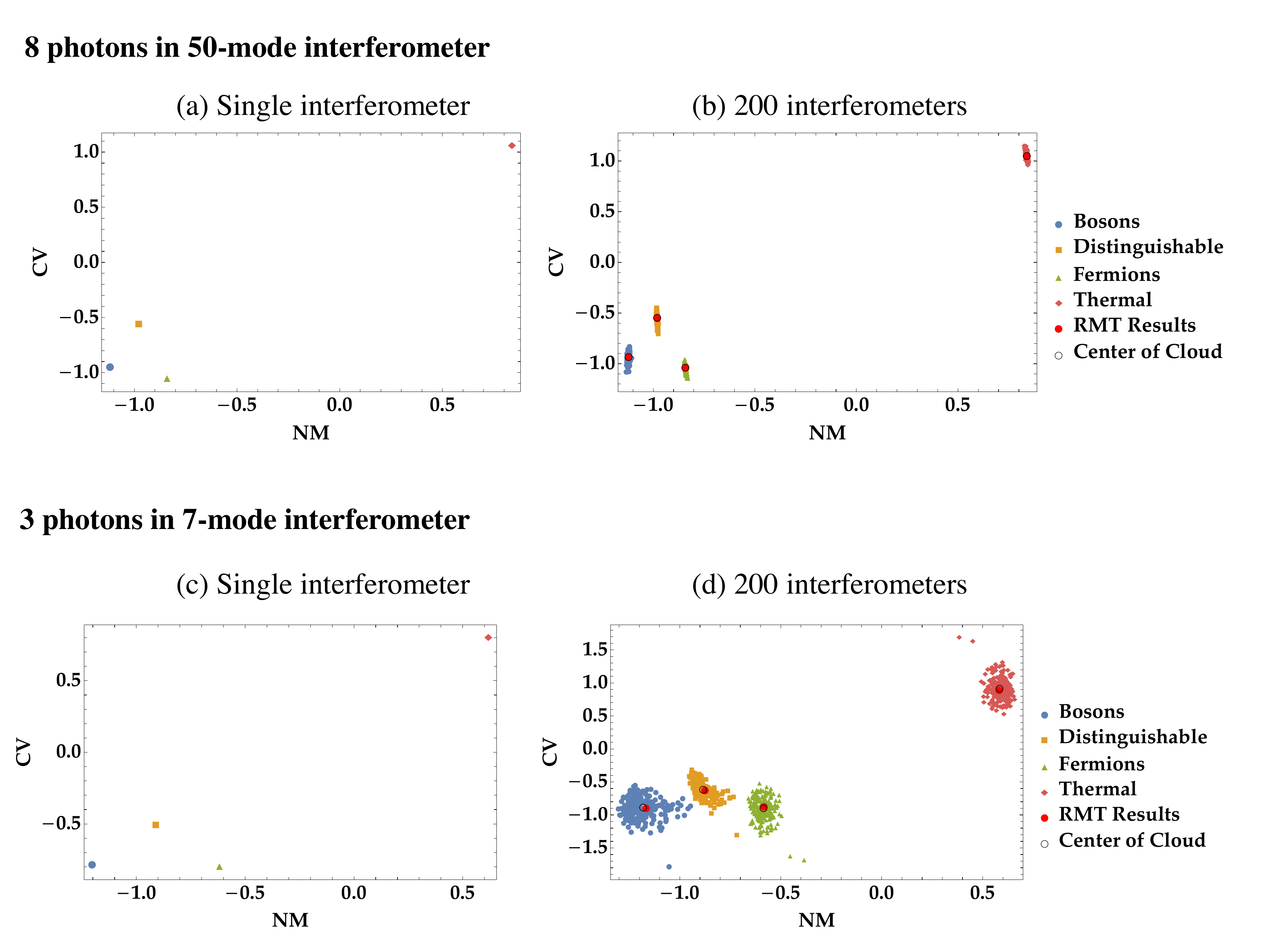}
\caption{Scatterplots depicting the normalised mean $NM$ (\ref{eq:NM}) on the horizontal axis and the coefficient of variation $CV$ (\ref{eq:CV}) on vertical axis. Panels (a) and (b) are obtained for a large 50-mode interferometer, in which $n=8$ particles are injected. Panels (c) and (d) are generated by injecting $n=3$ particles in a 7-mode interferometer. In (a) and (c), data are shown for one single interferometer, in which different particle types were injected: bosonic Fock states (blue dots), bosonic thermal states (red diamonds), fermionic number states (green triangles), and distinguishable particles (orange squares). On (b) and (d), the same type of data are shown for 200 different, randomly chosen interferometers; black circles indicate the centres of mass of the clouds of points that are obtained for different particle types. The random matrix predictions (\ref{eq:momentsB} - \ref{eq:momentsF}, \ref{eq:RMTmom2b} - \ref{eq:RMTmom2d}) for each particle type are shown by a large red dot.}
\label{fig:Cdatasets}
\end{figure}

In general, the values of the above moments differ for every interferometer, as we highlight in Fig.~\ref{fig:Cdatasets}. The figure shows the values of $NM$ and $CV$ for different particle types. In panels (a) and (c), we show the result for one single, randomly chosen interferometer. For different interferometers, it is natural to expect to find different values, and it is a natural question whether the variability in the observed values is dominated by the particle type or by the specificities of the interferometer. To investigate this aspect, we gather the results for 200 randomly chosen interferometers in panels (b) and (d). These results show an interesting feature: for a given particle type, the values for $NM$ and $CV$ generally fall in the same region of the plot, and these regions are distinct for different particle types. Moreover, it is clear from (c) and (d) that the method is more effective for larger systems with more modes (large $m$) and particles (large $n$). 

From the point of view of random matrix theory (RMT), it is not very surprising that such low-order moments of low-order correlations obtained from big random matrices are very much alike (think for example of the density of eigenvalues of one large random matrix that coincides with the ensemble average density of eigenvalues \cite{stockmann_quantum_2007}). Thus, it is observed that, for sufficiently large interferometers, the moments of the distribution of the $C$-dataset (and thus also $NM$ and $CV$) are very close to the average value over the ensemble of all possible interferometers (i.e. the average of $C_{o_1 o_2}$ over the Haar measure). This observation provides us with an interesting opportunity, since these averages can be evaluated analytically.\\

Let us denote the averaging over the set of all unitary matrices with respect to the Haar measure (i.e.~the uniform distribution over the ensemble of unitary matrices) by $\mathbb{E}_U(\dots)$, then we are interested in evaluating 
$\mathbb{E}_U(C^q_{o1 o_2})$
for an arbitrary choice of $o_1$ and $o_2$. First, we fix our attention on the first moment (q=1) for the bosonic case, where we directly see that
\begin{equation}\label{eq:195}
\mathbb{E}_U(C^B_{o1 o_2}) = -\sum_{k=1}^n  \mathbb{E}_U(\abs{U_{o_1 i_k}}^2\abs{U_{o_2 i_k}}^2)  +   \sum_{\substack{k,l = 1\\ k\neq l}}^n \mathbb{E}_U(U_{o_1 i_k}U_{o_2 i_l}U^*_{o_1 i_l}U^*_{o_2 i_k}).
\end{equation}
The technical task at hand is to evaluate $\mathbb{E}_U(U_{o_1 i_k}U_{o_2 i_l}U^*_{o_1 i_l}U^*_{o_2 i_k})$ for $k \neq l$ and $ \mathbb{E}_U(U_{o_1 i_k}U_{o_2 i_k}U^*_{o_1 i_k}U^*_{o_2 i_k})$, where in both cases $o_1 \neq o_2$. The core result at our disposal is the following identity for $m\times m$ random unitary matrices $U$ \cite{Creutz:1978aa,weingarten_asymptotic_1978,Samuel:1980aa}:
\begin{align}
	\label{eq:UnitaryAverage}
	         &\mathbb{E}_U(U_{a_1,b_1}\dots U_{a_n,b_n}U^*_{\alpha_1,\beta_1}\dots U^*_{\alpha_n,\beta_n}) 
	         \\
	         = & \sum_{\sigma,\pi \in S_n}\mathcal V_m(\sigma^{-1}\pi)\prod^n_{k=1}\delta(a_k-\alpha_{\sigma(k)})\delta(b_k-\beta_{\pi(k)}),\nonumber
\end{align}
where $\mathcal V_m(\sigma^{-1}\pi)$ are commonly referred to as the {\em Weingarten functions}. For low orders, the values of these functions can be obtained from tables that are available in literature \cite{brouwer_diagrammatic_1996}, or by using a direct, yet sophisticated approach based on the Schur-Weyl duality \cite{collins_integration_2006}. The calculations are simplified a little by the fact that these Weingarten functions only depend on the length of the cycles of the permutations.

To apply (\ref{eq:UnitaryAverage}), we define two permutations $e: (1, 2) \mapsto (2, 1)$ and $id: (1,2) \mapsto (1,2)$. Note that we must only consider cases where $\sigma = id$ and, thus obtain that
\begin{align}\label{eq:App122}
 \mathbb{E}_U\big( U_{o_1 i_k}U_{o_2 i_l}{U}^*_{o_1 i_l}{U}^*_{o_2 i_k}\big) &= V_m(id)\delta(i_k - i_l)\delta(i_l-i_k) +  V_m(e)\delta(i_k - i_k)\delta(i_l-i_l)\\
 & = V_m(e) = V_m(2),\nonumber
 \end{align}
where $V_m(2)$ refers to the Weingarten function for a permutation with one cycle of length 2. For the other term, we obtain that
\begin{align}\label{eq:App123}
\mathbb{E}_U\big( U_{o_1 i_k}U_{o_2 i_k}{U}^*_{o_1 i_k}{U}^*_{o_2 i_k} \big) &= V_m(id) \delta(i_k-i_k)\delta(i_k-i_k) + V_m(e) \delta(i_k-i_k)\delta(i_k-i_k) \\
&= V_m(1,1) + V_m(2),\nonumber
\end{align}
where $V_m(1,1)$ refers to the Weingarten function for a permutation with two cycle of length 1.
We can combine (\ref{eq:App122}) and (\ref{eq:App123}) to obtain that
\begin{align}
\mathbb{E}_U(C^B_{o_1 o_2}) &= \sum_{\substack{k,l = 1\\k \neq l}}^n V_m(2) - \sum_{k=1}^n \Big(V_m(1,1) + V_m(2)\Big)\\
&= -\frac{n (m+n-2)}{m \left(m^2-1\right)},\nonumber
\end{align}
where we used the tables of  \cite{brouwer_diagrammatic_1996} to obtain the final result. A completely analogous evaluation for the other particle types gives us the following random-matrix estimates for sufficiently large mode numbers $m$:
\begin{align}
m_1^B &\approx \mathbb{E}_U(C^B) = -\frac{n (m+n-2)}{m \left(m^2-1\right)},\label{eq:momentsB}\\
m_1^T &\approx \mathbb{E}_U(C^T) = \frac{n (m-n)}{m \left(m^2-1\right)},\\
m_1^D &\approx \mathbb{E}_U(C^D) = -\frac{n}{m (m+1)},\\
m_1^F &\approx \mathbb{E}_U(C^F) = -\frac{n (m-n)}{m \left(m^2-1\right)}.\label{eq:momentsF}
\end{align}
Note that these final results only depend on the number of modes and the number of particles, and not on the specific output and input modes which were chosen.\\

For the second moment, the situation is considerably more complicated, as we must evaluate
\begin{align}\label{eq:appblablabla}
&\mathbb{E}_U\Bigg( \bigg(-\sum_{k=1}^n  \abs{U_{o_1 i_k}}^2\abs{U_{o_2 i_k}}^2  +   \sum_{\substack{k,l = 1\\ k\neq l}}^n U_{o_1 i_k}U_{o_2 i_l}U^*_{o_1 i_l}U^*_{o_2 i_k}\bigg)\\
&\qquad\times \bigg( -\sum_{k'=1}^n  \abs{U_{o_1 i_{k'}}}^2\abs{U_{o_2 i_{k'}}}^2  +   \sum_{\substack{k',l' = 1\\ k\neq l'}}^n U_{o_1 i_{k'}}U_{o_2 i_{l'}}U^*_{o_1 i_{l'}}U^*_{o_2 i_{k'}} \bigg)\Bigg),\nonumber\\
& = \sum_{k,k' = 1}^n \mathbb{E}_U\bigg( \abs{U_{o_1 i_k}}^2\abs{U_{o_2 i_k}}^2\abs{U_{o_1 i_{k'}}}^2\abs{U_{o_2 i_{k'}}}^2\bigg) \nonumber\\
&\qquad \quad- 2 \sum_{\substack{k', k,l = 1\\ k\neq l}}^n \mathbb{E}_U\bigg(\abs{U_{o_1 i_{k'}}}^2\abs{U_{o_2 i_{k'}}}^2U_{o_1 i_k}U_{o_2 i_l}U^*_{o_1 i_l}U^*_{o_2 i_k}\bigg)\nonumber\\
&\qquad \quad+\sum_{\substack{k,k',l,l' = 1\\ k\neq l\\ k'\neq l'}}^n \mathbb{E}_U\bigg(U_{o_1 i_k}U_{o_2 i_l}U^*_{o_1 i_l}U^*_{o_2 i_k}U_{o_1 i_{k'}}U_{o_2 i_{l'}}U^*_{o_1 i_{l'}}U^*_{o_2 i_{k'}} \bigg).\nonumber
\end{align}
To understand the complexity of the problem, it is useful to return to (\ref{eq:App122}) and realise the importance of repeated indices. We must differentiate between all the cases where either $k'$ and/or $l'$ equals $k$ and/or $l$. For example, this means that we must split
\begin{align}
&\sum_{k,k' = 1}^n \mathbb{E}_U\bigg( \abs{U_{o_1 i_k}}^2\abs{U_{o_2 i_k}}^2\abs{U_{o_1 i_{k'}}}^2\abs{U_{o_2 i_{k'}}}^2\bigg)\\
&= \sum_{\substack{k,k' = 1\\k \neq k'}}^n \mathbb{E}_U\bigg( \abs{U_{o_1 i_k}}^2\abs{U_{o_2 i_k}}^2\abs{U_{o_1 i_{k'}}}^2\abs{U_{o_2 i_{k'}}}^2\bigg)\nonumber\\
&\qquad +\sum_{\substack{k= 1}}^n \mathbb{E}_U\bigg( \abs{U_{o_1 i_k}}^2\abs{U_{o_2 i_k}}^2\abs{U_{o_1 i_{k}}}^2\abs{U_{o_2 i_{k}}}^2\bigg).\nonumber
\end{align} 
For the other two terms in (\ref{eq:appblablabla}), we must perform an equal exercise, but with more ways of repeating the indices. In this Tutorial, we will not write down all of these terms explicitly, but the interested reader can find details in \cite{Walschaers:2018ab}. By evaluating (\ref{eq:App122}) for each different type of terms that appears in this expansion, we find the final results for the second moments:
\begin{align}
m^{B}_2 &\approx \mathbb{E}_U({C^B}^2) = \frac{2 n \left(m^2 n+m^2+9 m n-11 m+n^3-2 n^2+5 n-4\right)}{m^2 (m+2) (m+3) \left(m^2-1\right)},\label{eq:RMTmom2b}\\
m^{T}_2 = m^{F}_2 &\approx \mathbb{E}_U({C^F}^2) = \frac{2 n (n+1) (m-n) (m-n+1)}{m^2 (m+2) (m+3) \left(m^2-1\right)},\\
m^{D}_2 &\approx \mathbb{E}_U({C^D}^2) = \frac{n \left(m^2 n+3 m^2+m n-5 m+2 n-2\right)}{m^2 (m+2) (m+3) \left(m^2-1\right)}.\label{eq:RMTmom2d}
\end{align}
The obtained results (\ref{eq:momentsB} - \ref{eq:momentsF}) and (\ref{eq:RMTmom2b} - \ref{eq:RMTmom2d}) can now be used to obtain random-matrix estimates for the normalised mean $NM$ (\ref{eq:NM}) and the coefficient of variation $CV$ (\ref{eq:CV}) for every particle species. These estimates give us an analytical grasp on the expected statistical signatures of many-particle interference.

To highlight the accuracy of these analytical approximations, we pinpoint the random-matrix estimates in Fig.~\ref{fig:Cdatasets}. We clearly see a spread of the moments, obtained for individual randomly chosen interferometers, around the predictions from RMT. It is important to observe that the spread is much larger for the small seven-port interferometer. When the number of modes grows, the random-matrix estimates are expected to become more precise, and this is exactly the behaviour that manifests in this example. Furthermore, we see that the random-matrix results agree very well with the centre of the cloud of points, and can thus be seen as the average of the moments of many random interferometers.\\

Now that we have established a clear idea of the statistical signatures of different types of many-particle interference, it is important to understand how such a benchmark can serve in practice for the validation of a many-particle interference experiment such as Boson Sampling. One of the simplest techniques at our disposal is the evaluation of the distance from point $(NM, CV)$, as obtained from a specific interferometer, to the different random-matrix estimates. We can then conjecture that the observed probability distribution is associated with the particle type of the closest analytical value. Fig.~\ref{fig:Cdatasets} suggests that this method should be effective for large interferometers, but it may fail for smaller (more realistic) setups. Distance measures do not consider the characteristic size of the different the clusters of points for different particle types, even though the bosonic cluster of points is clearly more extended than for example the cluster for distinguishable particles. In principle, one can use RMT to estimate the spread of these clusters, since deviations around the estimates of moments are characterised by higher moments.

A more pragmatic solution corresponds to generating more statistics. It is clearly shown in Fig.~\ref{fig:Cdatasets} that the average over the $(NM, CV)$ of several interferometers coincides with the random-matrix results. Additional statistics can easily be acquired with a reconfigurable unitary circuit at one's disposal, but this is a technical challenge \cite{carolan_universal_2015,Taballione:2018aa}. However, when the particles are inserted in different input ports of one single interferometer, it acts almost as a new interferometer. Specifically when the number of modes is much larger than the number of particles ($m \gg n^2$), an entirely different set of input modes probes a different uncorrelated part of the unitary matrix that describes the entire interferometer \cite{aaronson_computational_2013}. In practice, this approach is natural in the problem of scattershot Boson Sampling \cite{lund_boson_2014,bentivegna_experimental_2015}.

An even more pragmatic take on benchmarking is to follow a data science approach and use ideas from machine learning. Indeed, the numerically simulated sampling data of Fig.~\ref{fig:Cdatasets} are easy to generate in large quantities, which means that we have ample data to train a supervised learning algorithm (e.g. a support vector machine) to associate regions in the  $(NM, CV)$ - plane with certain particle types. This approach have proven to be successful to validate many-particle interference experiments \cite{Giordani-2018}. Because also experimental imperfections in the interferometers can be included in simulations, the comparison to the random-matrix predictions allows to identify systematic errors that cause the statistics of the interferometers to deviate from the Haar measure, as was seen in \cite{Giordani-2018}. \\

There are several challenges related to the experimental observation of such a statistical signature of many-particle interference. First of all, there are practical difficulties introduced by the partial distinguishability of particles, which will be discussed in detail in the next section. Then, there is the need for particle-number-resolving detectors, which currently represents a formidable challenge for photonic systems. Furthermore, we are generally confronted with a more subtle class of finite-size effects in the evaluation of the correlations $C_{o_1 o_2}$ themselves. Generally, these correlations are evaluated based on a finite sample of output events, and therefore they only give an estimate to the real value of $C_{o_1 o_2}$. A recent work evaluated the impact of these finite-size effects \cite{Flamini-finite-2019}, which uses the Metropolis-type simulations \cite{neville2017} of Boson Sampling as an interesting tool to test the statistical benchmark.
%by comparing it to exact Boson Sampling data \cite{Clifford:2018:CCB:3174304.3175276} in a regime where the Metropolis algorithm has not yet fully converged. 

Finally, it must be emphasised that the above statistical signature, as presented here, can be generalised and extended in many different ways. A profound example is the analysis of higher moments of the C-dataset (e.g.~the skewness was considered in \cite{walschaers_statistical_2014} and made a more extended study of the information gained from different statistical quantifiers in \cite{Giordani-2018}). However, this does not undo the fact that the C-dataset built upon two-point correlations essentially probes two-particle interference processes, and it is bound to miss a considerable amount of physics. For instance, a more profound extension of the method that considers three-point correlations was explored in \cite{Walschaers:2018aa}. Finally, the approach based on the C-dataset is closely related to those that use the $g^{(2)}$ and $g^{(3)}$ functions \cite{PhysRevLett.117.213602,Rigovacca:2018aa}.\\

%An important ingredient of benchmarking is the availability of alternative models to compare to (the other particle types in our case). 
The goal of the above statistical benchmark, as well as the other validation protocols, is to rule out alternative physical models that may have given rise to the measured output data (the other particle types in our case). In this section, we explored four simple models, but one can also consider a more artificial sampling model, such as sampling from the uniform distribution of output events \cite{gogolin_boson-sampling_2013,spagnolo_experimental_2014,aaronson_bosonsampling_2014} (note that for such a sampler $CV$ vanishes). The so-called mean-field sampler \cite{10.1088/1367-2630/aaad92,tichy_stringent_2014,tichy_entanglement_2011,tichy_interference_2014} has been highly successful in the literature for reproducing sampling data that resemble Boson Sampling. It is interesting, from both a physical and a more application-oriented perspective, to investigate further ``error models'' for Boson Sampling experiments \cite{Hangleiter:2019aa,Shchesnovich-2019,Moylett-2019,Leverrier-2013,Kalai-2014}. Arguably, the most important error model to keep in mind is partial distinguishability. In the following section we explore how the statistical signature can indeed probe the distinguishability transition.

\subsection{The statistical Hong-Ou-Mandel effect}

In the previous section, we studied the applicability of the statistical benchmark with various particle types, among which we discussed distinguishable particles. In the light of Sections \ref{sec:Distinguish} and \ref{sec:partialdist}, we know that distinguishability is a subtle concept, in particular because particles can be partially distinguishable, as governed by their internal DOF. In this section, we explore the capability of the statistical benchmark to capture the distinguishability transition and, perhaps, quantify the degree of distinguishability between the particles.\\

To study the distinguishability transition the internal DOF of the particles must be taken into account. We will only consider two types of input states: those with a well-defined particle number, i.e.~bosonic and fermionic number states. Similar to Section \ref{sec:partialdist}, we now consider a Hilbert space ${\cal H} = \mathbb{C}^m \otimes {\cal H}_I$, where $\mathbb{C}^m$ describes the $m$ input ports, and $ {\cal H}_I$ describes the internal (non-observed) DOF. The input state is then, again, given by
\begin{equation}
\ket{\Psi} = a^{\dag}(e_{i_1}\otimes \psi_1)\dots a^{\dag}(e_{i_n}\otimes \psi_n)\ket{0},
\end{equation}
but detectors are blind to these internal DOF. Hence, we choose a basis $\{f_1, f_2, \dots\}$ of ${\cal H}_I$ (for simplicity, we assume the existence of a discrete basis) and define the observables
\begin{equation}
\hat N(e_j) = \sum_{k} a^{\dag}(e_j \otimes f_k)a(e_j \otimes f_k),
\end{equation}
which count the number of particles in the $j$th output detector, characterised by mode $e_j \in \mathbb{C}^m$, regardless of the internal DOF. Due to the action of the interferometer, we obtain an output state $\rho = E(U \otimes \mathbb{1}) \ket{\Psi}\bra{\Psi} E(U^{\dag}\otimes \mathbb{1})$, such that the C-dataset is given by
\begin{equation}
C_{o_1 o_2} = \tr[\hat N_{o_1}\hat N_{o_2}\rho] - \tr[\hat N_{o_1}\rho] \tr[\hat N_{o_2}\rho].
\end{equation}
Through exactly the same techniques used in (\ref{eq:corNNB}) and (\ref{eq:corNNF}), combined the identity $\sum_k \ket{f_k}\bra{f_k} = \mathbb{1}$, we find that
\begin{align}
C^{F}_{o_1 o_2} = &-\sum_{k=1}^n  \abs{U_{o_1 i_k}}^2\abs{U_{o_2 i_k}}^2 - \sum_{\substack{k,l = 1\\ k\neq l}}^n \abs{\langle \psi_k \mid \psi_l \rangle}^2 U_{o_1 i_k}U_{o_2 i_l}U^*_{o_1 i_l}U^*_{o_2 i_k},   \text{ (fermions)} \label{eq:corrDistTransB}\\
C^{B}_{o_1 o_2} = &-\sum_{k=1}^n  \abs{U_{o_1 i_k}}^2\abs{U_{o_2 i_k}}^2 +   \sum_{\substack{k,l = 1\\ k\neq l}}^n \abs{\langle \psi_k \mid \psi_l \rangle}^2 U_{o_1 i_k}U_{o_2 i_l}U^*_{o_1 i_l}U^*_{o_2 i_k}, \text{ (bosons)}.\label{eq:corrDistTransF}
\end{align}
As expected from Section~\ref{sec:partialdist}, we recover the result (\ref{eq:distCFinal}) for distinguishable particles when $\langle \psi_k \mid \psi_l \rangle = \delta_{k,l}$, and the results for fully indistinguishable bosons (\ref{eq:C2Bosons}) or fermions (\ref{eq:C2Fermions}) when $\langle \psi_k \mid \psi_l \rangle = 1$. In this section, though, we are interested in the intermediate regime.\\

We note that the correlations $C_{o_1 o_2}$ depend only on the overlaps of the states of the particles' internal DOF, i.e.~$\abs{\langle \psi_k \mid \psi_l \rangle}^2$. Here we see a strong resemblance to the Hong-Ou-Mandel effect (\ref{eq:HOMInterpret}), which highlights once more that the two-point correlations $C_{o_1 o_2}$ probe all the possible two-particle interference processes that take place.
%The more general many-particle interference results (\ref{eq:ManyPartIntB-PartDist}) and (\ref{eq:ManyPartIntF-PartDist}) show that the phases of these internal states can have an important impact on the final interference profile, but this information cannot be extracted from the C-dataset of two-point correlations. Nevertheless, the we can deduce a great deal of information about the distinguishability transition from the C-dataset.

The correlations (\ref{eq:corrDistTransB}, \ref{eq:corrDistTransF}) generally depend on the characteristics implemented by the interferometer, and on the chosen output modes $o_1, o_2$. As in the previous section, the value of a single $C_{o_1 o_2}$ does not provide insights in the exact degree of distinguishability, and it is more instructive to study moments $m_q$, see (\ref{eq:moment}), of the C-dataset. For a randomly chosen interferometer, the arguments from the previous section still apply; hence, we can surmise that for a sufficiently large numbers of modes, $m$, and particles, $n$, these moments can be approximated by RMT, such that $m_q \approx \mathbb{E}_U(C^q)$. We then start obtain the generalisation of (\ref{eq:195}) to include partial distinguishability:
\begin{equation}\label{eq:partialDistExpressionRMT}
\mathbb{E}_U(C_{o_1 o_2}) = -\sum_{k=1}^n  \mathbb{E}_U(\abs{U_{o_1 i_k}}^2\abs{U_{o_2 i_k}}^2)  \pm   \sum_{\substack{k,l = 1\\ k\neq l}}^n \abs{\langle \psi_k \mid \psi_l \rangle}^2 \mathbb{E}_U(U_{o_1 i_k}U_{o_2 i_l}U^*_{o_1 i_l}U^*_{o_2 i_k}),
\end{equation}
where ``$+$'' gives the result for bosons and ``$-$'' for fermions. The random-matrix averages to be evaluated in (\ref{eq:partialDistExpressionRMT}) are actually the same as those in (\ref{eq:195}). Using these results if Section~\ref{sec:RMT} following (\ref{eq:195}), we find
\begin{equation}\label{eq:m1PartialDist}
m_1 \approx \mathbb{E}_U(C) = - \frac{n}{m(m+1)} \mp \frac{1}{m(m^2-1)}\sum^n_{\substack{k,l = 1\\k \neq l}}\abs{\langle\psi_k \mid \psi_l\rangle}^2.
\end{equation}
By virtue of a computation analogous to (\ref{eq:appblablabla}), we can subsequently determine the result for the second moment:
\begin{align}\label{eq:PDTrainSecondMoment}
m_2 \approx \mathbb{E}_{U}(C^2) = &\frac{2 A - 2 B (m-5) +  C (10 + m + m^2) \pm 2 D (2 + 6 m -n + mn) }{(m-1) m^2 (m+1) (m+2) (m+3)}   
  \\& + \frac{(m-2) (1 + 3m)n + 2n^2 + m n^ 2 + 
      m^2n^2)}{(m-1) m^2 (m+1) (m+2) (m+3)},\nonumber
\end{align}
where ``$+$'' (``$-$'') gives the result for bosons (fermions).
%where ``$+$'' gives the result for bosons, whereas ``$-$'' shows the fermionic value.
\begin{align}
&A  = \sum^n_{\substack{k_1,k_2,l_1,l_2 = 0\\ k_1\neq k_2\neq l_1\neq l_2}}\abs{\langle\psi_{k_1} \mid \psi_{l_1}\rangle}^2\abs{\langle\psi_{k_2} \mid \psi_{l_2}\rangle}^2,\label{eq:ASecMom} \\
&B = \sum^n_{\substack{k ,l_1,l_2=0\\ k \neq l_1\neq l_2}} \abs{\langle\psi_{k} \mid \psi_{l_1}\rangle}^2\abs{\langle\psi_{k} \mid \psi_{l_2}\rangle}^2, \label{eq:BSecMom}\\
&C = \sum^n_{\substack{k ,l=0\\ k \neq l}} \abs{\langle\psi_k \mid \psi_l\rangle}^4, \label{eq:CSecMom} \\
&D = \sum^n_{\substack{k ,l=0\\ k \neq l}} \abs{\langle\psi_k \mid \psi_l\rangle}^2. \label{eq:DSecMom}
\end{align}\\

\begin{figure}
\centering
\includegraphics[width=0.99 \textwidth]{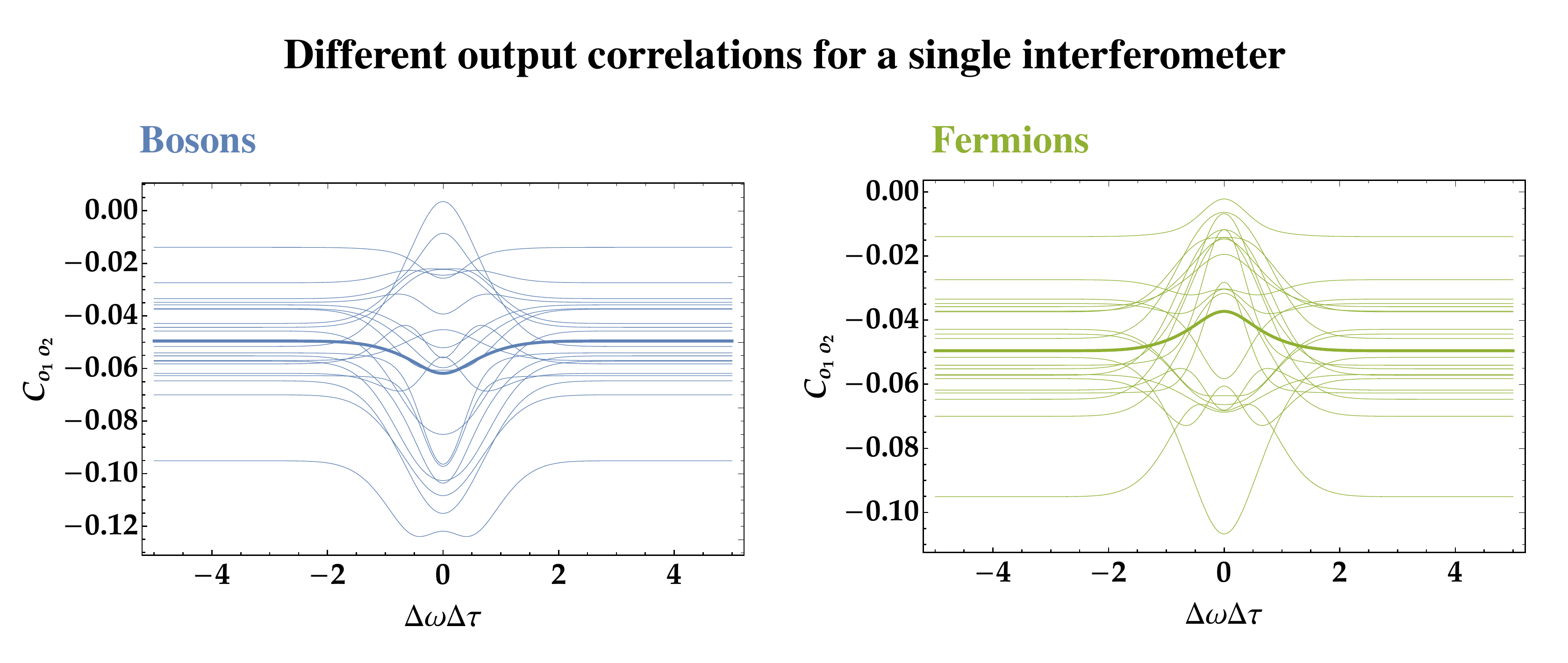}
\caption{Pair-correlations $C_{o_1 o_2}$ (\ref{eq:corrDistTransB}, \ref{eq:corrDistTransF}) for varying values of $\Delta \omega \Delta \tau$. Each curve represents a different pair of output detectors $o_1, o_2$, for three bosons (left) or fermions (right) injected in a randomly chosen 7-mode interferometer. }
\label{fig:DistTransCdatasets}
\end{figure}

In principle, there is a lot of freedom to choose the states of the particles for various internal DOF \cite{walschaers-from-2016}. As our go-to example, we will consider a train of wave packets, characterised by overlaps (\ref{eq:DistParameter}), as shown in Fig.~\ref{fig:WavePackets}, and attempt to use the statistical signatures to explore the indistinguishability transition. First of all, we explore the behaviour of individual correlators for a single interferometer in Fig.~\ref{fig:DistTransCdatasets}. The only thing that can be concluded from these data is that neither for fermions, nor for bosons there is a global pattern in the indistinguishability transition; in some cases the correlations increase, whereas in others they decrease. Furthermore, the correlations do not even necessarily reach a maximum or minimum value in $\Delta \tau \Delta \omega = 0$. This clearly emphasises the need from more robust quantifiers.

\begin{figure}
\centering
\includegraphics[width=0.99 \textwidth]{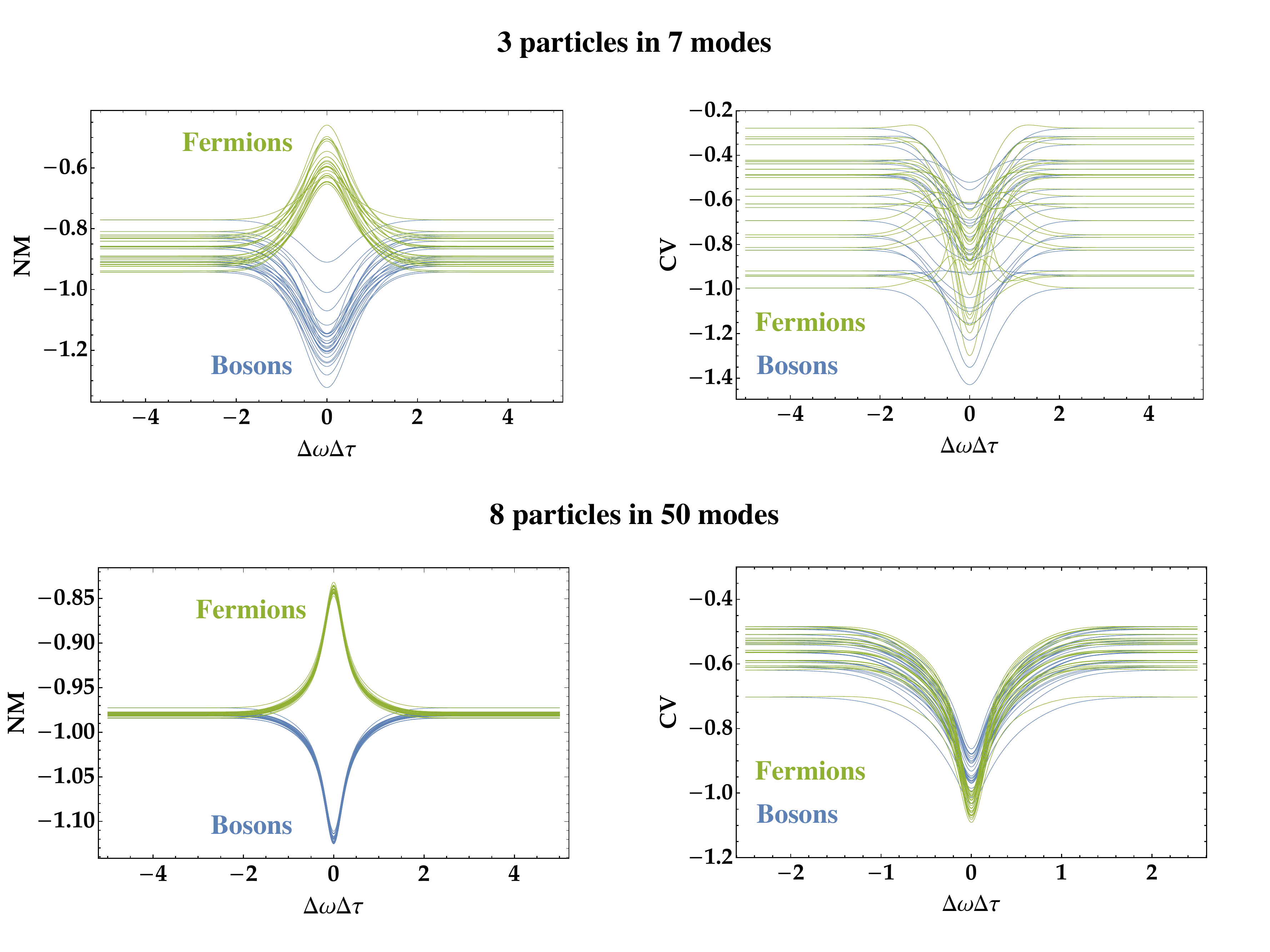}
\caption{Distinguishability transition as seen by the normalised mean $NM$ (\ref{eq:NM}) and the coefficient of variation $CV$ (\ref{eq:CV}), by varying the time delay relative to the spectral width of the wave packets $\Delta \omega \Delta \tau$. Top panels show the case where three bosonic (blue curves) and fermionic (green curves) particles were injected in 25 randomly chosen seven-mode interferometers. Bottom panels show eight bosonic (blue curves) and fermionic (green curves) particles were injected in 25 randomly chosen 50-mode interferometers. Data are obtained by averaging the correlations for all possible output ports of each interferometer.}
\label{fig:Dist-Trans}
\end{figure}

In Fig.~\ref{fig:Dist-Trans} we observe a much more systematic behaviour when we evaluate the normalised mean $NM$ for many different interferometers. For bosonic particles, $NM$ increases monotonically when the particles become more distinguishable, i.e.~when the values of $\Delta \omega \Delta \tau$ divert further from zero. In the fermionic case, the behaviour is exactly the opposite, as is to be expected from (\ref{eq:corrDistTransB}, \ref{eq:corrDistTransF}). For small interferometers, the feature is qualitatively robust, although we still see significant fluctuations between the different curves (i.e.~different interferometers). However, as we increase the size of the system, we gradually acquire a more robust quantitative behaviour, which manifests by the results for eight particles in several different 50-mode interferometers. Indeed, the different curves are in such a good agreement that it becomes hard to distinguish them. 

For the coefficient of variation, $CV$, we see a much wilder behaviour for the small interferometers. Generally, $CV$ reaches a minimal value for maximal indistinguishability, i.e.~$\Delta \tau \Delta \omega = 0$. It is more remarkable that the bosonic and fermionic curves seem to behave in a qualitatively similar way. Again, when we explore the regime of more particles in larger interferometers ($n=8$ and $m=50$ in this case), we observe an increased robustness. The observed curves for $CV$ seem to be far less dependent on the specific details of the realised interferometers, and depend more on some coarse-grained parameters.

\begin{figure}
\centering
\includegraphics[width=0.79 \textwidth]{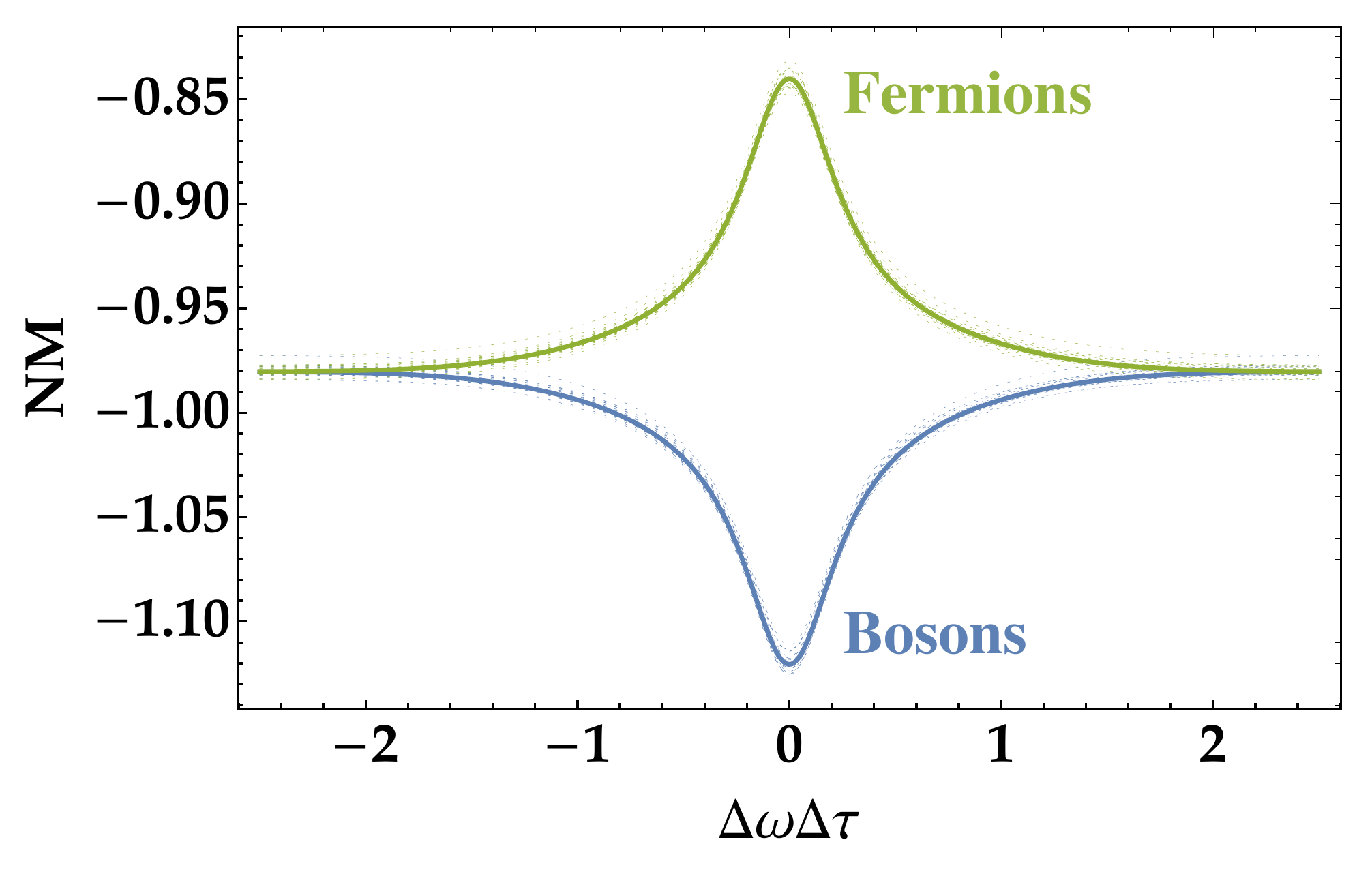}
\caption{Random matrix approximation (solid lines) for the normalised mean $NM$ (\ref{eq:NM}, \ref{eq:m1PartialDist}), compared to the data of Fig.~\ref{fig:Dist-Trans} that were obtained for individual interferometers (dashed lines). The distinguishability transition is seen by varying the time delay between consecutive particles relative to the spectral width of the wave packets, i.e.~$\Delta \omega \Delta \tau$.}
\label{fig:Dist-Trans-NM-RMT}
\end{figure}

\begin{figure}
\centering
\includegraphics[width=0.79 \textwidth]{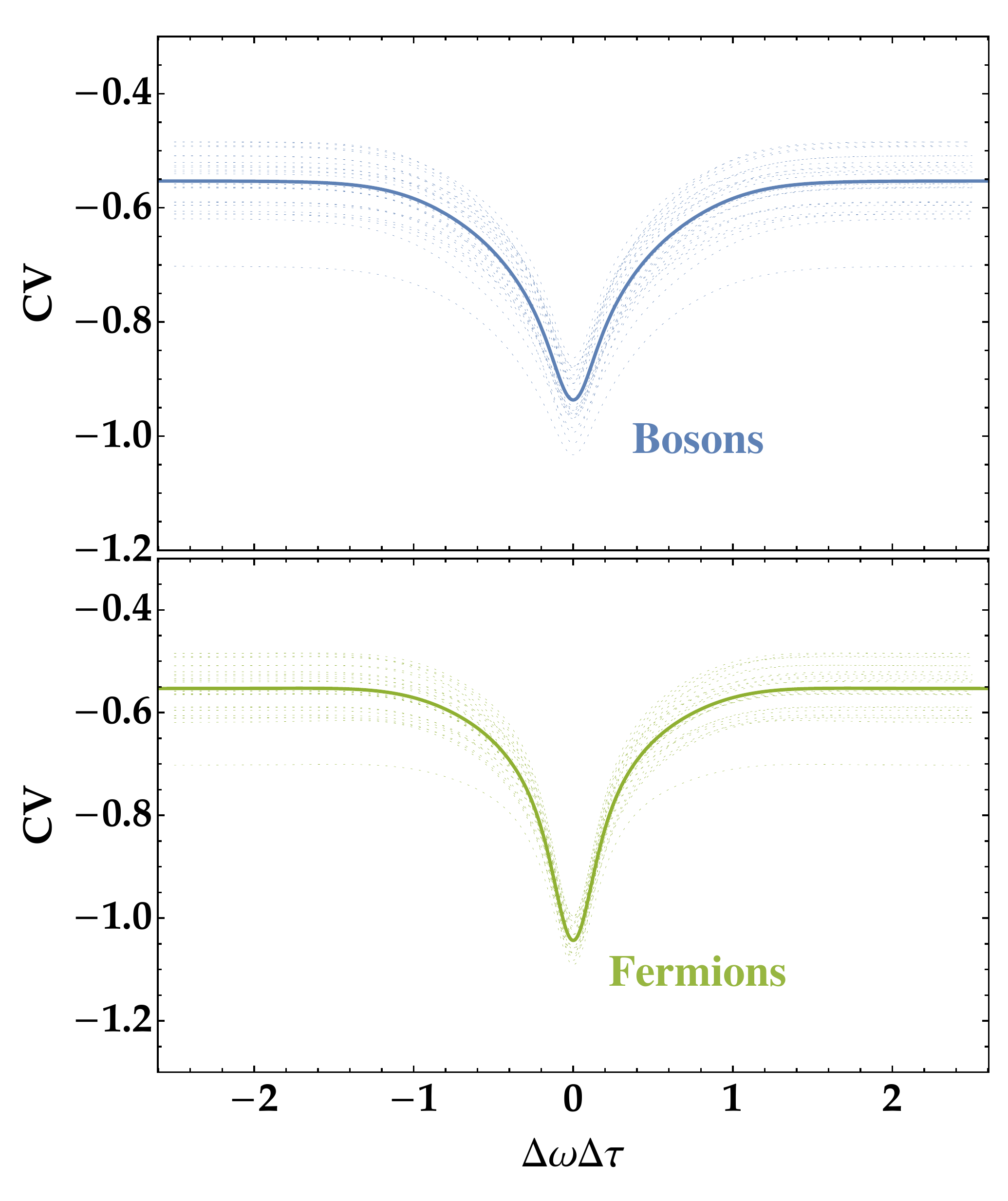}
\caption{Random matrix approximation (solid lines) for the coefficient of variation $CV$ (\ref{eq:CV}, \ref{eq:PDTrainSecondMoment}), compared to the data of Fig.~\ref{fig:Dist-Trans} that were obtained for individual interferometers (dashed lines). The distinguishability transition is seen by varying the time delay between consecutive particles relative to the spectral width of the wave packets, i.e.~$\Delta \omega \Delta \tau$.}
\label{fig:DistTransNMCV}
\end{figure}

The increased robustness of $NM$ and $CV$ to differences between various randomly chosen interferometers indicates clearly that random matrix methods should be virtuous to reproduce these results and obtain an analytical understanding. In other words, we expect (\ref{eq:m1PartialDist}) and (\ref{eq:PDTrainSecondMoment}) to be good approximations for the observed curves. This expectation is clearly lived up to in Figs.~\ref{fig:Dist-Trans-NM-RMT} and \ref{fig:DistTransNMCV}, where the dashed line indicate single random realisations of the interferometers, and the solid line shows the random matrix prediction. %Finally, Fig.~\ref{fig:DistTransNMCV} shows more clearly that there is, indeed, only a small difference between the fermionic and bosonic indistinguishability transition as seen by $CV$.

%\begin{figure}
%\centering
%\includegraphics[width=0.69 \textwidth]{Dist-Trans-CV-RMT-BF-n8-m50.pdf}
%\caption{Random matrix approximation for the coefficient of variation $CV$ (\ref{eq:CV}, \ref{eq:PDTrainSecondMoment}) for bosons (blue solid curve), compared to that for fermions (green dashed curve).}
%\label{fig:DistTransNMCV}
%\end{figure}

\begin{figure}
\centering
\includegraphics[width=0.99 \textwidth]{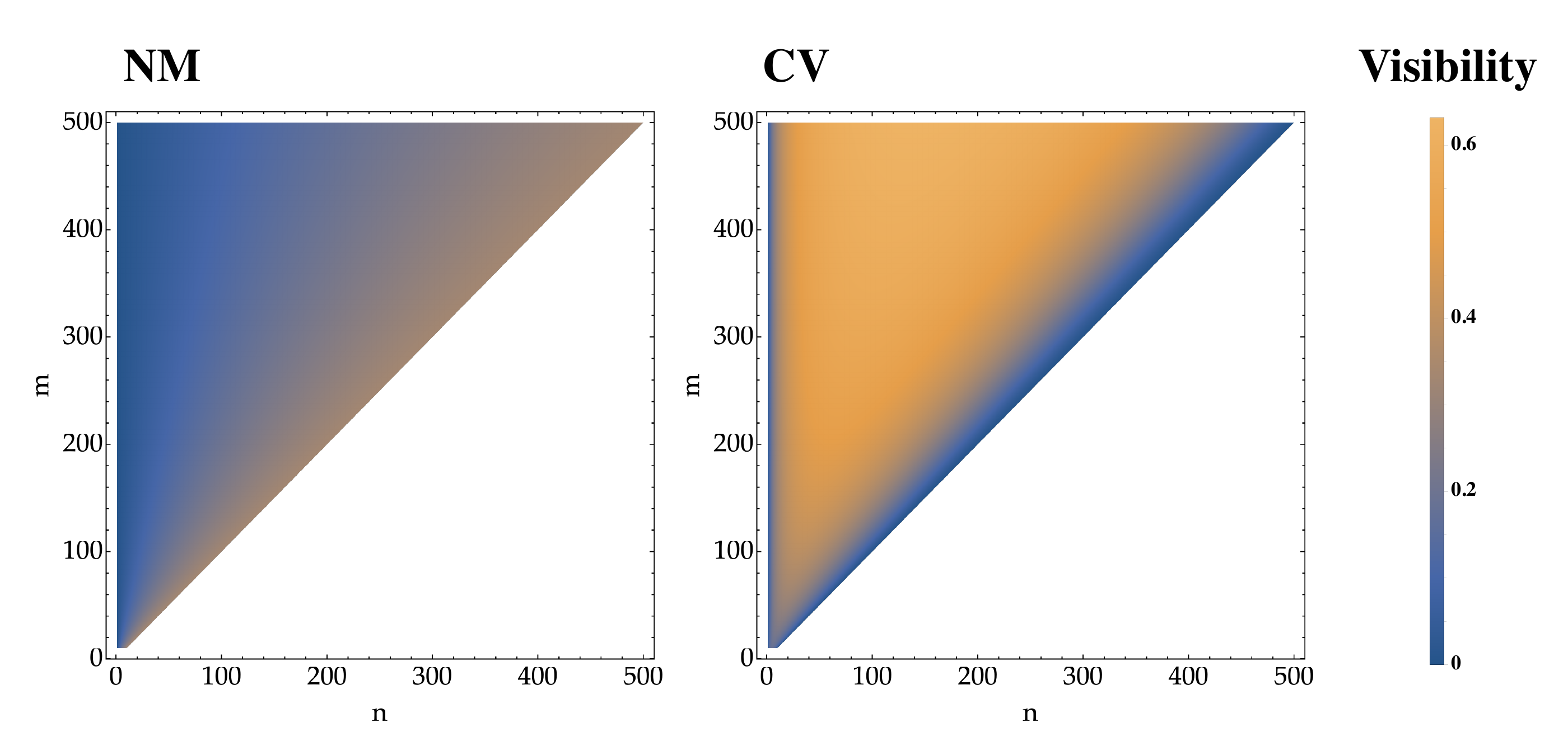}
\caption{Density plot for the visibility of the distinguishability transition for the normalised mean $NM$ (\ref{eq:VisNM}) and for the coefficient of variation $CV$ (\ref{eq:VisCV}), for varying numbers $n$ of input particles, and sizes $m$ of the interferometers. Brighter (more orange) colours indicate higher visibility. All data are obtained from the bosonic random matrix approximations (\ref{eq:m1PartialDist}) and (\ref{eq:PDTrainSecondMoment}).}
\label{fig:VisibPlots}
\end{figure}

Finally, one may wonder why it is necessary to consider the coefficient of variation $CV$, when $NM$ clearly gives a more robust result. The answer is provided in Fig.~\ref{fig:VisibPlots}, which shows that $CV$ has a far greater interferometric visibility in the regions of interest ($n\ll m$). The interferometric visibility is defined as 
\begin{align}
{\cal V}_{NM} &= \frac{\abs{NM^{\rm ind}_{B/F} - NM^{\rm dis}_{B/F}}}{\abs{NM^{\rm ind}_{B/F} + NM^{\rm dis}_{B/F}}},\label{eq:VisNM}\\
{\cal V}_{CV} &= \frac{\abs{CV^{\rm ind}_{B/F} - CV^{\rm dis}_{B/F}}}{\abs{CV^{\rm ind}_{B/F} + CV^{\rm dis}_{B/F}}},\label{eq:VisCV}
\end{align}
and serves to quantify the size of the observed indistinguishability transition. In Fig.~\ref{fig:VisibPlots} only the bosonic case is shown, but, from the random matrix predictions, it is easy to see that the result for the fermionic case is qualitatively similar. One could argue that these visibilities should in the first place be compared to the size of the statistical fluctuation around the random matrix results. However, a typical experiment does contain other sources of errors, for instance due to the fact that the measured correlations are extracted from a finite set of measurements \cite{Flamini-finite-2019}. Such additional errors might make it hard to observe a statistically significant feature of small visibility, such as for the normalised mean $NM$. \\

\begin{figure}
\centering
\includegraphics[width=0.69 \textwidth]{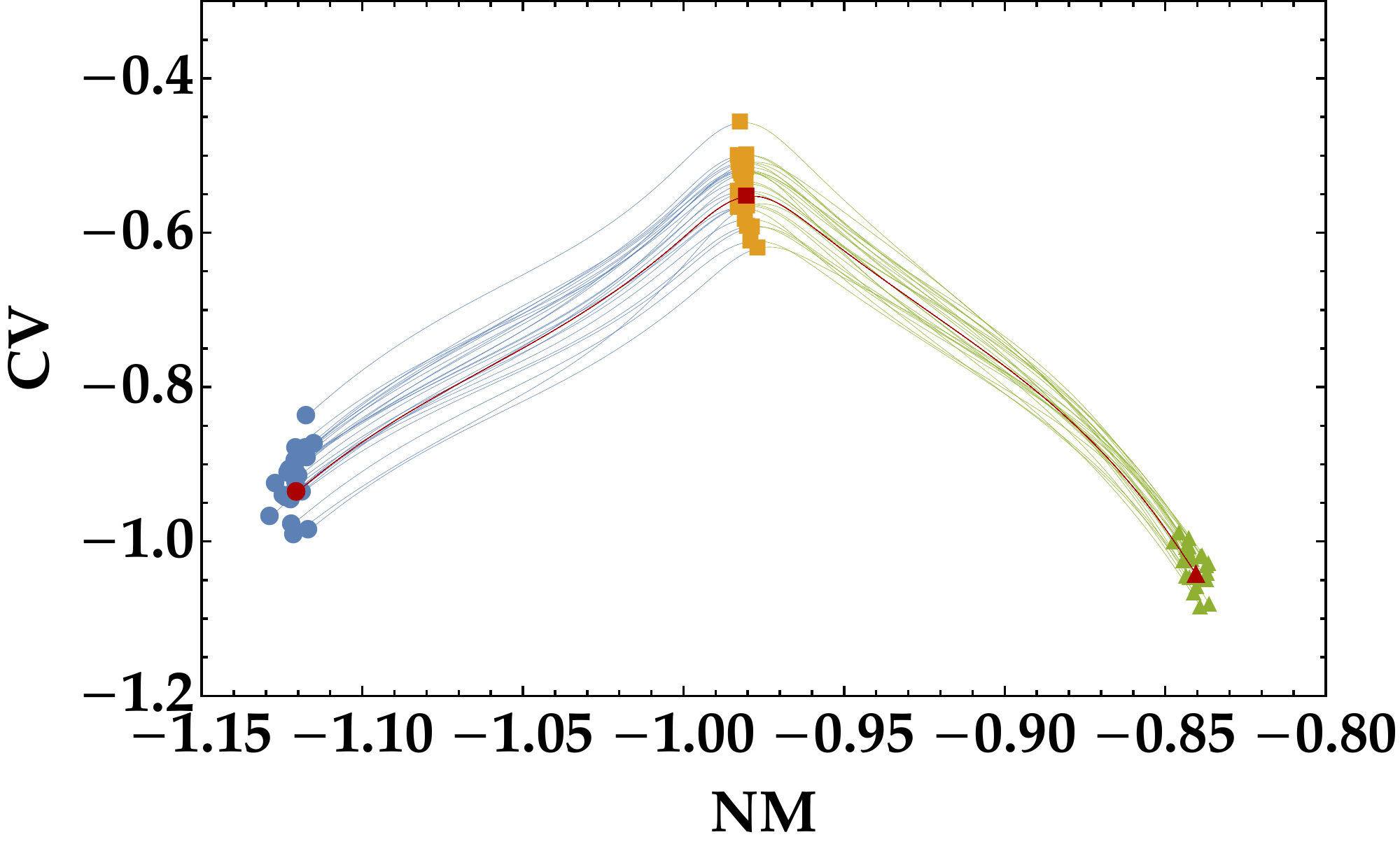}
\caption{Scatter plot showing the same type of data as Fig.~\ref{fig:Cdatasets}. The normalised mean $NM$ (\ref{eq:NM}) is represented on the horizontal axis and the coefficient of variation $CV$ (\ref{eq:CV}) on vertical axis. Points were obtained by injecting eight bosonic Fock states (blue dots), fermionic number states (green triangles), and distinguishable particles (orange squares) into a 50-mode interferometer that was randomly chosen from the Haar measure. The random matrix predictions (\ref{eq:momentsB} - \ref{eq:momentsF}, \ref{eq:RMTmom2b} - \ref{eq:RMTmom2d}) for each particle type are shown by a large red dot. Curves connecting the different points show how the position of the point changes by varying $\Delta \omega \Delta \tau$, hence making the particles more distinguishable.}
\label{fig:CloudPlot_Trajectories}
\end{figure}

The phenomenology discussed throughout this section can be retrieved within the final Fig.~\ref{fig:CloudPlot_Trajectories}, where we show the different trajectories that are followed by the points in Fig.~\ref{fig:Cdatasets} as we gradually make the particles more distinguishable. This figure nicely captures that $CV$ is a good quantifier for the indistinguishability transition, whereas it is far less suitable to discriminate between bosons and fermions. Hence, a combined study of $NM$ and $CV$ provides more information.\\

These results are only the tip of the iceberg. After all, we did only consider separable number states at the input, and only studied the lowest order correlations. It is to be expected that one may reach a far richer zoo of phenomena, once one allows entanglement between the particles. It is likely that we will observe a different behaviour when one includes entanglement in the external DOF and entanglement in the internal DOF. Furthermore, higher order correlations may allow us to explore much more intricate aspects of partial distinguishability, such as those discussed in \cite{PhysRevLett.118.153603,Shchesnovich:2018aa}.

Partial distinguishability is known to be an important factor in the breakdown of the computational complexity of Boson Sampling \cite{Shchesnovich-2019,Moylett-2019,Renema:2018aa,Moylett:2018aa}, and, therefore, it is indispensable to have good benchmarks at our disposal to characterise these effects. In this light, more general measures of indistinguishability such as \cite{Dittel-2019} can provide additional insights that may help improve and extend the statistical signatures of many-particle interference.

\section{Conclusions and outlook}

Throughout this Tutorial, we have extensively introduced the mathematical framework that describes many-particle quantum systems. This framework was then used to introduce the phenomenon of many-particle interference, for which we finally developed several observable signatures. Nevertheless, it should be stressed that this Tutorial is only intended to equip the interested reader with a toolbox to approach the field. The results that are presented and the literature that is cited is certainly not all-embracing. For a broader overview, we refer the reader to recent reviews on photonic quantum technologies \cite{Pan:2012aa,Flamini:2018aa} and quantum supremacy \cite{Lund:2017aa,Harrow:2017aa,Brod-2019}.

The signatures of many-particle interference that were described in this Tutorial are likely to be just scratch the surface of what many-particle interference has to offer. On the level of fundamental physics, there is a dire need for a general framework to characterise the distinguishability of particles. Even though we generally know how transfer probabilities change due to the gradual onset of distinguishability \cite{tillmann_generalized_2015,tamma_multiboson_2015,shchesnovich_partial_2015,tichy_sampling_2015}, it is also known that distinguishability can appear in many forms \cite{Shchesnovich:2018aa}. Even though progress is being made \cite{Dittel-2019}, it remains an open question whether there are tractable measures for understanding how partial distinguishability distorts the many-particle interference pattern. It is conjectured that higher-order correlations will play a role in answering this question. This question is of particular relevance since we are gradually starting to understand how partial distinguishability distorts the computation complexity of Boson Sampling \cite{Shchesnovich-2019,Moylett-2019,Renema:2018aa,Moylett:2018aa}.

Apart from the questions that still surround the standard Boson Sampling setup, there is by now also a whole range of extensions to the framework. A first set of protocols have explored Boson Sampling with continuous variable detection \cite{PhysRevA.96.032326,PhysRevA.96.062307}, also demonstrating the computational complexity in this setting. Most attention has been devoted to the so-called Gaussian Boson Sampling \cite{PhysRevLett.119.170501}, where one injects non-classical light into a linear-optical interferometer, rather than photonic Fock states. This setup is gradually becoming more prominent, due to its inherent scalability and potential applications beyond merely proving a quantum computational advantage \cite{Huh:2015aa,Clements:2018aa,Arrazola:2018aa,Arrazola:2018ab,Bradler:2018aa}. It was shown that the statistical signature for many-particle interference can be extended to serve as a benchmark for Gaussian Boson Sampling \cite{Phillips:2019aa}. However, it is currently unclear which physical phenomenon really makes Gaussian Boson Sampling hard. It is currently still an open question whether these extended setups can be mapped to the notion of many-particle interference as presented throughout this Tutorial .

What all these different setups have in common, however, is the presence of at least one non-Gaussian element in the setup. This can be either the state \cite{PhysRevA.96.032326,PhysRevA.96.062307}, the measurement process \cite{PhysRevLett.119.170501}, or both \cite{aaronson_computational_2013}. This observation makes a connection to an essential fact in continuous-variable quantum information processing: non-Gaussian features are necessary to reach a quantum computational advantage \cite{Bartlett:2002aa,Mari:2012aa,rahimi-keshari_sufficient_2016}. This connection became even more pronounced when it was pointed out that the scenario of fermionic many-particle interference does, in fact, only contain Gaussian elements. However, non-Gaussian features represent only a necessary condition for reaching a quantum advantage. What other elements are required, and how non-Gaussianity should be used to reach a regime that is intractable for a classical computer, both remain an important open question.\\

Finally, let us ponder upon the limitations of the setting that was considered in this Tutorial. We initially stressed the important lack of interactions between the particles we study. Interactions would also be a potential source of non-Gaussian effects, but the downside is that they are hard to control (both theoretically and experimentally). Nevertheless, statistical signatures of many-particle interference have been used in the context of interacting particles \cite{1367-2630-19-12-125015,Brunner:2018aa}. Hence, it is reasonable to expect that also sampling from such setups is computationally hard, but turning this conjecture into a formal proof is far from evident.

Another aspect that was not considered throughout the Tutorial is the possibility of adding many-particle entanglement to the mix. The formalism that was introduced in Section \ref{sec:formalism}---and in particular the discussion of Section \ref{sec:Distinguish}---provides a good starting point for studying the phenomenon of many-particle entanglement. The identity (\ref{eq:isoHilbert}) is crucial for understanding many of the subtleties in the ongoing debate; for some recent developments, see for example \cite{Benatti:2012aa,Benatti:2012ab,tichy_entanglement_2013,Benatti:2017aa,Lo-Franco:2018aa}. How many-particle interference processes are affected by the presence of many-particle entanglement, be it either in the internal or external DOF, is still a largely open question. It has been shown \cite{Beenakker:2009aa} that many-particle interference effects can serve to detect entanglement between a pair of particles by using methods that are closely related to the statistical signatures of Section \ref{sec:RMT}. Furthermore, one may wonder whether adding entanglement between the incoming bosons can increase the computational complexity of Boson Sampling, which may allow to reach a computational advantage with a smaller number of photons (at the cost of having to entangle them).\\

It is hard to deduce what exactly Dirac had in mind when he wrote that {\em ``Interference between two different photons never occurs.''} \cite{Dirac-1930}. Therefore, it is crude to go as far as to say that Dirac was blatantly mistaken. However, if the reader is to take away one main message from this Tutorial, it is that photons (and all identical particles for that matter) can be made to interfere, and the signatures of these interferences are gradually being unveiled.

\section*{Acknowledgements}
First of all, I profoundly thank Fulvio Flamini for his careful reading and useful feedback, which significantly improved this Tutorial. A next word of gratitude goes out to Jack Kuipers, for teaching me how to perform the random matrix calculation that appear throughout this Tutorial. Over the years, I greatly enjoyed many stimulating discussions with colleagues, among which I want to explicitly thank Juan-Diego Urbina, Gabriel Dufour, Eric Brunner, Fabio Sciarrino, David Philips, Jan Sperling, Ulysse Chabaud, Raul Garcia-Patron, Jelmer Renema, Juliane Klatt, and Chahan Kropf. Their combined insights and questions have helped shape this Tutorial. Last but not least, I thank my PhD supervisors Andreas Buchleitner and Mark Fannes for inspiration, guidance, and patience.

This work was made possible by the financial support of research fellowship WA 3969/2-1 from the German Research Foundation (DFG).

\section*{References}

%\bibliography{Bib_Thesis_Update}
%\bibliographystyle{unsrt-abbrv}

\appendix

\section{Direct sums and tensor products}\label{sec:appSumProd}

Here we take a moment to briefly introduce some basic notions of the direct sum ``$\oplus$'' and the tensor product ``$\otimes$''. We first go over their mathematical structure, and subsequently discuss the physical meaning of these constructs.

\subsection{Mathematical structure}

Both the direct sum and the tensor product are operations that act on a pair of vectors in a Hilbert space, which result in a vector in a larger space. When defined on the level of Hilbert space, these operations serve to create a larger sapce with a certain structure.

\subsubsection{The direct sum} 

First, we consider the direct sum for finite-dimensional Hilbert spaces ${\cal H}_1 = \mathbb{C}^{d_1}$ and ${\cal H}_2 =\mathbb{C}^{d_2}$. Let us start by choosing vectors $\ket v \in \mathbb{C}^{d_1}$ and $\ket w \in \mathbb{C}^{d_2}$. In some arbitrary basis, these vectors can be represented as $\ket v = (v_1, \dots, v_{d_1})^t$ and $\ket w = (w_1, \dots, w_{d_2})^t$. The general rule is than that we can represent 
\begin{equation}\label{eq:vplusw}
\ket v \oplus \ket w = (v_1, \dots, v_{d_1}, w_1, \dots, w_{d_2})^t.
\end{equation}
On immediately seems that $\ket v \oplus \ket w$ is a vector of dimension $d_1 + d_2$. 

The direct sum for vectors can then be generalised to a direct sum for Hilbert spaces by defining
\begin{equation}
{\cal H}_1 \oplus {\cal H}_2 = \{\ket v\oplus \ket w \mid \ket v \in {\cal H}_1, \ket w \in {\cal H}_2\}.
\end{equation}
We see that ${\rm dim} {\cal H}_1 \oplus {\cal H}_2 = d_1+d_2$, and as a consequence we find the isomorphism $\mathbb{C}^{d_1} \oplus \mathbb{C}^{d_2} \cong \mathbb{C}^{d_1 + d_2}.$ Note that (\ref{eq:vplusw}) provides an explicit construction of this isomorphism. Furthermore, there is a natural basis of ${\cal H}_1 \oplus {\cal H}_2$ with respect to the direct sum structure. When we choose a basis ${\cal E}$ of ${\cal H}_1$ and ${\cal F}$ of ${\cal H}_2$, we can construct the basis 
\begin{equation}
\{\ket e\oplus \ket 0 \mid \ket e \in {\cal E}\} \cup \{\ket 0\oplus \ket f \mid \ket f \in {\cal F}\},
\end{equation}
where $0$ represents the zero vector. 

Importantly, we can also use direct sums to decompose a Hilbert space and give it more structure. For example, we can consider ${\cal H} = \mathbb{C}^{d}$, and use the direct sum structure to decompose it as 
\begin{equation}\label{eq:decompDirectSumApp}
\mathbb{C}^{d} \cong \underbrace{\mathbb{C} \oplus \dots \oplus \mathbb{C}}_{\times d}.
\end{equation}
As a simple example, let us consider a qubit $\mathbb{C}^{2}$. We can then choose the basis $\{\ket{0}, \ket{1}\}$, such that a general state vector $\ket{\phi}$ can be represented as 
\begin{equation}
\ket{\phi} = \alpha \ket{0} + \beta \ket{1} = \begin{pmatrix}\alpha \\ \beta \end{pmatrix} = \alpha \oplus \beta,
\end{equation}
with $\alpha, \beta \in \mathbb{C}$. Note that this decomposition depends on our chosen basis $\{\ket{0}, \ket{1}\}$, were we to choose another basis, the definition would be different. This leads to the more general observation that the decomposition (\ref{eq:decompDirectSumApp}) must be interpreted in an associated basis.\\

Finally, we note that the above structures can also be defined when ${\cal H}_1$ and ${\cal H}_2$ are infinite-dimensional. We can then take $\ket f \in {\cal H}_1$ and $\ket g \in {\cal H}_2$, and generally represent the elements of the Hilbert space as functions $f(x)$ and $g(x)$, respectively, over some domain $\Lambda_1$ (for ${\cal H}_1$) and $\Lambda_2$ (for ${\cal H}_2$), with $\Lambda_1 \cap \Lambda_2 = \emptyset$. We can then define the function
\begin{equation}
f \oplus g : \Lambda_1\cup \Lambda_2 \rightarrow \mathbb{C}: x \mapsto  \begin{cases}
f(x), &x \in \Lambda_1\\
g(x), &x \in \Lambda_2.
\end{cases}
\end{equation}
Just like in the finite-dimensional case we can use then define ${\cal H}_1 \oplus {\cal H}_2$, construct a natural basis, et cetera. Here, we will limit ourselves to the important example of square integrable functions. Let us consider the square integrable functions in a volume of space $\Lambda_1$, i.e.~${\cal L}^2(\Lambda_1)$, and in a second volume of space $\Lambda_2$, i.e.~${\cal L}^2(\Lambda_2)$. We now find the identity
\begin{equation}
{\cal L}^2(\Lambda_1) \oplus {\cal L}^2(\Lambda_2) \cong {\cal L}^2(\Lambda_1\cup \Lambda_2).
\end{equation}
We have clearly increased the space through the direct sum, but the dimension of the domain has remained unchanged.

\subsubsection{The tensor product} 

A second natural construct to increase the size of a Hilbert space is the tensor product. We again start by considering finite-dimensional Hilbert spaces ${\cal H}_1 = \mathbb{C}^{d_1}$ and ${\cal H}_2 =\mathbb{C}^{d_2}$, choosing vectors $\ket v \in \mathbb{C}^{d_1}$ and $\ket w \in \mathbb{C}^{d_2}$. We, again, represent the vectors in some basis $\ket v = (v_1, \dots, v_{d_1})^t$ and $\ket w = (w_1, \dots, w_{d_2})^t$ to describe the general rule for the tensor product
\begin{equation}\label{eq:vplusw}
\ket v \otimes \ket w = (v_1 w_1, \dots, v_1 w_{d_2},v_2 w_1, \dots, v_2 w_{d_2}, \dots, v_{d_1} w_1, \dots, v_{d_1} w_{d_2} )^t.
\end{equation}
The resulting vector is now of dimension $d_1d_2$. We can use this construction to define a new Hilbert space
\begin{equation}
{\cal H}_1 \otimes {\cal H}_2 = \{\ket v\otimes \ket w \mid \ket v \in {\cal H}_1, \ket w \in {\cal H}_2\},
\end{equation}
with ${\rm dim}{\cal H}_1 \otimes {\cal H}_2 = d_1d_2$. This space comes with an associated natural basis that is considerably different from that of the direct sum structure; when we choose a basis ${\cal E}$ of ${\cal H}_1$ and ${\cal F}$ of ${\cal H}_2$, we obtain the basis 
\begin{equation}
\{\ket e \otimes \ket f \mid \ket e \in {\cal E}, \ket f \in {\cal F}\},
\end{equation}
for ${\cal H}_1 \otimes {\cal H}_2$. Contrary to the direct sum, we cannot decompose any possible finite-dimensional Hilbert space as a long tensor product.\\

The construction for infinite dimensional spaces is also considerably different from the direct sum. When we consider $\ket f \in {\cal H}_1$ and $\ket g \in {\cal H}_2$, we can represent the vectors as functions $f: \Lambda_1 \rightarrow \mathbb{C}$ and $g: \Lambda_2 \rightarrow \mathbb{C}$, respectively. We can then define 
\begin{equation}
f \oplus g : \Lambda_1 \times \Lambda_2 \rightarrow \mathbb{C}: (x_1,x_2) \mapsto f(x_1)g(x_2).
\end{equation}
Note that the domain of the function $f \oplus g$ is of a higher dimensions than the domains of $f$ and $g$. A particularly important example is found, again, for the square integrable functions, where we now consider the cases ${\cal L}^2(\mathbb{R}^{d_1})$ and ${\cal L}^2(\mathbb{R}^{d_2})$. It can then be shown that
\begin{equation}
{\cal L}^2(\mathbb{R}^{d_1}) \otimes {\cal L}^2(\mathbb{R}^{d_2}) \cong {\cal L}^2(\mathbb{R}^{d_1+d_2}).
\end{equation}

\subsection{Physical interpretation}
Physically, the direct sum and the tensor product have a very different meaning, even though they are often intertwined in one way or the other. The golden rule rule to keep in mind is that, generally, {\em tensor products indicate different DOF}. This may be the case for different distinguishable particles, or different spins in a spin chain. Each spin or particle comes with its own small Hilbert space, and the total system is then described by the tensor product of all these small Hilbert spaces. Furthermore, one also uses the tensor product to combine the wide range of different DOF for a single particle. In the case of a photon, one may for example take a tensor product of its polarisation DOF, spatial DOF, and spectral DOF (note that for a photon one typically refers to these DOF as {\em modes}).\\

The interpretation of the direct sum is less straightforward, since it does not increase the number of DOF, but rather the values that these DOF can take. This may sound somewhat exotic, but in some settings this idea is quite natural. When, for example, we use a tight-binding model to describe an atom in a one-dimensional optical lattice, we have essentially one spatial dimension, with a variety of different possible measurement outcomes (i.e.~the different lattice sites). We can divide the set of lattice sites in two groups, for example ``left'' and ``right''. This gives us a Hilbert space for the left part of the system, and another one for the right part of the system. The Hilbert space of the total lattice can than be retrieved by taking the direct sum of the left and right Hilbert spaces. 

Let us consider, as an additional example, the case of a photon with some spectral DOF. Imagine that we have a frequency comb (light with a discrete set of equidistantly spaced frequencies) at our disposal, with central frequency $\omega_0$. We can separately describe the frequencies that are closer to the red, and those that are closer to the blue, resulting in two independent Hilbert space (also referred to as mode spaces in the optics context). To describe the whole frequency comb, we then take the direct sum of these different mode spaces.\\

\section{Moments of correlations}
\subsection{The Fourier interferometer}\label{sec:fourier}
A notable example for which moments of the correlations between the interferometer's output ports can be calculated explicitly is the Fourier interferometer. This circuit implements a discrete Fourier transformation and has played in important role in the development of suppression laws \cite{tichy_many-particle_2012}. This interferometer is given by the unitary matrix $F$ with components 
\begin{equation}
F_{o j} = \frac{1}{\sqrt{m}} \exp \left(2\pi i \frac{(o-1)(j-1)}{m}\right),
\end{equation}
such that we find correlations
\begin{align}
&C^B_{o_1 o_2} = -\frac{n}{m^2} + \frac{1}{m^2} \sum^n_{\substack{k,l = 1\\ k\neq l}}\exp \left(2\pi i \frac{(o_2-o_1)(j_k-j_l)}{m}\right),\\
&C^T_{o_1 o_2}  = \frac{n}{m^2} + \frac{1}{m^2} \sum^n_{\substack{k,l = 1\\ k\neq l}}\exp \left(2\pi i \frac{(o_2-o_1)(j_k-j_l)}{m}\right),\\
&C^F_{o_1 o_2}  = -\frac{n}{m^2} - \frac{1}{m^2} \sum^n_{\substack{k,l = 1\\ k\neq l}}\exp \left(2\pi i \frac{(o_2-o_1)(j_k-j_l)}{m}\right),\\
&C^D_{o_1 o_2}  = -\frac{n}{m^2}
\end{align}
Because $C_{o_1 o_2} = C_{o_2 o_1},$ we can rewrite 
\begin{equation}
m_q = \frac{1}{m(m-1)} \sum^m_{\substack{o_1,o_2 = 1\\ o_1\neq o_2}} \left(C_{o_1 o_2}\right)^q,
\end{equation}
which is particularly convenient, because it allows us to use the identity
\begin{equation}\label{eq:usefullIdentityAppendix}
\sum^m_{\substack{o_1,o_2 = 1\\ o_1\neq o_2}} \exp \left(2\pi i \frac{(o_2-o_1)(j_k-j_l)}{m}\right) = -m,
\end{equation}
which holds regardless of the values of the integers $j_k$ and $j_l$. Invoking this identity leads to 
\begin{align}
&m_1^B= -\frac{n}{m^2} - \frac{n(n-1)}{m^2(m-1)},\\
&m_1^T = \frac{n}{m^2} - \frac{n(n-1)}{m^2(m-1)},\\
&m_1^F  = -\frac{n}{m^2} + \frac{n(n-1)}{m^2(m-1)},\\
&m_1^D  = -\frac{n}{m^2}.
\end{align}

A similar argument can be used to calculate the second moment, which we will only present explicitly for the bosonic case. Note first of all that
\begin{align}
\left(C^B_{o_1 o_2}\right)^2& = -\frac{n^2}{m^4} - \frac{2 n}{m^4} \sum^n_{\substack{k,l = 1\\ k\neq l}}\exp \left(2\pi i \frac{(o_2-o_1)(j_k-j_l)}{m}\right)\\ 
&+ \frac{1}{m^4} \sum^n_{\substack{k,l,p,q = 1\\ k\neq l \\ p\neq q}}\exp \left(2\pi i \frac{(o_2-o_1)(j_k-j_l + j_p - j_q)}{m}\right),\nonumber
\end{align}
The calculation of 
\begin{equation}
\frac{1}{m^5(m-1)} \sum^m_{\substack{o_1,o_2 = 1\\ o_1\neq o_2}} \sum^n_{\substack{k,l,p,q = 1\\ k\neq l \\ p\neq q}}\exp \left(2\pi i \frac{(o_2-o_1)(j_k-j_l + j_p - j_q)}{m}\right)
\end{equation}
is rather cumbersome, since we have to count all the terms where $j_k-j_l + j_p - j_q = 0$, a number which is found to be $(2n-1)(n-1)n/3$. For all of these terms, we find that $\exp \left(2\pi i (o_2-o_1)(j_k-j_l + j_p - j_q)/m\right) = 1$. For all other terms, we can again invoke the identity (\ref{eq:usefullIdentityAppendix}). This leads to the result
\begin{align}
&\frac{1}{m^5(m-1)} \sum^m_{\substack{o_1,o_2 = 1\\ o_1\neq o_2}} \sum^n_{\substack{k,l,p,q = 1\\ k\neq l \\ p\neq q}}\exp \left(2\pi i \frac{(o_2-o_1)(j_k-j_l + j_p - j_q)}{m}\right)\nonumber\\
&=\frac{(2n-1)(n-1)n}{3m^4} - \frac{(n-1) n (n (3 n-5)+1)}{3m^4(m-1)},
\end{align}
which can be used to reach the following result for the second moment:
\begin{equation}
m_2^B =-\frac{n^2}{m^4} + \frac{2 n^2 (n-1)}{m^4(m-1)} +  \frac{(2n-1)(n-1)n}{3m^4} - \frac{(n-1) n (n (3 n-5)+1)}{3m^4(m-1)}.
\end{equation}
The calculation of $m_2^B$ shows how the expressions for these moments gradually get more complicated.

\subsection{The special case of the first moment}\label{sec:first_moment}
It is generally possible to simplify the first moment of the correlation between output detectors of a single interferometer, i.e.~the case where $q = 1$ in (\ref{eq:moment}). At the heart of this simplification lies the following identity for a unitary $m \times m$ matrix $U$:
\begin{equation}
\sum_{\substack{o_2 = 1\\ o_2 \neq o_1}}^m U_{o_2 i}U^*_{o_2 j} = \delta_{i,j} - U_{o_1 i}U^*_{o_1 j}.
\end{equation}
We can directly use this result to calculate 
\begin{align}
\nonumber\frac{1}{m(m-1)}&\sum_{\substack{o_2, o_1 = 1\\ o_2 \neq o_1}}^m \sum_{k=1}^n  \abs{U_{o_1 i_k}}^2\abs{U_{o_2 i_k}}^2\\ &= \frac{1}{m(m-1)}\sum_{k=1}^n \sum_{o_1=1}^m \abs{U_{o_1 i_k}}^2(1-\abs{U_{o_1 i_k}}^2)\\
& = \frac{n}{m(m-1)} - \frac{1}{m(m-1)} \sum_{k=1}^n\sum_{o_1=1}^m \abs{U_{o_1 i_k}}^4,
\end{align}
and
\begin{align}
\nonumber \frac{1}{m(m-1)}&\sum_{\substack{o_2, o_1 = 1\\ o_2 \neq o_1}}\sum_{\substack{k,l = 1\\ k\neq l}}^n U_{o_1 i_k}U_{o_2 i_l}U^*_{o_1 i_l}U^*_{o_2 i_k}\\ &= - \frac{1}{m(m-1)}\sum_{\substack{k,l = 1\\ k\neq l}}^n \sum_{o_1}^m \abs{U_{o_1 i_k}}^2\abs{U_{o_1 i_l}}^2\\
& = - \frac{1}{m(m-1)} \sum_{o_1=1}^m \left(\sum_{k=1}^n \abs{U_{o_1 i_k}}^2\right)^2 + \frac{1}{m(m-1)}\sum_{\substack{k=1}}^n \sum_{o_1=1}^m \abs{U_{o_1 i_k}}^4
\end{align}
A direct calculation then shows that
\begin{align}
m_1^B=& -\frac{n}{m(m-1)}  - \frac{1}{m(m-1)} \sum_{o_1=1}^m \left(\sum_{k=1}^n \abs{U_{o_1 i_k}}^2\right)^2 \\ &\quad+  \frac{2}{m(m-1)}\sum_{\substack{k=1}}^n \sum_{o_1=1}^m \abs{U_{o_1 i_k}}^4 ,\nonumber\\
m_1^T=& \frac{n}{m(m-1)}  - \frac{1}{m(m-1)} \sum_{o_1=1}^m \left(\sum_{k=1}^n \abs{U_{o_1 i_k}}^2\right)^2,\\
m_1^F =& -\frac{n}{m(m-1)}  + \frac{1}{m(m-1)} \sum_{o_1=1}^m \left(\sum_{k=1}^n \abs{U_{o_1 i_k}}^2\right)^2,\\
m_1^D =& -\frac{n}{m(m-1)} +  \frac{1}{m(m-1)}\sum_{\substack{k=1}}^n \sum_{o_1=1}^m \abs{U_{o_1 i_k}}^4.
\end{align}
Interestingly, this result directly shows a hierarchy that holds for every interferometer:
\begin{equation}
\frac{n}{m(m-1)} > m_1^T > m_1^F > m_1^D > m_1^B > -\frac{n}{m(m-1)}.
\end{equation}
We also uncover an interesting relation between the case of bosons in a number states, thermal bosons, and distinguishable particles:
\begin{equation}
m_1^B = m_1^T + 2 m_1^D.
\end{equation}
Note, finally, that \begin{equation} m_1^F = -m_1^T \end{equation}, which is a consequence of the general fact that $C_{o_1 o_2}^F = -C^T_{o_1 o_2}$ for any pair of output correlators in any interferometer. This fact can be traced back to (\ref{eq:pefectMatch}) and (\ref{eq:pefectMatchF}), since both cases result from a Gaussian state. In absence of any squeezing, we find that
\begin{align}
C^{T}_{o_1,o_2} &= \tr[\rho^T a^{\dag}(e_{o_1})a^{\dag}(e_{o_2})a(e_{o_2})a(e_{o_1})] - \tr[\rho^T a^{\dag}(e_{o_1})a(e_{o_1})] \tr[\rho^T a^{\dag}(e_{o_2})a(e_{o_2})]  \nonumber\\
&=  \abs{\tr[\rho^T a^{\dag}(e_{o_1})a(e_{o_2})]}^2\\
C^{F}_{o_1,o_2} &= \tr[\rho^F a^{\dag}(e_{o_1})a^{\dag}(e_{o_2})a(e_{o_2})a(e_{o_1})] - \tr[\rho^F a^{\dag}(e_{o_1})a(e_{o_1})] \tr[\rho^F a^{\dag}(e_{o_2})a(e_{o_2})]  \nonumber\\ 
&=  -\abs{\tr[\rho^F a^{\dag}(e_{o_1})a(e_{o_2})]}^2,
\end{align}
where we use the bosonic rule (\ref{eq:pefectMatch}) for $C^{T}_{o_1,o_2}$, and the fermionic rule (\ref{eq:pefectMatchF}) for $C^{F}_{o_1,o_2}$. To finally conclude that $C_{o_1 o_2}^F = -C^T_{o_1 o_2}$, note that the thermal states are constructed in a way such that $\tr[\rho^F a^{\dag}(\phi)a(\psi)] = \tr[\rho^T a^{\dag}(\phi)a(\psi)]$ for all $\phi$ and $\psi \in {\cal H}$.

\end{document}